

\catcode`\@=11


\message{Loading jyTeX fonts...}



\font\vptrm=cmr5
\font\vptmit=cmmi5
\font\vptsy=cmsy5
\font\vptbf=cmbx5

\skewchar\vptmit='177 \skewchar\vptsy='60
\fontdimen16 \vptsy=\the\fontdimen17 \vptsy

\def\vpt{\ifmmode\err@badsizechange\else
     \@mathfontinit
     \textfont0=\vptrm  \scriptfont0=\vptrm  \scriptscriptfont0=\vptrm
     \textfont1=\vptmit \scriptfont1=\vptmit \scriptscriptfont1=\vptmit
     \textfont2=\vptsy  \scriptfont2=\vptsy  \scriptscriptfont2=\vptsy
     \textfont3=\xptex  \scriptfont3=\xptex  \scriptscriptfont3=\xptex
     \textfont\bffam=\vptbf
     \scriptfont\bffam=\vptbf
     \scriptscriptfont\bffam=\vptbf
     \@fontstyleinit
     \def\rm{\vptrm\fam=\z@}%
     \def\bf{\vptbf\fam=\bffam}%
     \def\oldstyle{\vptmit\fam=\@ne}%
     \rm\fi}


\font\viptrm=cmr6
\font\viptmit=cmmi6
\font\viptsy=cmsy6
\font\viptbf=cmbx6

\skewchar\viptmit='177 \skewchar\viptsy='60
\fontdimen16 \viptsy=\the\fontdimen17 \viptsy

\def\vipt{\ifmmode\err@badsizechange\else
     \@mathfontinit
     \textfont0=\viptrm  \scriptfont0=\vptrm  \scriptscriptfont0=\vptrm
     \textfont1=\viptmit \scriptfont1=\vptmit \scriptscriptfont1=\vptmit
     \textfont2=\viptsy  \scriptfont2=\vptsy  \scriptscriptfont2=\vptsy
     \textfont3=\xptex   \scriptfont3=\xptex  \scriptscriptfont3=\xptex
     \textfont\bffam=\viptbf
     \scriptfont\bffam=\vptbf
     \scriptscriptfont\bffam=\vptbf
     \@fontstyleinit
     \def\rm{\viptrm\fam=\z@}%
     \def\bf{\viptbf\fam=\bffam}%
     \def\oldstyle{\viptmit\fam=\@ne}%
     \rm\fi}


\font\viiptrm=cmr7
\font\viiptmit=cmmi7
\font\viiptsy=cmsy7
\font\viiptit=cmti7
\font\viiptbf=cmbx7

\skewchar\viiptmit='177 \skewchar\viiptsy='60
\fontdimen16 \viiptsy=\the\fontdimen17 \viiptsy

\def\viipt{\ifmmode\err@badsizechange\else
     \@mathfontinit
     \textfont0=\viiptrm  \scriptfont0=\vptrm  \scriptscriptfont0=\vptrm
     \textfont1=\viiptmit \scriptfont1=\vptmit \scriptscriptfont1=\vptmit
     \textfont2=\viiptsy  \scriptfont2=\vptsy  \scriptscriptfont2=\vptsy
     \textfont3=\xptex    \scriptfont3=\xptex  \scriptscriptfont3=\xptex
     \textfont\itfam=\viiptit
     \scriptfont\itfam=\viiptit
     \scriptscriptfont\itfam=\viiptit
     \textfont\bffam=\viiptbf
     \scriptfont\bffam=\vptbf
     \scriptscriptfont\bffam=\vptbf
     \@fontstyleinit
     \def\rm{\viiptrm\fam=\z@}%
     \def\it{\viiptit\fam=\itfam}%
     \def\bf{\viiptbf\fam=\bffam}%
     \def\oldstyle{\viiptmit\fam=\@ne}%
     \rm\fi}


\font\viiiptrm=cmr8
\font\viiiptmit=cmmi8
\font\viiiptsy=cmsy8
\font\viiiptit=cmti8
\font\viiiptbf=cmbx8

\skewchar\viiiptmit='177 \skewchar\viiiptsy='60
\fontdimen16 \viiiptsy=\the\fontdimen17 \viiiptsy

\def\viiipt{\ifmmode\err@badsizechange\else
     \@mathfontinit
     \textfont0=\viiiptrm  \scriptfont0=\viptrm  \scriptscriptfont0=\vptrm
     \textfont1=\viiiptmit \scriptfont1=\viptmit \scriptscriptfont1=\vptmit
     \textfont2=\viiiptsy  \scriptfont2=\viptsy  \scriptscriptfont2=\vptsy
     \textfont3=\xptex     \scriptfont3=\xptex   \scriptscriptfont3=\xptex
     \textfont\itfam=\viiiptit
     \scriptfont\itfam=\viiptit
     \scriptscriptfont\itfam=\viiptit
     \textfont\bffam=\viiiptbf
     \scriptfont\bffam=\viptbf
     \scriptscriptfont\bffam=\vptbf
     \@fontstyleinit
     \def\rm{\viiiptrm\fam=\z@}%
     \def\it{\viiiptit\fam=\itfam}%
     \def\bf{\viiiptbf\fam=\bffam}%
     \def\oldstyle{\viiiptmit\fam=\@ne}%
     \rm\fi}


\def\getixpt{%
     \font\ixptrm=cmr9
     \font\ixptmit=cmmi9
     \font\ixptsy=cmsy9
     \font\ixptit=cmti9
     \font\ixptbf=cmbx9
     \skewchar\ixptmit='177 \skewchar\ixptsy='60
     \fontdimen16 \ixptsy=\the\fontdimen17 \ixptsy}

\def\ixpt{\ifmmode\err@badsizechange\else
     \@mathfontinit
     \textfont0=\ixptrm  \scriptfont0=\viiptrm  \scriptscriptfont0=\vptrm
     \textfont1=\ixptmit \scriptfont1=\viiptmit \scriptscriptfont1=\vptmit
     \textfont2=\ixptsy  \scriptfont2=\viiptsy  \scriptscriptfont2=\vptsy
     \textfont3=\xptex   \scriptfont3=\xptex    \scriptscriptfont3=\xptex
     \textfont\itfam=\ixptit
     \scriptfont\itfam=\viiptit
     \scriptscriptfont\itfam=\viiptit
     \textfont\bffam=\ixptbf
     \scriptfont\bffam=\viiptbf
     \scriptscriptfont\bffam=\vptbf
     \@fontstyleinit
     \def\rm{\ixptrm\fam=\z@}%
     \def\it{\ixptit\fam=\itfam}%
     \def\bf{\ixptbf\fam=\bffam}%
     \def\oldstyle{\ixptmit\fam=\@ne}%
     \rm\fi}


\font\xptrm=cmr10
\font\xptmit=cmmi10
\font\xptsy=cmsy10
\font\xptex=cmex10
\font\xptit=cmti10
\font\xptsl=cmsl10
\font\xptbf=cmbx10
\font\xpttt=cmtt10
\font\xptss=cmss10
\font\xptsc=cmcsc10
\font\xptbfs=cmb10
\font\xptbmit=cmmib10

\skewchar\xptmit='177 \skewchar\xptbmit='177 \skewchar\xptsy='60
\fontdimen16 \xptsy=\the\fontdimen17 \xptsy

\def\xpt{\ifmmode\err@badsizechange\else
     \@mathfontinit
     \textfont0=\xptrm  \scriptfont0=\viiptrm  \scriptscriptfont0=\vptrm
     \textfont1=\xptmit \scriptfont1=\viiptmit \scriptscriptfont1=\vptmit
     \textfont2=\xptsy  \scriptfont2=\viiptsy  \scriptscriptfont2=\vptsy
     \textfont3=\xptex  \scriptfont3=\xptex    \scriptscriptfont3=\xptex
     \textfont\itfam=\xptit
     \scriptfont\itfam=\viiptit
     \scriptscriptfont\itfam=\viiptit
     \textfont\bffam=\xptbf
     \scriptfont\bffam=\viiptbf
     \scriptscriptfont\bffam=\vptbf
     \textfont\bfsfam=\xptbfs
     \scriptfont\bfsfam=\viiptbf
     \scriptscriptfont\bfsfam=\vptbf
     \textfont\bmitfam=\xptbmit
     \scriptfont\bmitfam=\viiptmit
     \scriptscriptfont\bmitfam=\vptmit
     \@fontstyleinit
     \def\rm{\xptrm\fam=\z@}%
     \def\it{\xptit\fam=\itfam}%
     \def\sl{\xptsl}%
     \def\bf{\xptbf\fam=\bffam}%
     \def\tt{\xpttt}%
     \def\ss{\xptss}%
     \def\sc{\xptsc}%
     \def\bfs{\xptbfs\fam=\bfsfam}%
     \def\bmit{\fam=\bmitfam}%
     \def\oldstyle{\xptmit\fam=\@ne}%
     \rm\fi}


\def\getxipt{%
     \font\xiptrm=cmr10  scaled\magstephalf
     \font\xiptmit=cmmi10 scaled\magstephalf
     \font\xiptsy=cmsy10 scaled\magstephalf
     \font\xiptex=cmex10 scaled\magstephalf
     \font\xiptit=cmti10 scaled\magstephalf
     \font\xiptsl=cmsl10 scaled\magstephalf
     \font\xiptbf=cmbx10 scaled\magstephalf
     \font\xipttt=cmtt10 scaled\magstephalf
     \font\xiptss=cmss10 scaled\magstephalf
     \skewchar\xiptmit='177 \skewchar\xiptsy='60
     \fontdimen16 \xiptsy=\the\fontdimen17 \xiptsy}

\def\xipt{\ifmmode\err@badsizechange\else
     \@mathfontinit
     \textfont0=\xiptrm  \scriptfont0=\viiiptrm  \scriptscriptfont0=\viptrm
     \textfont1=\xiptmit \scriptfont1=\viiiptmit \scriptscriptfont1=\viptmit
     \textfont2=\xiptsy  \scriptfont2=\viiiptsy  \scriptscriptfont2=\viptsy
     \textfont3=\xiptex  \scriptfont3=\xptex     \scriptscriptfont3=\xptex
     \textfont\itfam=\xiptit
     \scriptfont\itfam=\viiiptit
     \scriptscriptfont\itfam=\viiptit
     \textfont\bffam=\xiptbf
     \scriptfont\bffam=\viiiptbf
     \scriptscriptfont\bffam=\viptbf
     \@fontstyleinit
     \def\rm{\xiptrm\fam=\z@}%
     \def\it{\xiptit\fam=\itfam}%
     \def\sl{\xiptsl}%
     \def\bf{\xiptbf\fam=\bffam}%
     \def\tt{\xipttt}%
     \def\ss{\xiptss}%
     \def\oldstyle{\xiptmit\fam=\@ne}%
     \rm\fi}


\font\xiiptrm=cmr12
\font\xiiptmit=cmmi12
\font\xiiptsy=cmsy10  scaled\magstep1
\font\xiiptex=cmex10  scaled\magstep1
\font\xiiptit=cmti12
\font\xiiptsl=cmsl12
\font\xiiptbf=cmbx12
\font\xiipttt=cmtt12
\font\xiiptss=cmss12
\font\xiiptsc=cmcsc10 scaled\magstep1
\font\xiiptbfs=cmb10  scaled\magstep1
\font\xiiptbmit=cmmib10 scaled\magstep1

\skewchar\xiiptmit='177 \skewchar\xiiptbmit='177 \skewchar\xiiptsy='60
\fontdimen16 \xiiptsy=\the\fontdimen17 \xiiptsy

\def\xiipt{\ifmmode\err@badsizechange\else
     \@mathfontinit
     \textfont0=\xiiptrm  \scriptfont0=\viiiptrm  \scriptscriptfont0=\viptrm
     \textfont1=\xiiptmit \scriptfont1=\viiiptmit \scriptscriptfont1=\viptmit
     \textfont2=\xiiptsy  \scriptfont2=\viiiptsy  \scriptscriptfont2=\viptsy
     \textfont3=\xiiptex  \scriptfont3=\xptex     \scriptscriptfont3=\xptex
     \textfont\itfam=\xiiptit
     \scriptfont\itfam=\viiiptit
     \scriptscriptfont\itfam=\viiptit
     \textfont\bffam=\xiiptbf
     \scriptfont\bffam=\viiiptbf
     \scriptscriptfont\bffam=\viptbf
     \textfont\bfsfam=\xiiptbfs
     \scriptfont\bfsfam=\viiiptbf
     \scriptscriptfont\bfsfam=\viptbf
     \textfont\bmitfam=\xiiptbmit
     \scriptfont\bmitfam=\viiiptmit
     \scriptscriptfont\bmitfam=\viptmit
     \@fontstyleinit
     \def\rm{\xiiptrm\fam=\z@}%
     \def\it{\xiiptit\fam=\itfam}%
     \def\sl{\xiiptsl}%
     \def\bf{\xiiptbf\fam=\bffam}%
     \def\tt{\xiipttt}%
     \def\ss{\xiiptss}%
     \def\sc{\xiiptsc}%
     \def\bfs{\xiiptbfs\fam=\bfsfam}%
     \def\bmit{\fam=\bmitfam}%
     \def\oldstyle{\xiiptmit\fam=\@ne}%
     \rm\fi}


\def\getxiiipt{%
     \font\xiiiptrm=cmr12  scaled\magstephalf
     \font\xiiiptmit=cmmi12 scaled\magstephalf
     \font\xiiiptsy=cmsy9  scaled\magstep2
     \font\xiiiptit=cmti12 scaled\magstephalf
     \font\xiiiptsl=cmsl12 scaled\magstephalf
     \font\xiiiptbf=cmbx12 scaled\magstephalf
     \font\xiiipttt=cmtt12 scaled\magstephalf
     \font\xiiiptss=cmss12 scaled\magstephalf
     \skewchar\xiiiptmit='177 \skewchar\xiiiptsy='60
     \fontdimen16 \xiiiptsy=\the\fontdimen17 \xiiiptsy}

\def\xiiipt{\ifmmode\err@badsizechange\else
     \@mathfontinit
     \textfont0=\xiiiptrm  \scriptfont0=\xptrm  \scriptscriptfont0=\viiptrm
     \textfont1=\xiiiptmit \scriptfont1=\xptmit \scriptscriptfont1=\viiptmit
     \textfont2=\xiiiptsy  \scriptfont2=\xptsy  \scriptscriptfont2=\viiptsy
     \textfont3=\xivptex   \scriptfont3=\xptex  \scriptscriptfont3=\xptex
     \textfont\itfam=\xiiiptit
     \scriptfont\itfam=\xptit
     \scriptscriptfont\itfam=\viiptit
     \textfont\bffam=\xiiiptbf
     \scriptfont\bffam=\xptbf
     \scriptscriptfont\bffam=\viiptbf
     \@fontstyleinit
     \def\rm{\xiiiptrm\fam=\z@}%
     \def\it{\xiiiptit\fam=\itfam}%
     \def\sl{\xiiiptsl}%
     \def\bf{\xiiiptbf\fam=\bffam}%
     \def\tt{\xiiipttt}%
     \def\ss{\xiiiptss}%
     \def\oldstyle{\xiiiptmit\fam=\@ne}%
     \rm\fi}


\font\xivptrm=cmr12   scaled\magstep1
\font\xivptmit=cmmi12  scaled\magstep1
\font\xivptsy=cmsy10  scaled\magstep2
\font\xivptex=cmex10  scaled\magstep2
\font\xivptit=cmti12  scaled\magstep1
\font\xivptsl=cmsl12  scaled\magstep1
\font\xivptbf=cmbx12  scaled\magstep1
\font\xivpttt=cmtt12  scaled\magstep1
\font\xivptss=cmss12  scaled\magstep1
\font\xivptsc=cmcsc10 scaled\magstep2
\font\xivptbfs=cmb10  scaled\magstep2
\font\xivptbmit=cmmib10 scaled\magstep2

\skewchar\xivptmit='177 \skewchar\xivptbmit='177 \skewchar\xivptsy='60
\fontdimen16 \xivptsy=\the\fontdimen17 \xivptsy

\def\xivpt{\ifmmode\err@badsizechange\else
     \@mathfontinit
     \textfont0=\xivptrm  \scriptfont0=\xptrm  \scriptscriptfont0=\viiptrm
     \textfont1=\xivptmit \scriptfont1=\xptmit \scriptscriptfont1=\viiptmit
     \textfont2=\xivptsy  \scriptfont2=\xptsy  \scriptscriptfont2=\viiptsy
     \textfont3=\xivptex  \scriptfont3=\xptex  \scriptscriptfont3=\xptex
     \textfont\itfam=\xivptit
     \scriptfont\itfam=\xptit
     \scriptscriptfont\itfam=\viiptit
     \textfont\bffam=\xivptbf
     \scriptfont\bffam=\xptbf
     \scriptscriptfont\bffam=\viiptbf
     \textfont\bfsfam=\xivptbfs
     \scriptfont\bfsfam=\xptbfs
     \scriptscriptfont\bfsfam=\viiptbf
     \textfont\bmitfam=\xivptbmit
     \scriptfont\bmitfam=\xptbmit
     \scriptscriptfont\bmitfam=\viiptmit
     \@fontstyleinit
     \def\rm{\xivptrm\fam=\z@}%
     \def\it{\xivptit\fam=\itfam}%
     \def\sl{\xivptsl}%
     \def\bf{\xivptbf\fam=\bffam}%
     \def\tt{\xivpttt}%
     \def\ss{\xivptss}%
     \def\sc{\xivptsc}%
     \def\bfs{\xivptbfs\fam=\bfsfam}%
     \def\bmit{\fam=\bmitfam}%
     \def\oldstyle{\xivptmit\fam=\@ne}%
     \rm\fi}


\font\xviiptrm=cmr17
\font\xviiptmit=cmmi12 scaled\magstep2
\font\xviiptsy=cmsy10 scaled\magstep3
\font\xviiptex=cmex10 scaled\magstep3
\font\xviiptit=cmti12 scaled\magstep2
\font\xviiptbf=cmbx12 scaled\magstep2
\font\xviiptbfs=cmb10 scaled\magstep3

\skewchar\xviiptmit='177 \skewchar\xviiptsy='60
\fontdimen16 \xviiptsy=\the\fontdimen17 \xviiptsy

\def\xviipt{\ifmmode\err@badsizechange\else
     \@mathfontinit
     \textfont0=\xviiptrm  \scriptfont0=\xiiptrm  \scriptscriptfont0=\viiiptrm
     \textfont1=\xviiptmit \scriptfont1=\xiiptmit \scriptscriptfont1=\viiiptmit
     \textfont2=\xviiptsy  \scriptfont2=\xiiptsy  \scriptscriptfont2=\viiiptsy
     \textfont3=\xviiptex  \scriptfont3=\xiiptex  \scriptscriptfont3=\xptex
     \textfont\itfam=\xviiptit
     \scriptfont\itfam=\xiiptit
     \scriptscriptfont\itfam=\viiiptit
     \textfont\bffam=\xviiptbf
     \scriptfont\bffam=\xiiptbf
     \scriptscriptfont\bffam=\viiiptbf
     \textfont\bfsfam=\xviiptbfs
     \scriptfont\bfsfam=\xiiptbfs
     \scriptscriptfont\bfsfam=\viiiptbf
     \@fontstyleinit
     \def\rm{\xviiptrm\fam=\z@}%
     \def\it{\xviiptit\fam=\itfam}%
     \def\bf{\xviiptbf\fam=\bffam}%
     \def\bfs{\xviiptbfs\fam=\bfsfam}%
     \def\oldstyle{\xviiptmit\fam=\@ne}%
     \rm\fi}


\font\xxiptrm=cmr17  scaled\magstep1


\def\xxipt{\ifmmode\err@badsizechange\else
     \@mathfontinit
     \@fontstyleinit
     \def\rm{\xxiptrm\fam=\z@}%
     \rm\fi}


\font\xxvptrm=cmr17  scaled\magstep2


\def\xxvpt{\ifmmode\err@badsizechange\else
     \@mathfontinit
     \@fontstyleinit
     \def\rm{\xxvptrm\fam=\z@}%
     \rm\fi}




\message{Loading jyTeX macros...}

\message{modifications to plain.tex,}


\def\newcount{\alloc@0\count\countdef\insc@unt}
\def\newdimen{\alloc@1\dimen\dimendef\insc@unt}
\def\newskip{\alloc@2\skip\skipdef\insc@unt}
\def\newmuskip{\alloc@3\muskip\muskipdef\@cclvi}
\def\newbox{\alloc@4\box\chardef\insc@unt}
\def\newtoks{\alloc@5\toks\toksdef\@cclvi}
\def\newhelp#1#2{\newtoks#1\global#1\expandafter{\csname#2\endcsname}}
\def\newread{\alloc@6\read\chardef\sixt@@n}
\def\newwrite{\alloc@7\write\chardef\sixt@@n}
\def\newfam{\alloc@8\fam\chardef\sixt@@n}
\def\newinsert#1{\global\advance\insc@unt by\m@ne
     \ch@ck0\insc@unt\count
     \ch@ck1\insc@unt\dimen
     \ch@ck2\insc@unt\skip
     \ch@ck4\insc@unt\box
     \allocationnumber=\insc@unt
     \global\chardef#1=\allocationnumber
     \wlog{\string#1=\string\insert\the\allocationnumber}}
\def\newif#1{\count@\escapechar \escapechar\m@ne
     \expandafter\expandafter\expandafter
          \xdef\@if#1{true}{\let\noexpand#1=\noexpand\iftrue}%
     \expandafter\expandafter\expandafter
          \xdef\@if#1{false}{\let\noexpand#1=\noexpand\iffalse}%
     \global\@if#1{false}\escapechar=\count@}


\newlinechar=`\^^J
\overfullrule=0pt




\let\itfam=\undefined

\let\bffam=\undefined

\count18=3


\chardef\sharps="19


\mathchardef\alpha="710B
\mathchardef\beta="710C
\mathchardef\gamma="710D
\mathchardef\delta="710E
\mathchardef\epsilon="710F
\mathchardef\zeta="7110
\mathchardef\eta="7111
\mathchardef\theta="7112
\mathchardef\iota="7113
\mathchardef\kappa="7114
\mathchardef\lambda="7115
\mathchardef\mu="7116
\mathchardef\nu="7117
\mathchardef\xi="7118
\mathchardef\pi="7119
\mathchardef\rho="711A
\mathchardef\sigma="711B
\mathchardef\tau="711C
\mathchardef\upsilon="711D
\mathchardef\phi="711E
\mathchardef\chi="711F
\mathchardef\psi="7120
\mathchardef\omega="7121
\mathchardef\varepsilon="7122
\mathchardef\vartheta="7123
\mathchardef\varpi="7124
\mathchardef\varrho="7125
\mathchardef\varsigma="7126
\mathchardef\varphi="7127
\mathchardef\imath="717B
\mathchardef\jmath="717C
\mathchardef\ell="7160
\mathchardef\wp="717D
\mathchardef\partial="7140
\mathchardef\flat="715B
\mathchardef\natural="715C
\mathchardef\sharp="715D
\def\vec{\mathaccent"717E }



\def\angle{{\vbox{\ialign{$\m@th\scriptstyle##$\crcr
     \not\mathrel{\mkern14mu}\crcr
     \noalign{\nointerlineskip}
     \mkern2.5mu\leaders\hrule height.34\rp@\hfill\mkern2.5mu\crcr}}}}
\def\vdots{\vbox{\baselineskip4\rp@ \lineskiplimit\z@
     \kern6\rp@\hbox{.}\hbox{.}\hbox{.}}}
\def\ddots{\mathinner{\mkern1mu\raise7\rp@\vbox{\kern7\rp@\hbox{.}}\mkern2mu
     \raise4\rp@\hbox{.}\mkern2mu\raise\rp@\hbox{.}\mkern1mu}}
\def\overrightarrow#1{\vbox{\ialign{##\crcr
     \rightarrowfill\crcr
     \noalign{\kern-\rp@\nointerlineskip}
     $\hfil\displaystyle{#1}\hfil$\crcr}}}
\def\overleftarrow#1{\vbox{\ialign{##\crcr
     \leftarrowfill\crcr
     \noalign{\kern-\rp@\nointerlineskip}
     $\hfil\displaystyle{#1}\hfil$\crcr}}}
\def\overbrace#1{\mathop{\vbox{\ialign{##\crcr
     \noalign{\kern3\rp@}
     \downbracefill\crcr
     \noalign{\kern3\rp@\nointerlineskip}
     $\hfil\displaystyle{#1}\hfil$\crcr}}}\limits}
\def\underbrace#1{\mathop{\vtop{\ialign{##\crcr
     $\hfil\displaystyle{#1}\hfil$\crcr
     \noalign{\kern3\rp@\nointerlineskip}
     \upbracefill\crcr
     \noalign{\kern3\rp@}}}}\limits}
\def\big#1{{\hbox{$\left#1\vbox to8.5\rp@ {}\right.\n@space$}}}
\def\Big#1{{\hbox{$\left#1\vbox to11.5\rp@ {}\right.\n@space$}}}
\def\bigg#1{{\hbox{$\left#1\vbox to14.5\rp@ {}\right.\n@space$}}}
\def\Bigg#1{{\hbox{$\left#1\vbox to17.5\rp@ {}\right.\n@space$}}}
\def\@vereq#1#2{\lower.5\rp@\vbox{\baselineskip\z@skip\lineskip-.5\rp@
     \ialign{$\m@th#1\hfil##\hfil$\crcr#2\crcr=\crcr}}}
\def\rlh@#1{\vcenter{\hbox{\ooalign{\raise2\rp@
     \hbox{$#1\rightharpoonup$}\crcr
     $#1\leftharpoondown$}}}}
\def\bordermatrix#1{\begingroup\m@th
     \setbox\z@\vbox{%
          \def\cr{\crcr\noalign{\kern2\rp@\global\let\cr\endline}}%
          \ialign{$##$\hfil\kern2\rp@\kern\p@renwd
               &\thinspace\hfil$##$\hfil&&\quad\hfil$##$\hfil\crcr
               \omit\strut\hfil\crcr
               \noalign{\kern-\baselineskip}%
               #1\crcr\omit\strut\cr}}%
     \setbox\tw@\vbox{\unvcopy\z@\global\setbox\@ne\lastbox}%
     \setbox\tw@\hbox{\unhbox\@ne\unskip\global\setbox\@ne\lastbox}%
     \setbox\tw@\hbox{$\kern\wd\@ne\kern-\p@renwd\left(\kern-\wd\@ne
          \global\setbox\@ne\vbox{\box\@ne\kern2\rp@}%
          \vcenter{\kern-\ht\@ne\unvbox\z@\kern-\baselineskip}%
          \,\right)$}%
     \null\;\vbox{\kern\ht\@ne\box\tw@}\endgroup}
\def\endinsert{\egroup
     \if@mid\dimen@\ht\z@
          \advance\dimen@\dp\z@
          \advance\dimen@12\rp@
          \advance\dimen@\pagetotal
          \ifdim\dimen@>\pagegoal\@midfalse\p@gefalse\fi
     \fi
     \if@mid\bigskip\box\z@
          \bigbreak
     \else\insert\topins{\penalty100 \splittopskip\z@skip
               \splitmaxdepth\maxdimen\floatingpenalty\z@
               \ifp@ge\dimen@\dp\z@
                    \vbox to\vsize{\unvbox\z@\kern-\dimen@}%
               \else\box\z@\nobreak\bigskip
               \fi}%
     \fi
     \endgroup}


\def\cases#1{\left\{\,\vcenter{\m@th
     \ialign{$##\hfil$&\quad##\hfil\crcr#1\crcr}}\right.}
\def\matrix#1{\null\,\vcenter{\m@th
     \ialign{\hfil$##$\hfil&&\quad\hfil$##$\hfil\crcr
          \mathstrut\crcr
          \noalign{\kern-\baselineskip}
          #1\crcr
          \mathstrut\crcr
          \noalign{\kern-\baselineskip}}}\,}


\newif\ifraggedbottom

\def\raggedbottom{\ifraggedbottom\else
     \advance\topskip by\z@ plus60pt \raggedbottomtrue\fi}%
\def\normalbottom{\ifraggedbottom
     \advance\topskip by\z@ plus-60pt \raggedbottomfalse\fi}

\message{hacks,}


\toksdef\toks@i=1
\toksdef\toks@ii=2


\def\TeX{T\kern-.1667em \lower.5ex \hbox{E}\kern-.125em X\null}
\def\jyTeX{{\leavevmode
     \raise.587ex \hbox{\it\j}\kern-.1em \lower.048ex \hbox{\it y}\kern-.12em
     \TeX}}

\let\then=\iftrue
\def\ifnoarg#1\then{\def\hack@{#1}\ifx\hack@\empty}
\def\ifundefined#1\then{%
     \expandafter\ifx\csname\expandafter\blank\string#1\endcsname\relax}
\def\useif#1\then{\csname#1\endcsname}
\def\usename#1{\csname#1\endcsname}
\def\useafter#1#2{\expandafter#1\csname#2\endcsname}

\long\def\loop#1\repeat{\def\@iterate{#1\expandafter\@iterate\fi}\@iterate
     \let\@iterate=\relax}

\let\TeXend=\end
\def\begin#1{\begingroup\def\@@blockname{#1}\usename{begin#1}}
\def\end#1{\usename{end#1}\def\hack@{#1}%
     \ifx\@@blockname\hack@
          \endgroup
     \else\err@badgroup\hack@\@@blockname
     \fi}
\def\@@blockname{}

\def\defaultoption[#1]#2{%
     \def\hack@{\ifx\hack@ii[\toks@={#2}\else\toks@={#2[#1]}\fi\the\toks@}%
     \futurelet\hack@ii\hack@}

\def\markup#1{\let\@@marksf=\empty
     \ifhmode\edef\@@marksf{\spacefactor=\the\spacefactor\relax}\/\fi
     ${}^{\hbox{\subscriptfonts#1}}$\@@marksf}


\newtoks\shortyear
\newtoks\militaryhour
\newtoks\standardhour
\newtoks\minute
\newtoks\amorpm

\def\settime{\count@=\time\divide\count@ by60
     \militaryhour=\expandafter{\number\count@}%
     {\multiply\count@ by-60 \advance\count@ by\time
          \xdef\hack@{\ifnum\count@<10 0\fi\number\count@}}%
     \minute=\expandafter{\hack@}%
     \ifnum\count@<12
          \amorpm={am}
     \else\amorpm={pm}
          \ifnum\count@>12 \advance\count@ by-12 \fi
     \fi
     \standardhour=\expandafter{\number\count@}%
     \def\hack@19##1##2{\shortyear={##1##2}}%
          \expandafter\hack@\the\year}

\def\monthword#1{%
     \ifcase#1
          $\bullet$\err@badcountervalue{monthword}%
          \or January\or February\or March\or April\or May\or June%
          \or July\or August\or September\or October\or November\or December%
     \else$\bullet$\err@badcountervalue{monthword}%
     \fi}

\def\monthabbr#1{%
     \ifcase#1
          $\bullet$\err@badcountervalue{monthabbr}%
          \or Jan\or Feb\or Mar\or Apr\or May\or Jun%
          \or Jul\or Aug\or Sep\or Oct\or Nov\or Dec%
     \else$\bullet$\err@badcountervalue{monthabbr}%
     \fi}

\def\militarytime{\the\militaryhour:\the\minute}
\def\standardtime{\the\standardhour:\the\minute}


\def\@setnumstyle#1#2{\expandafter\global\expandafter\expandafter
     \expandafter\let\expandafter\expandafter
     \csname @\expandafter\blank\string#1style\endcsname
     \csname#2\endcsname}
\def\numstyle#1{\usename{@\expandafter\blank\string#1style}#1}
\def\ifblank#1\then{\useafter\ifx{@\expandafter\blank\string#1}\blank}

\def\blank#1{}

\def\Roman#1{\expandafter\uppercase\expandafter{\romannumeral#1}}
\def\alphabetic#1{%
     \ifcase#1
          $\bullet$\err@badcountervalue{alphabetic}%
          \or a\or b\or c\or d\or e\or f\or g\or h\or i\or j\or k\or l\or m%
          \or n\or o\or p\or q\or r\or s\or t\or u\or v\or w\or x\or y\or z%
     \else$\bullet$\err@badcountervalue{alphabetic}%
     \fi}
\def\Alphabetic#1{\expandafter\uppercase\expandafter{\alphabetic{#1}}}
\def\symbols#1{%
     \ifcase#1
          $\bullet$\err@badcountervalue{symbols}%
          \or*\or\dag\or\ddag\or\S\or$\|$%
          \or**\or\dag\dag\or\ddag\ddag\or\S\S\or$\|\|$%
     \else$\bullet$\err@badcountervalue{symbols}%
     \fi}


\catcode`\^^?=13 \def^^?{\relax}

\def\trimleading#1\to#2{\edef#2{#1}%
     \expandafter\@trimleading\expandafter#2#2^^?^^?}
\def\@trimleading#1#2#3^^?{\ifx#2^^?\def#1{}\else\def#1{#2#3}\fi}

\def\trimtrailing#1\to#2{\edef#2{#1}%
     \expandafter\@trimtrailing\expandafter#2#2^^? ^^?\relax}
\def\@trimtrailing#1#2 ^^?#3{\ifx#3\relax\toks@={}%
     \else\def#1{#2}\toks@={\trimtrailing#1\to#1}\fi
     \the\toks@}

\def\trim#1\to#2{\trimleading#1\to#2\trimtrailing#2\to#2}

\catcode`\^^?=15


\long\def\additemL#1\to#2{\toks@={\^^\{#1}}\toks@ii=\expandafter{#2}%
     \xdef#2{\the\toks@\the\toks@ii}}

\long\def\additemR#1\to#2{\toks@={\^^\{#1}}\toks@ii=\expandafter{#2}%
     \xdef#2{\the\toks@ii\the\toks@}}

\def\getitemL#1\to#2{\expandafter\@getitemL#1\hack@#1#2}
\def\@getitemL\^^\#1#2\hack@#3#4{\def#4{#1}\def#3{#2}}

\message{font macros,}


\newdimen\rp@
\newcount\@@sizeindex \@@sizeindex=0
\newcount\@@factori
\newcount\@@factorii
\newcount\@@factoriii
\newcount\@@factoriv

\countdef\maxfam=18
\newfam\itfam
\newfam\bffam
\newfam\bfsfam
\newfam\bmitfam

\def\@mathfontinit{\count@=4
     \loop\textfont\count@=\nullfont
          \scriptfont\count@=\nullfont
          \scriptscriptfont\count@=\nullfont
          \ifnum\count@<\maxfam\advance\count@ by\@ne
     \repeat}

\def\@fontstyleinit{%
     \def\it{\err@fontnotavailable\it}%
     \def\bf{\err@fontnotavailable\bf}%
     \def\bfs{\err@bfstobf}%
     \def\bmit{\err@fontnotavailable\bmit}%
     \def\sc{\err@fontnotavailable\sc}%
     \def\sl{\err@sltoit}%
     \def\ss{\err@fontnotavailable\ss}%
     \def\tt{\err@fontnotavailable\tt}}

\def\@parameterinit#1{\rm\rp@=.1em \@getscaling{#1}%
     \let\^^\=\@doscaling\scalingskipslist
     \setbox\strutbox=\hbox{\vrule
          height.708\baselineskip depth.292\baselineskip width\z@}}

\def\@getfactor#1#2#3#4{\@@factori=#1 \@@factorii=#2
     \@@factoriii=#3 \@@factoriv=#4}

\def\@getscaling#1{\count@=#1 \advance\count@ by-\@@sizeindex\@@sizeindex=#1
     \ifnum\count@<0
          \let\@mulordiv=\divide
          \let\@divormul=\multiply
          \multiply\count@ by\m@ne
     \else\let\@mulordiv=\multiply
          \let\@divormul=\divide
     \fi
     \edef\@@scratcha{\ifcase\count@                {1}{1}{1}{1}\or
          {1}{7}{23}{3}\or     {2}{5}{3}{1}\or      {9}{89}{13}{1}\or
          {6}{25}{6}{1}\or     {8}{71}{14}{1}\or    {6}{25}{36}{5}\or
          {1}{7}{53}{4}\or     {12}{125}{108}{5}\or {3}{14}{53}{5}\or
          {6}{41}{17}{1}\or    {13}{31}{13}{2}\or   {9}{107}{71}{2}\or
          {11}{139}{124}{3}\or {1}{6}{43}{2}\or     {10}{107}{42}{1}\or
          {1}{5}{43}{2}\or     {5}{69}{65}{1}\or    {11}{97}{91}{2}\fi}%
     \expandafter\@getfactor\@@scratcha}

\def\@doscaling#1{\@mulordiv#1by\@@factori\@divormul#1by\@@factorii
     \@mulordiv#1by\@@factoriii\@divormul#1by\@@factoriv}


\newskip\headskip
\newskip\footskip

\def\typesize=#1pt{\count@=#1 \advance\count@ by-10
     \ifcase\count@
          \@setsizex\or\err@badtypesize\or
          \@setsizexii\or\err@badtypesize\or
          \@setsizexiv
     \else\err@badtypesize
     \fi}

\def\@setsizex{\getixpt
     \def\subsubscriptfonts{\vpt}%
          \def\subsubscriptsize{\vpt\@parameterinit{-8}}%
     \def\subscriptfonts{\viipt}\def\subscriptsize{\viipt\@parameterinit{-4}}%
     \def\footnotefonts{\viiipt}\def\footnotesize{\viiipt\@parameterinit{-2}}%
     \def\smallfonts{\ixpt}\def\smallsize{\ixpt\@parameterinit{-1}}%
     \def\normalfonts{\xpt}\def\normalsize{\xpt\@parameterinit{0}}%
     \def\bigfonts{\xiipt}\def\bigsize{\xiipt\@parameterinit{2}}%
     \def\Bigfonts{\xivpt}\def\Bigsize{\xivpt\@parameterinit{4}}%
     \def\biggfonts{\xviipt}\def\biggsize{\xviipt\@parameterinit{6}}%
     \def\Biggfonts{\xxipt}\def\Biggsize{\xxipt\@parameterinit{8}}%
     \def\tinyfonts{\vpt}\def\tinysize{\vpt\@parameterinit{-8}}%
     \def\HUGEFONTS{\xxvpt}\def\HUGESIZE{\xxvpt\@parameterinit{10}}%
     \normalsize\fixedskipslist}

\def\@setsizexii{\getxipt
     \def\subsubscriptfonts{\vipt}%
          \def\subsubscriptsize{\vipt\@parameterinit{-6}}%
     \def\subscriptfonts{\viiipt}%
          \def\subscriptsize{\viiipt\@parameterinit{-2}}%
     \def\footnotefonts{\xpt}\def\footnotesize{\xpt\@parameterinit{0}}%
     \def\smallfonts{\xipt}\def\smallsize{\xipt\@parameterinit{1}}%
     \def\normalfonts{\xiipt}\def\normalsize{\xiipt\@parameterinit{2}}%
     \def\bigfonts{\xivpt}\def\bigsize{\xivpt\@parameterinit{4}}%
     \def\Bigfonts{\xviipt}\def\Bigsize{\xviipt\@parameterinit{6}}%
     \def\biggfonts{\xxipt}\def\biggsize{\xxipt\@parameterinit{8}}%
     \def\Biggfonts{\xxvpt}\def\Biggsize{\xxvpt\@parameterinit{10}}%
     \def\tinyfonts{\vpt}\def\tinysize{\vpt\@parameterinit{-8}}%
     \def\HUGEFONTS{\xxvpt}\def\HUGESIZE{\xxvpt\@parameterinit{10}}%
     \normalsize\fixedskipslist}

\def\@setsizexiv{\getxiiipt
     \def\subsubscriptfonts{\viipt}%
          \def\subsubscriptsize{\viipt\@parameterinit{-4}}%
     \def\subscriptfonts{\xpt}\def\subscriptsize{\xpt\@parameterinit{0}}%
     \def\footnotefonts{\xiipt}\def\footnotesize{\xiipt\@parameterinit{2}}%
     \def\smallfonts{\xiiipt}\def\smallsize{\xiiipt\@parameterinit{3}}%
     \def\normalfonts{\xivpt}\def\normalsize{\xivpt\@parameterinit{4}}%
     \def\bigfonts{\xviipt}\def\bigsize{\xviipt\@parameterinit{6}}%
     \def\Bigfonts{\xxipt}\def\Bigsize{\xxipt\@parameterinit{8}}%
     \def\biggfonts{\xxvpt}\def\biggsize{\xxvpt\@parameterinit{10}}%
     \def\Biggfonts{\err@sizetoolarge\Biggfonts\HUGEFONTS}%
          \def\Biggsize{\err@sizetoolarge\Biggsize\HUGESIZE}%
     \def\tinyfonts{\vpt}\def\tinysize{\vpt\@parameterinit{-8}}%
     \def\HUGEFONTS{\xxvpt}\def\HUGESIZE{\xxvpt\@parameterinit{10}}%
     \normalsize\fixedskipslist}

\def\subsubscriptfonts{\vpt} \def\subsubscriptsize{\vpt\@parameterinit{-8}}
\def\subscriptfonts{\viipt}  \def\subscriptsize{\viipt\@parameterinit{-4}}
\def\footnotefonts{\viiipt}  \def\footnotesize{\viiipt\@parameterinit{-2}}
\def\smallfonts{\err@sizenotavailable\smallfonts}
                             \def\smallsize{\ixpt\@parameterinit{-1}}
\def\normalfonts{\xpt}       \def\normalsize{\xpt\@parameterinit{0}}
\def\bigfonts{\xiipt}        \def\bigsize{\xiipt\@parameterinit{2}}
\def\Bigfonts{\xivpt}        \def\Bigsize{\xivpt\@parameterinit{4}}
\def\biggfonts{\xviipt}      \def\biggsize{\xviipt\@parameterinit{6}}
\def\Biggfonts{\xxipt}       \def\Biggsize{\xxipt\@parameterinit{8}}
\def\tinyfonts{\vpt}         \def\tinysize{\vpt\@parameterinit{-8}}
\def\HUGEFONTS{\xxvpt}       \def\HUGESIZE{\xxvpt\@parameterinit{10}}

\message{document layout,}


\newtoks\everyoutput \everyoutput={}
\newdimen\depthofpage
\newcount\pagenum \pagenum=0

\newdimen\oddtopmargin  \newdimen\eventopmargin
\newdimen\oddleftmargin \newdimen\evenleftmargin
\newtoks\oddhead        \newtoks\evenhead
\newtoks\oddfoot        \newtoks\evenfoot

\def\topmargin{\afterassignment\@seteventop\oddtopmargin}
\def\leftmargin{\afterassignment\@setevenleft\oddleftmargin}
\def\head{\afterassignment\@setevenhead\oddhead}
\def\foot{\afterassignment\@setevenfoot\oddfoot}

\def\@seteventop{\eventopmargin=\oddtopmargin}
\def\@setevenleft{\evenleftmargin=\oddleftmargin}
\def\@setevenhead{\evenhead=\oddhead}
\def\@setevenfoot{\evenfoot=\oddfoot}

\def\pagenumstyle#1{\@setnumstyle\pagenum{#1}}

\newif\ifdraft
\def\draft{\drafttrue\leftmargin=.5in \overfullrule=5pt }

\def\outputstyle#1{\global\expandafter\let\expandafter
          \@outputstyle\csname#1output\endcsname
     \usename{#1setup}}

\output={\@outputstyle}

\def\normaloutput{\the\everyoutput
     \global\advance\pagenum by\@ne
     \ifodd\pagenum
          \voffset=\oddtopmargin \hoffset=\oddleftmargin
     \else\voffset=\eventopmargin \hoffset=\evenleftmargin
     \fi
     \advance\voffset by-1in  \advance\hoffset by-1in
     \count0=\pagenum
     \expandafter\shipout\pagebox
     \ifnum\outputpenalty>-\@MM\else\dosupereject\fi}

\newdimen\fullhsize
\newbox\leftpage
\newcount\leftpagenum
\newcount\outputpagenum \outputpagenum=0
\let\leftorright=L

\def\twoupoutput{\the\everyoutput
     \global\advance\pagenum by\@ne
     \if L\leftorright
          \global\setbox\leftpage=\leftline{\pagebox}%
          \global\leftpagenum=\pagenum
          \global\let\leftorright=R%
     \else\global\advance\outputpagenum by\@ne
          \ifodd\outputpagenum
               \voffset=\oddtopmargin \hoffset=\oddleftmargin
          \else\voffset=\eventopmargin \hoffset=\evenleftmargin
          \fi
          \advance\voffset by-1in  \advance\hoffset by-1in
          \count0=\leftpagenum \count1=\pagenum
          \shipout\vbox{\hbox to\fullhsize
               {\box\leftpage\hfil\leftline{\pagebox}}}%
          \global\let\leftorright=L%
     \fi
     \ifnum\outputpenalty>-\@MM
     \else\dosupereject
          \if R\leftorright
               \globaldefs=\@ne\head={\hfil}\foot={\hfil}\globaldefs=\z@
               \null\newpage
          \fi
     \fi}

\def\pagebox{\vbox{\makeheadline\pagebody\makefootline}}

\def\makeheadline{%
     \vbox to\z@{\baselinestretch=\@m
          \vskip\topskip\vskip-.708\baselineskip\vskip-\headskip
          \line{\vbox to\ht\strutbox{}%
               \ifodd\pagenum\the\oddhead\else\the\evenhead\fi}%
          \vss}%
     \nointerlineskip}

\def\pagebody{\vbox to\vsize{%
     \boxmaxdepth\maxdepth
     \ifvoid\topins\else\unvbox\topins\fi
     \depthofpage=\dp255
     \unvbox255
     \ifraggedbottom\kern-\depthofpage\vfil\fi
     \ifvoid\footins
     \else\vskip\skip\footins
          \footnoterule
          \unvbox\footins
          \vskip-\footnoteskip
     \fi}}

\def\makefootline{\baselineskip=\footskip
     \line{\ifodd\pagenum\the\oddfoot\else\the\evenfoot\fi}}


\newskip\abovechapterskip
\newskip\belowchapterskip
\newskip\abovesectionskip
\newskip\belowsectionskip
\newskip\abovesubsectionskip
\newskip\belowsubsectionskip

\def\chapterstyle#1{\global\expandafter\let\expandafter\@chapterstyle
     \csname#1text\endcsname}
\def\sectionstyle#1{\global\expandafter\let\expandafter\@sectionstyle
     \csname#1text\endcsname}
\def\subsectionstyle#1{\global\expandafter\let\expandafter\@subsectionstyle
     \csname#1text\endcsname}

\def\chapter#1{%
     \ifdim\lastskip=17sp \else\chapterbreak\vskip\abovechapterskip\fi
     \@chapterstyle{\ifblank\chapternumstyle\then
          \else\newchapternum=\next\chapternumformat\ \fi#1}%
     \nobreak\vskip\belowchapterskip\vskip17sp }

\def\section#1{%
     \ifdim\lastskip=17sp \else\sectionbreak\vskip\abovesectionskip\fi
     \@sectionstyle{\ifblank\sectionnumstyle\then
          \else\newsectionnum=\next\sectionnumformat\ \fi#1}%
     \nobreak\vskip\belowsectionskip\vskip17sp }

\def\subsection#1{%
     \ifdim\lastskip=17sp \else\subsectionbreak\vskip\abovesubsectionskip\fi
     \@subsectionstyle{\ifblank\subsectionnumstyle\then
          \else\newsubsectionnum=\next\subsectionnumformat\ \fi#1}%
     \nobreak\vskip\belowsubsectionskip\vskip17sp }


\let\TeXunderline=\underline
\let\TeXoverline=\overline
\def\underline#1{\relax\ifmmode\TeXunderline{#1}\else
     $\TeXunderline{\hbox{#1}}$\fi}
\def\overline#1{\relax\ifmmode\TeXoverline{#1}\else
     $\TeXoverline{\hbox{#1}}$\fi}

\def\baselinestretch{\afterassignment\@baselinestretch\count@}
\def\@baselinestretch{\baselineskip=\normalbaselineskip
     \divide\baselineskip by\@m\baselineskip=\count@\baselineskip
     \setbox\strutbox=\hbox{\vrule
          height.708\baselineskip depth.292\baselineskip width\z@}%
     \bigskipamount=\the\baselineskip
          plus.25\baselineskip minus.25\baselineskip
     \medskipamount=.5\baselineskip
          plus.125\baselineskip minus.125\baselineskip
     \smallskipamount=.25\baselineskip
          plus.0625\baselineskip minus.0625\baselineskip}

\def\\{\ifhmode\ifnum\lastpenalty=-\@M\else\hfil\penalty-\@M\fi\fi
     \ignorespaces}
\def\newpage{\vfil\break}

\def\lefttext#1{\par{\@text\leftskip=\z@\rightskip=\centering
     \noindent#1\par}}
\def\righttext#1{\par{\@text\leftskip=\centering\rightskip=\z@
     \noindent#1\par}}
\def\centertext#1{\par{\@text\leftskip=\centering\rightskip=\centering
     \noindent#1\par}}
\def\@text{\parindent=\z@ \parfillskip=\z@ \everypar={}%
     \spaceskip=.3333em \xspaceskip=.5em
     \def\\{\ifhmode\ifnum\lastpenalty=-\@M\else\penalty-\@M\fi\fi
          \ignorespaces}}

\def\beginleft{\par\@text\leftskip=\z@ \rightskip=\centering}
     
\def\beginright{\par\@text\leftskip=\centering\rightskip=\z@ }
     
\def\begincenter{\par\@text\leftskip=\centering\rightskip=\centering}

\def\beginnarrow{\defaultoption[\parindent]\@beginnarrow}
\def\@beginnarrow[#1]{\par\advance\leftskip by#1\advance\rightskip by#1}

\begingroup
\catcode`\[=1 \catcode`\{=11
\gdef\beginignore[\endgroup\bgroup
     \catcode`\e=0 \catcode`\\=12 \catcode`\{=11 \catcode`\f=12 \let\or=\relax
     \let\nd{ignor=\fi \let\}=\egroup
     \iffalse}
\endgroup

\long\def\marginnote#1{\leavevmode
     \edef\@marginsf{\spacefactor=\the\spacefactor\relax}%
     \ifdraft\strut\vadjust{%
          \hbox to\z@{\hskip\hsize\hskip.1in
               \vbox to\z@{\vskip-\dp\strutbox
                    \marginnoteformat
                    \vskip-\ht\strutbox
                    \noindent\strut#1\par
                    \vss}%
               \hss}}%
     \fi
     \@marginsf}


\newtoks\everybye \everybye={\par\vfil}
\outer\def\bye{\the\everybye
     \footnotecheck
     \prelabelcheck
     \streamcheck
     \supereject
     \TeXend}

\message{footnotes,}

\newcount\footnotenum \footnotenum=0
\newskip\footnoteskip
\let\@footnotelist=\empty

\def\footnotenumstyle#1{\@setnumstyle\footnotenum{#1}%
     \useafter\ifx{@footnotenumstyle}\symbols
          \global\let\@footup=\empty
     \else\global\let\@footup=\markup
     \fi}

\def\footnote{\footnotecheck\defaultoption[]\@footnote}
\def\@footnote[#1]{\@footnotemark[#1]\@footnotetext}

\def\footnotemark{\defaultoption[]\@footnotemark}
\def\@footnotemark[#1]{\let\@footsf=\empty
     \ifhmode\edef\@footsf{\spacefactor=\the\spacefactor\relax}\/\fi
     \ifnoarg#1\then
          \global\advance\footnotenum by\@ne
          \@footup{\footnotenumformat}%
          \edef\@@foota{\footnotenum=\the\footnotenum\relax}%
          \expandafter\additemR\expandafter\@footup\expandafter
               {\@@foota\footnotenumformat}\to\@footnotelist
          \global\let\@footnotelist=\@footnotelist
     \else\markup{#1}%
          \additemR\markup{#1}\to\@footnotelist
          \global\let\@footnotelist=\@footnotelist
     \fi
     \@footsf}

\def\footnotetext{%
     \ifx\@footnotelist\empty\err@extrafootnotetext\else\@footnotetext\fi}
\def\@footnotetext{%
     \getitemL\@footnotelist\to\@@foota
     \global\let\@footnotelist=\@footnotelist
     \insert\footins\bgroup
     \footnoteformat
     \splittopskip=\ht\strutbox\splitmaxdepth=\dp\strutbox
     \interlinepenalty=\interfootnotelinepenalty\floatingpenalty=\@MM
     \noindent\llap{\@@foota}\strut
     \bgroup\aftergroup\@footnoteend
     \let\@@scratcha=}
\def\@footnoteend{\strut\par\vskip\footnoteskip\egroup}

\def\footnoterule{\normalfonts
     \kern-.3em \hrule width2in height.04em \kern .26em }

\def\footnotecheck{%
     \ifx\@footnotelist\empty
     \else\err@extrafootnotemark
          \global\let\@footnotelist=\empty
     \fi}

\message{labels,}

\let\@@labeldef=\xdef
\newif\if@labelfile
\newwrite\@labelfile
\let\@prelabellist=\empty

\def\label#1#2{\trim#1\to\@@labarg\edef\@@labtext{#2}%
     \edef\@@labname{lab@\@@labarg}%
     \useafter\ifundefined\@@labname\then\else\@yeslab\fi
     \useafter\@@labeldef\@@labname{#2}%
     \ifstreaming
          \expandafter\toks@\expandafter\expandafter\expandafter
               {\csname\@@labname\endcsname}%
          \immediate\write\streamout{\noexpand\label{\@@labarg}{\the\toks@}}%
     \fi}
\def\@yeslab{%
     \useafter\ifundefined{if\@@labname}\then
          \err@labelredef\@@labarg
     \else\useif{if\@@labname}\then
               \err@labelredef\@@labarg
          \else\global\usename{\@@labname true}%
               \useafter\ifundefined{pre\@@labname}\then
               \else\useafter\ifx{pre\@@labname}\@@labtext
                    \else\err@badlabelmatch\@@labarg
                    \fi
               \fi
               \if@labelfile
               \else\global\@labelfiletrue
                    \immediate\write\sixt@@n{--> Creating file \jobname.lab}%
                    \immediate\openout\@labelfile=\jobname.lab
               \fi
               \immediate\write\@labelfile
                    {\noexpand\prelabel{\@@labarg}{\@@labtext}}%
          \fi
     \fi}

\def\putlab#1{\trim#1\to\@@labarg\edef\@@labname{lab@\@@labarg}%
     \useafter\ifundefined\@@labname\then\@nolab\else\usename\@@labname\fi}
\def\@nolab{%
     \useafter\ifundefined{pre\@@labname}\then
          \undefinedlabelformat
          \err@needlabel\@@labarg
          \useafter\xdef\@@labname{\undefinedlabelformat}%
     \else\usename{pre\@@labname}%
          \useafter\xdef\@@labname{\usename{pre\@@labname}}%
     \fi
     \useafter\newif{if\@@labname}%
     \expandafter\additemR\@@labarg\to\@prelabellist}

\def\prelabel#1{\useafter\gdef{prelab@#1}}

\def\ifundefinedlabel#1\then{%
     \expandafter\ifx\csname lab@#1\endcsname\relax}
\def\useiflab#1\then{\csname iflab@#1\endcsname}

\def\prelabelcheck{{%
     \def\^^\##1{\useiflab{##1}\then\else\err@undefinedlabel{##1}\fi}%
     \@prelabellist}}

\message{equation numbering,}

\newcount\chapternum
\newcount\sectionnum
\newcount\subsectionnum
\newcount\equationnum
\newcount\subequationnum
\newcount\figurenum
\newcount\subfigurenum
\newcount\tablenum
\newcount\subtablenum

\newif\if@subeqncount
\newif\if@subfigcount
\newif\if@subtblcount

\def\newchapternum{\newsectionnum=\z@\@resetnum\chapternum}
\def\newsectionnum{\newsubsectionnum=\z@\@resetnum\sectionnum}
\def\newsubsectionnum{\newequationnum=\z@\newfigurenum=\z@\newtablenum=\z@
     \@resetnum\subsectionnum}
\def\newequationnum{\newsubequationnum=\z@\@resetnum\equationnum}
\def\newsubequationnum{\@resetnum\subequationnum}
\def\newfigurenum{\newsubfigurenum=\z@\@resetnum\figurenum}
\def\newsubfigurenum{\@resetnum\subfigurenum}
\def\newtablenum{\newsubtablenum=\z@\@resetnum\tablenum}
\def\newsubtablenum{\@resetnum\subtablenum}

\def\@resetnum#1{\global\advance#1by1 \edef\next{\the#1\relax}\global#1}

\newchapternum=0

\def\chapternumstyle#1{\@setnumstyle\chapternum{#1}}
\def\sectionnumstyle#1{\@setnumstyle\sectionnum{#1}}
\def\subsectionnumstyle#1{\@setnumstyle\subsectionnum{#1}}
\def\equationnumstyle#1{\@setnumstyle\equationnum{#1}}
\def\subequationnumstyle#1{\@setnumstyle\subequationnum{#1}%
     \ifblank\subequationnumstyle\then\global\@subeqncountfalse\fi
     \ignorespaces}
\def\figurenumstyle#1{\@setnumstyle\figurenum{#1}}
\def\subfigurenumstyle#1{\@setnumstyle\subfigurenum{#1}%
     \ifblank\subfigurenumstyle\then\global\@subfigcountfalse\fi
     \ignorespaces}
\def\tablenumstyle#1{\@setnumstyle\tablenum{#1}}
\def\subtablenumstyle#1{\@setnumstyle\subtablenum{#1}%
     \ifblank\subtablenumstyle\then\global\@subtblcountfalse\fi
     \ignorespaces}

\def\eqnlabel#1{%
     \if@subeqncount
          \newsubequationnum=\next
     \else\newequationnum=\next
          \ifblank\subequationnumstyle\then
          \else\global\@subeqncounttrue
               \newsubequationnum=\@ne
          \fi
     \fi
     \label{#1}{\puteqnformat}(\puteqn{#1})%
     \ifdraft\rlap{\hskip.1in{\tt#1}}\fi}

\let\puteqn=\putlab

\def\equation#1#2{\useafter\gdef{eqn@#1}{#2\eqno\eqnlabel{#1}}}
\def\Equation#1{\useafter\gdef{eqn@#1}}

\def\putequation#1{\useafter\ifundefined{eqn@#1}\then
     \err@undefinedeqn{#1}\else\usename{eqn@#1}\fi}

\def\eqnseriesstyle#1{\gdef\@eqnseriesstyle{#1}}
\def\begineqnseries{\subequationnumstyle{\@eqnseriesstyle}%
     \defaultoption[]\@begineqnseries}
\def\@begineqnseries[#1]{\edef\@@eqnname{#1}}
\def\endeqnseries{\subequationnumstyle{blank}%
     \expandafter\ifnoarg\@@eqnname\then
     \else\label\@@eqnname{\puteqnformat}%
     \fi
     \aftergroup\ignorespaces}

\def\figlabel#1{%
     \if@subfigcount
          \newsubfigurenum=\next
     \else\newfigurenum=\next
          \ifblank\subfigurenumstyle\then
          \else\global\@subfigcounttrue
               \newsubfigurenum=\@ne
          \fi
     \fi
     \label{#1}{\putfigformat}\putfig{#1}%
     {\def\marginnoteformat{\tt}\marginnote{#1}}}

\let\putfig=\putlab

\def\figseriesstyle#1{\gdef\@figseriesstyle{#1}}
\def\beginfigseries{\subfigurenumstyle{\@figseriesstyle}%
     \defaultoption[]\@beginfigseries}
\def\@beginfigseries[#1]{\edef\@@figname{#1}}
\def\endfigseries{\subfigurenumstyle{blank}%
     \expandafter\ifnoarg\@@figname\then
     \else\label\@@figname{\putfigformat}%
     \fi
     \aftergroup\ignorespaces}

\def\tbllabel#1{%
     \if@subtblcount
          \newsubtablenum=\next
     \else\newtablenum=\next
          \ifblank\subtablenumstyle\then
          \else\global\@subtblcounttrue
               \newsubtablenum=\@ne
          \fi
     \fi
     \label{#1}{\puttblformat}\puttbl{#1}%
     {\def\marginnoteformat{\tt}\marginnote{#1}}}

\let\puttbl=\putlab

\def\tblseriesstyle#1{\gdef\@tblseriesstyle{#1}}
\def\begintblseries{\subtablenumstyle{\@tblseriesstyle}%
     \defaultoption[]\@begintblseries}
\def\@begintblseries[#1]{\edef\@@tblname{#1}}
\def\endtblseries{\subtablenumstyle{blank}%
     \expandafter\ifnoarg\@@tblname\then
     \else\label\@@tblname{\puttblformat}%
     \fi
     \aftergroup\ignorespaces}

\message{reference numbering,}

\newcount\referencenum \referencenum=0
\newcount\@@prerefcount \@@prerefcount=0
\newcount\@@thisref
\newcount\@@lastref
\newcount\@@loopref
\newcount\@@refseq
\newdimen\refnumindent
\let\@undefreflist=\empty

\def\referencenumstyle#1{\@setnumstyle\referencenum{#1}}

\def\referencestyle#1{\usename{@ref#1}}

\def\@refsequential{%
     \gdef\@refpredef##1{\global\advance\referencenum by\@ne
          \let\^^\=0\label{##1}{\^^\{\the\referencenum}}%
          \useafter\gdef{ref@\the\referencenum}{{##1}{\undefinedlabelformat}}}%
     \gdef\@reference##1##2{%
          \ifundefinedlabel##1\then
          \else\def\^^\####1{\global\@@thisref=####1\relax}\putlab{##1}%
               \useafter\gdef{ref@\the\@@thisref}{{##1}{##2}}%
          \fi}%
     \gdef\endputreferences{%
          \loop\ifnum\@@loopref<\referencenum
                    \advance\@@loopref by\@ne
                    \expandafter\expandafter\expandafter\@printreference
                         \csname ref@\the\@@loopref\endcsname
          \repeat
          \par}}

\def\@refpreordered{%
     \gdef\@refpredef##1{\global\advance\referencenum by\@ne
          \additemR##1\to\@undefreflist}%
     \gdef\@reference##1##2{%
          \ifundefinedlabel##1\then
          \else\global\advance\@@loopref by\@ne
               {\let\^^\=0\label{##1}{\^^\{\the\@@loopref}}}%
               \@printreference{##1}{##2}%
          \fi}
     \gdef\endputreferences{%
          \def\^^\####1{\useiflab{####1}\then
               \else\reference{####1}{\undefinedlabelformat}\fi}%
          \@undefreflist
          \par}}

\def\beginprereferences{\par
     \def\reference##1##2{\global\advance\referencenum by1\@ne
          \let\^^\=0\label{##1}{\^^\{\the\referencenum}}%
          \useafter\gdef{ref@\the\referencenum}{{##1}{##2}}}}
\def\endprereferences{\global\@@prerefcount=\the\referencenum\par}

\def\beginputreferences{\par
     \refnumindent=\z@\@@loopref=\z@
     \loop\ifnum\@@loopref<\referencenum
               \advance\@@loopref by\@ne
               \setbox\z@=\hbox{\referencenum=\@@loopref
                    \referencenumformat\enskip}%
               \ifdim\wd\z@>\refnumindent\refnumindent=\wd\z@\fi
     \repeat
     \putreferenceformat
     \@@loopref=\z@
     \loop\ifnum\@@loopref<\@@prerefcount
               \advance\@@loopref by\@ne
               \expandafter\expandafter\expandafter\@printreference
                    \csname ref@\the\@@loopref\endcsname
     \repeat
     \let\reference=\@reference}

\def\@printreference#1#2{\ifx#2\undefinedlabelformat\err@undefinedref{#1}\fi
     \noindent\ifdraft\rlap{\hskip\hsize\hskip.1in \tt#1}\fi
     \llap{\referencenum=\@@loopref\referencenumformat\enskip}#2\par}

\def\reference#1#2{{\par\refnumindent=\z@\putreferenceformat\noindent#2\par}}

\def\putref#1{\trim#1\to\@@refarg
     \expandafter\ifnoarg\@@refarg\then
          \toks@={\relax}%
     \else\@@lastref=-\@m\def\@@refsep{}\def\@more{\@nextref}%
          \toks@={\@nextref#1,,}%
     \fi\the\toks@}
\def\@nextref#1,{\trim#1\to\@@refarg
     \expandafter\ifnoarg\@@refarg\then
          \let\@more=\relax
     \else\ifundefinedlabel\@@refarg\then
               \expandafter\@refpredef\expandafter{\@@refarg}%
          \fi
          \def\^^\##1{\global\@@thisref=##1\relax}%
          \global\@@thisref=\m@ne
          \setbox\z@=\hbox{\putlab\@@refarg}%
     \fi
     \advance\@@lastref by\@ne
     \ifnum\@@lastref=\@@thisref\advance\@@refseq by\@ne\else\@@refseq=\@ne\fi
     \ifnum\@@lastref<\z@
     \else\ifnum\@@refseq<\thr@@
               \@@refsep\def\@@refsep{,}%
               \ifnum\@@lastref>\z@
                    \advance\@@lastref by\m@ne
                    {\referencenum=\@@lastref\putrefformat}%
               \else\undefinedlabelformat
               \fi
          \else\def\@@refsep{--}%
          \fi
     \fi
     \@@lastref=\@@thisref
     \@more}

\message{streaming,}

\newif\ifstreaming

\def\streamto{\defaultoption[\jobname]\@streamto}
\def\@streamto[#1]{\global\streamingtrue
     \immediate\write\sixt@@n{--> Streaming to #1.str}%
     \newwrite\streamout\immediate\openout\streamout=#1.str }

\def\streamfrom{\defaultoption[\jobname]\@streamfrom}
\def\@streamfrom[#1]{\newread\streamin\openin\streamin=#1.str
     \ifeof\streamin
          \expandafter\err@nostream\expandafter{#1.str}%
     \else\immediate\write\sixt@@n{--> Streaming from #1.str}%
          \let\@@labeldef=\gdef
          \ifstreaming
               \edef\@elc{\endlinechar=\the\endlinechar}%
               \endlinechar=\m@ne
               \loop\read\streamin to\@@scratcha
                    \ifeof\streamin
                         \streamingfalse
                    \else\toks@=\expandafter{\@@scratcha}%
                         \immediate\write\streamout{\the\toks@}%
                    \fi
                    \ifstreaming
               \repeat
               \@elc
               \input #1.str
               \streamingtrue
          \else\input #1.str
          \fi
          \let\@@labeldef=\xdef
     \fi}

\def\streamcheck{\ifstreaming
     \immediate\write\streamout{\pagenum=\the\pagenum}%
     \immediate\write\streamout{\footnotenum=\the\footnotenum}%
     \immediate\write\streamout{\referencenum=\the\referencenum}%
     \immediate\write\streamout{\chapternum=\the\chapternum}%
     \immediate\write\streamout{\sectionnum=\the\sectionnum}%
     \immediate\write\streamout{\subsectionnum=\the\subsectionnum}%
     \immediate\write\streamout{\equationnum=\the\equationnum}%
     \immediate\write\streamout{\subequationnum=\the\subequationnum}%
     \immediate\write\streamout{\figurenum=\the\figurenum}%
     \immediate\write\streamout{\subfigurenum=\the\subfigurenum}%
     \immediate\write\streamout{\tablenum=\the\tablenum}%
     \immediate\write\streamout{\subtablenum=\the\subtablenum}%
     \immediate\closeout\streamout
     \fi}


\def\err@badtypesize{%
     \errhelp={The limited availability of certain fonts requires^^J%
          that the base type size be 10pt, 12pt, or 14pt.^^J}%
     \errmessage{--> Illegal base type size}}

\def\err@badsizechange{\immediate\write\sixt@@n
     {--> Size change not allowed in math mode, ignored}}

\def\err@sizetoolarge#1{\immediate\write\sixt@@n
     {--> \noexpand#1 too big, substituting HUGE}}

\def\err@sizenotavailable#1{\immediate\write\sixt@@n
     {--> Size not available, \noexpand#1 ignored}}

\def\err@fontnotavailable#1{\immediate\write\sixt@@n
     {--> Font not available, \noexpand#1 ignored}}

\def\err@sltoit{\immediate\write\sixt@@n
     {--> Style \noexpand\sl not available, substituting \noexpand\it}%
     \it}

\def\err@bfstobf{\immediate\write\sixt@@n
     {--> Style \noexpand\bfs not available, substituting \noexpand\bf}%
     \bf}

\def\err@badgroup#1#2{%
     \errhelp={The block you have just tried to close was not the one^^J%
          most recently opened.^^J}%
     \errmessage{--> \noexpand\end{#1} doesn't match \noexpand\begin{#2}}}

\def\err@badcountervalue#1{\immediate\write\sixt@@n
     {--> Counter (#1) out of bounds}}

\def\err@extrafootnotemark{\immediate\write\sixt@@n
     {--> \noexpand\footnotemark command
          has no corresponding \noexpand\footnotetext}}

\def\err@extrafootnotetext{%
     \errhelp{You have given a \noexpand\footnotetext command without first
          specifying^^Ja \noexpand\footnotemark.^^J}%
     \errmessage{--> \noexpand\footnotetext command has no corresponding
          \noexpand\footnotemark}}

\def\err@labelredef#1{\immediate\write\sixt@@n
     {--> Label "#1" redefined}}

\def\err@badlabelmatch#1{\immediate\write\sixt@@n
     {--> Definition of label "#1" doesn't match value in \jobname.lab}}

\def\err@needlabel#1{\immediate\write\sixt@@n
     {--> Label "#1" cited before its definition}}

\def\err@undefinedlabel#1{\immediate\write\sixt@@n
     {--> Label "#1" cited but never defined}}

\def\err@undefinedeqn#1{\immediate\write\sixt@@n
     {--> Equation "#1" not defined}}

\def\err@undefinedref#1{\immediate\write\sixt@@n
     {--> Reference "#1" not defined}}

\def\err@nostream#1{%
     \errhelp={You have tried to input a stream file that doesn't exist.^^J}%
     \errmessage{--> Stream file #1 not found}}

\message{jyTeX initialization}

\everyjob{\immediate\write16{--> jyTeX version \fmtversion}%
     \edef\@@jobname{\jobname}%
     \edef\jobname{\@@jobname}%
     \settime
     \openin0=\jobname.lab
     \ifeof0
     \else\closein0
          \immediate\write16{--> Getting labels from file \jobname.lab}%
          \input\jobname.lab
     \fi}


\def\fixedskipslist{%
     \^^\{\topskip}%
     \^^\{\splittopskip}%
     \^^\{\maxdepth}%
     \^^\{\skip\topins}%
     \^^\{\skip\footins}%
     \^^\{\headskip}%
     \^^\{\footskip}}

\def\scalingskipslist{%
     \^^\{\p@renwd}%
     \^^\{\delimitershortfall}%
     \^^\{\nulldelimiterspace}%
     \^^\{\scriptspace}%
     \^^\{\jot}%
     \^^\{\normalbaselineskip}%
     \^^\{\normallineskip}%
     \^^\{\normallineskiplimit}%
     \^^\{\baselineskip}%
     \^^\{\lineskip}%
     \^^\{\lineskiplimit}%
     \^^\{\bigskipamount}%
     \^^\{\medskipamount}%
     \^^\{\smallskipamount}%
     \^^\{\parskip}%
     \^^\{\parindent}%
     \^^\{\abovedisplayskip}%
     \^^\{\belowdisplayskip}%
     \^^\{\abovedisplayshortskip}%
     \^^\{\belowdisplayshortskip}%
     \^^\{\abovechapterskip}%
     \^^\{\belowchapterskip}%
     \^^\{\abovesectionskip}%
     \^^\{\belowsectionskip}%
     \^^\{\abovesubsectionskip}%
     \^^\{\belowsubsectionskip}}


\def\twoupsetup{
     \topmargin=.75in
     \leftmargin=.5in
     \vsize=6.9in
     \hsize=4.75in
     \fullhsize=10in
     \let\draft=\relax}

\outputstyle{normal}                             

\def\marginnoteformat{\subscriptsize             
     \hsize=1in \baselinestretch=1000 \everypar={}%
     \tolerance=5000 \hbadness=5000 \parskip=0pt \parindent=0pt
     \leftskip=0pt \rightskip=0pt \raggedright}

\head={\ifdraft\normalfonts\it\hfil DRAFT\hfil   
     \llap{\number\day\ \monthword\month\ \militarytime}\else\hfil\fi}
\foot={\hfil\normalfonts\numstyle\pagenum\hfil}  

\normalbaselineskip=12pt                         
\normallineskip=0pt                              
\normallineskiplimit=0pt                         
\normalbaselines                                 

\topskip=.85\baselineskip
\splittopskip=\topskip
\headskip=2\baselineskip
\footskip=\headskip

\pagenumstyle{arabic}                            

\parskip=0pt                                     
\parindent=20pt                                  

\baselinestretch=1000                            


\chapterstyle{left}                              
\chapternumstyle{blank}                          
\def\chapterbreak{\newpage}                      
\abovechapterskip=0pt                            
\belowchapterskip=1.5\baselineskip               
     plus.38\baselineskip minus.38\baselineskip
\def\chapternumformat{\numstyle\chapternum.}     

\sectionstyle{left}                              
\sectionnumstyle{blank}                          
\def\sectionbreak{\vskip0pt plus4\baselineskip\penalty-100
     \vskip0pt plus-4\baselineskip}              
\abovesectionskip=1.5\baselineskip               
     plus.38\baselineskip minus.38\baselineskip
\belowsectionskip=\the\baselineskip              
     plus.25\baselineskip minus.25\baselineskip
\def\sectionnumformat{
     \ifblank\chapternumstyle\then\else\numstyle\chapternum.\fi
     \numstyle\sectionnum.}

\subsectionstyle{left}                           
\subsectionnumstyle{blank}                       
\def\subsectionbreak{\vskip0pt plus4\baselineskip\penalty-100
     \vskip0pt plus-4\baselineskip}              
\abovesubsectionskip=\the\baselineskip           
     plus.25\baselineskip minus.25\baselineskip
\belowsubsectionskip=.75\baselineskip            
     plus.19\baselineskip minus.19\baselineskip
\def\subsectionnumformat{
     \ifblank\chapternumstyle\then\else\numstyle\chapternum.\fi
     \ifblank\sectionnumstyle\then\else\numstyle\sectionnum.\fi
     \numstyle\subsectionnum.}


\footnotenumstyle{symbols}                       
\footnoteskip=0pt                                
\def\footnotenumformat{\numstyle\footnotenum}    
\def\footnoteformat{\footnotesize                
     \everypar={}\parskip=0pt \parfillskip=0pt plus1fil
     \leftskip=1em \rightskip=0pt
     \spaceskip=0pt \xspaceskip=0pt
     \def\\{\ifhmode\ifnum\lastpenalty=-10000
          \else\hfil\penalty-10000 \fi\fi\ignorespaces}}


\def\undefinedlabelformat{$\bullet$}             


\equationnumstyle{arabic}                        
\subequationnumstyle{blank}                      
\figurenumstyle{arabic}                          
\subfigurenumstyle{blank}                        
\tablenumstyle{arabic}                           
\subtablenumstyle{blank}                         

\eqnseriesstyle{alphabetic}                      
\figseriesstyle{alphabetic}                      
\tblseriesstyle{alphabetic}                      

\def\puteqnformat{\hbox{
     \ifblank\chapternumstyle\then\else\numstyle\chapternum.\fi
     \ifblank\sectionnumstyle\then\else\numstyle\sectionnum.\fi
     \ifblank\subsectionnumstyle\then\else\numstyle\subsectionnum.\fi
     \numstyle\equationnum
     \numstyle\subequationnum}}
\def\putfigformat{\hbox{
     \ifblank\chapternumstyle\then\else\numstyle\chapternum.\fi
     \ifblank\sectionnumstyle\then\else\numstyle\sectionnum.\fi
     \ifblank\subsectionnumstyle\then\else\numstyle\subsectionnum.\fi
     \numstyle\figurenum
     \numstyle\subfigurenum}}
\def\puttblformat{\hbox{
     \ifblank\chapternumstyle\then\else\numstyle\chapternum.\fi
     \ifblank\sectionnumstyle\then\else\numstyle\sectionnum.\fi
     \ifblank\subsectionnumstyle\then\else\numstyle\subsectionnum.\fi
     \numstyle\tablenum
     \numstyle\subtablenum}}


\referencestyle{sequential}                      
\referencenumstyle{arabic}                       
\def\putrefformat{\numstyle\referencenum}        
\def\referencenumformat{\numstyle\referencenum.} 
\def\putreferenceformat{
     \everypar={\hangindent=1em \hangafter=1 }%
     \def\\{\hfil\break\null\hskip-1em \ignorespaces}%
     \leftskip=\refnumindent\parindent=0pt \interlinepenalty=1000 }


\normalsize


\def\fmtversion{2.6M (June 1992)}

\catcode`\@=12

\topmargin= .8in\vsize=9.4in
\baselinestretch=1500


\footnotenum= 0


\def\n{\noindent}

\def\c#1:#2{c^{#1}_{#2}}
\def\desc{:\phi^1_0 \phi^1_0:}

\def\t{\vartheta}
\def\bt{\bar \vartheta}

\def\undbib{\underbar {\hbox to 2cm{\hfill}}, }
\def\subon{\subequationnumstyle{alphabetic}}
\def\suboff{\subequationnumstyle{blank}}
\def\h{\hbox to .5cm{\hfill}}
\def\hof{\hbox to .15cm{\hfill}}
\def\htf{\hbox to .35cm{\hfill}}
\def\hso{\hskip 1cm}
\def\hsf{\hskip .7cm}
\def\hve{\hfill\vfill\eject }
\def\referencenumformat{[\numstyle\referencenum]}

\def\theta{\vartheta}
\def\mpr#1{\markup{[\putref{#1}]}}

\referencestyle{preordered}

\begin{ignore}

\font\twelveBbb=msym10 scaled \magstep1
\font\nineBbb=msym9
\font\sevenBbb=msym7
\newfam\Bbbfam
\textfont\Bbbfam=\twelveBbb
\scriptfont\Bbbfam=\nineBbb
\scriptscriptfont\Bbbfam=\sevenBbb
\def\Bbb{\fam\Bbbfam\twelveBbb}

  \def\S{{\Bbb S}}

\def\Z{{\Bbb Z}}
\end{ignore}
\def\Z{Z\!\!\!Z}
\footnotenumstyle{arabic}
\pagenumstyle{arabic}\pagenum=0

\prelabel {ref9d}{\^^\{1}}
\prelabel {ref9b}{\^^\{2}}
\prelabel {ref15a}{\^^\{3}}
\prelabel {ref9c}{\^^\{4}}
\prelabel {ref3}{\^^\{5}}
\prelabel {ref61}{\^^\{6}}
\prelabel {ref62}{\^^\{7}}
\prelabel {argyres91f}{\^^\{8}}
\prelabel {ref64}{\^^\{9}}
\prelabel {ref65}{\^^\{10}}
\prelabel {ref12}{\^^\{11}}
\prelabel {ref14d}{\^^\{12}}
\prelabel {cd}{\^^\{13}}
\prelabel {ad93}{\^^\{14}}
\prelabel {ref68}{\^^\{15}}
\prelabel {ref63}{\^^\{16}}
\prelabel {new1}{\^^\{17}}
\prelabel {ref7}{\^^\{18}}
\prelabel {ref13}{\^^\{19}}
\prelabel {ref1a}{\^^\{20}}
\prelabel {ref15b}{\^^\{21}}
\prelabel {ref1d}{\^^\{22}}
\prelabel {ref8b}{\^^\{23}}
\prelabel {ref8a}{\^^\{24}}
\prelabel {ref1c}{\^^\{25}}
\prelabel {ref9a}{\^^\{26}}
\prelabel {ref14a}{\^^\{27}}
\prelabel {ref1b}{\^^\{28}}
\prelabel {new2}{\^^\{29}}
\prelabel {ref11}{\^^\{30}}
\prelabel {ref4}{\^^\{31}}

\bigfonts

{\pagenumstyle{blank}
\hbox to 1cm{\hfill}
\vskip .3 truecm
\footnotenumstyle{symbols}
{\rightline{\hbox to 4.5cm{\vtop{\hsize= 4.5cm
\baselinestretch=960\footnotefonts
\hfill\\
CALT-68-1756\\
DOE RESEARCH AND\\
DEVELOPMENT REPORT}}{\hbox to .35cm{\hfill}}}}
\vskip 2.0 truecm

\centerline{{{\biggfonts Aspects of Fractional
Superstrings}}\footnote{Work supported in part
by the U.S. Department of Energy under Contract no. DEAC-03-81ER40050.}}
\vskip 1.4 truecm
\centerline{{\Bigfonts Gerald B.
Cleaver}\footnote{gcleaver@theory3.caltech.edu}}
\vskip .6 truecm
\centerline{\it and}
\vskip .6 truecm
\centerline{{\Bigfonts Philip J. Rosenthal}\footnote{phil@theory3.caltech.edu}}
\vskip 1.0 truecm
\centerline{{\it California Institute of Technology, Pasadena,} CA, 91125}
\vskip 1.5 truecm
\centerline{\bf Abstract}
\vskip .3 truecm
We investigate some issues relating to recently
proposed fractional superstring theories with $D_{\rm critical}<10$.
Using the factorization
approach of Gepner and Qiu, we systematically rederive
the partition functions of the $K=4,\, 8,$ and $16$ theories
and examine their spacetime
supersymmetry. Generalized GSO projection operators for the $K=4$ model
are found. Uniqueness of the twist field, $\phi^{K/4}_{K/4}$,
as source of spacetime fermions is demonstrated.
Last, we derive a linear (rather than quadratic) relationship
between  the required conformal anomaly and the
conformal dimension of the supercurrent ghost.
\hfill\vfill\eject}

\baselinestretch=1500

\pagenumstyle{arabic}
\section{\bf Section 1: Introduction}
\sectionnumstyle{arabic}\sectionnum=1
\footnotenumstyle{arabic}
\footnotenum=0

In the last few years, several generalizations of standard (supersymmetric)
string
theory have been proposed.\markup{[\putref{ref1a,ref1b,ref1c,ref1d}]}
One of
them\markup{[\putref{ref63,ref62,ref65,ref61,ref64,ad93,ref68,cd}]} uses the
(fractional spin)
parafermions introduced from the perspective of
2-D conformal field
theory (CFT) by Zamolodchikov and Fateev\markup{[\putref{ref4}]} in 1985
and further developed by Gepner and
Qiu\markup{[\putref{ref7}]}.\footnote{This is not to be confused with the
original definition of ``parafermions.''
The term ``parafermion'' was introduced by H. S. Green in
1953.\markup{[\putref{ref13}]}  Green's parafermions are defined as
spin-1/2 particles that do not obey standard
anticommutation rules, but instead follow more general trilinear
relations.\markup{[\putref{ref3,ref14a,ref14d,ref15a,ref15b}]}}
In a series of papers, possible
new string theories with local parafermionic world sheet
currents (of fractional conformal spin) giving critical dimensions
\hbox{$D=6$, $4$, $3, {\rm ~and~}2$} have been
proposed.\markup{[\putref{ref63,ref62,ref65,ref61,ref64}]}

At the heart of these new ``fractional superstrings'' are $\Z_K$
parafermion conformal field theories (PCFT's) with central charge
$c= {2(K-1)\over K+2}$. (Equivalently, these are $SU(2)_K/U(1)$ conformal
field theories.)
The (integer) level-$K$
PCFT contains a set of unitary primary fields $\phi^j_m$, where
$0\leq j$, $\vert m\vert\leq K/2$; $j,\, m\in \Z/2$, and
$j-m = 0 \pmod{1}$,
which have the identifications
$$\phi^j_m = \phi^j_{m+K} = \phi^{{K\over 2}-j}_{m-{K\over
2}}.\eqno\eqnlabel{phidents}$$
In the range  $\vert m\vert\leq j$, the conformal dimension is
$h(\phi^j_m) = {j(j+1)\over K+2} - {m^2\over K}$. At a given level
the fusion rules are
$$\phi^{j_1}_{m_2}\times\phi^{j_2}_{m_2} = \sum^r_{j=\vert j_1 - j_2\vert}
\phi^j_{m_1+m_2}\, ,\eqno\eqnlabel{fusion}$$
where $r\equiv {\rm min} (j_1 + j_2, K-j_1-j_2)$.\footnote{For clarity, we use
``$\times$'' to denote fusion of two fields or operators,
in contrast to choosing ``$\otimes$'' (or no operator at all)
to denote a tensor product.}
This CFT contains a
subset of primary fields,
$$\{\phi_i\,\equiv\phi^0_i\equiv\phi^{K/2}_{-K/2 + i}\,\, ;\,\, 0\leq i\leq
K-1\}
\eqno\eqnlabel{subset}$$
$(\phi^{\dag}_i\equiv\phi_{K-i})$
which,  under fusion, form a closed subalgebra  possessing
a $\Z_K$ Abelian symmetry:
$$\phi_i\times\phi_j=\phi_{(i+j)} \pmod{K}\,.\eqno\eqnlabel{Zk}$$
The conformal dimensions, $h(\phi_i)$, of the fields in this subgroup have
the form
$$ h(\phi_i)= {i(K-i)\over K}\,\, . \eqno\eqnlabel{subalgcd}$$

It has been proposed that string models based on tensor products of a
level-$K$ PCFT
are generalizations of the Type II $D=10$
superstring.\markup{[\putref{ref63,ref62,ref65,ref61,ref64}]}
the standard $c={1\over 2}$ fermionic superpartner, $\psi(z)$, of the
holomorphic world sheet
scalar, $X(z)$, is replaced by the ``energy operator,''
$\epsilon(z)\equiv\phi^1_0(z)$,
of the $\Z_K$ PCFT.
(Similar substitution occurs in the antiholomorphic sector.)
Note that $\epsilon$ is not in
the $\Z_K$ Abelian subgroup, and thus is not a $\Z_K$
parafermion, except
for the degenerate $K=2$ superstring case, where $\phi^1_0\equiv\phi^0_1$.
$\epsilon$ has conformal dimension (spin) $2\over K+2$,
which is ``fractional'' ({\it i.e.}, neither integer nor half-integer)
for $K\neq 2$. This accounts for the name of these models.
Each $\epsilon-X$ pair has a total conformal anomaly (or central charge)
$c={3K\over K+2}$.

The naive generalization of the (holomorphic) supercurrent (SC) of the
standard superstring,
$J_{\rm SC}(z)= \psi(z)\cdot \partial_z X(z)$,
(where $\psi$ is a real world sheet fermion) to
$J(z)= \phi^1_0(z)\cdot \partial_z X(z)$
proves to be inadequate.\markup{[\putref{ref64}]}
Instead, the proposed ``fractional supercurrent'' (FSC) is
$$J_{\rm FSC}(z)= \phi^1_0(z)\cdot \partial_z X(z)
+ :\phi^1_0(z)\phi^1_0 (z): \, .
\eqno\eqnlabel{current}$$
$\desc$ (which vanishes for $K=2$) is the first
descendent field of $\phi^1_0$ .
$J_{\rm FSC}(z)$ is the generator of a local ``fractional'' world sheet
supersymmetry between
$\epsilon(z)$  and $X(z)$,
extending the Virasoro algebra of the stress-energy
tensor $T(z)$.  This local current of spin
\hbox{$h(J)= 1 + {2\over K+2}$}
has fractional powers of $1\over (z-w)$ in
the OPE with itself, implying a non-local world sheet interaction and,
hence, producing cuts on the world sheet.
The corresponding chiral ``fractional superconformal
algebra''\mpr{ref64} is,
\subon
$$T(z)T(w)={{1\over 2}c\over (z-w)^4}+{2T(w)\over
(z-w)^2}+...\eqno\eqnlabel{FSCalgebra-a}$$
$$T(z)J_{\rm FSC}(w)={h J_{\rm FSC}(w)\over (z-w)^2}+
{\partial J_{\rm FSC}(w)\over
(z-w)}+...\eqno\eqnlabel{FSCalgebra-b}$$
$$J_{\rm FSC}(z)J_{\rm FSC}(w)={1\over (z-w)^{2h}}+{{2h\over c}T(w)\over
(z-w)^{2h-2}}+{\lambda_K(c_0)J_{\rm FSC}(w)\over (z-w)^{h}}+{{1\over
2}\lambda_K(c_0)\partial J_{\rm FSC}(w)\over
(z-w)^{h-1}}+...\eqno\eqnlabel{FSCalgebra-c}$$
\suboff
where $c=Dc_0$. $D$ is the critical dimension,
$c_0= {3K\over K+2}$ is the central charge for one dimension,
and $\lambda_K$ is a constant.\mpr{argyres91f}

The relationship between critical dimension, $D$,  and the level, $K$,
of the parafermion CFT
may be shown to be
$$ D= 2 + {16\over K}\, , \eqno\eqnlabel{dimeqn}$$
for $K=2,\, 4,\, 8,\, 16,\, {\rm and~} \infty$.
(The $K=2$ theory
is the standard Type II superstring theory with its partition
function expressed in terms of string functions rather than
theta-functions, which implies a set of identities between these two
classes of functions.) In [\putref{ref63,ref62,ref65,ref61,ref64}]
the relationship (\puteqn{dimeqn}) is
derived by requiring a massless spin-1 particle in the open string
spectrum, produced by $\phi^1_0 (z)^{\mu}$ (where $\mu$ is the spacetime
index) operating on the vacuum.

The purpose of this paper is to examine a number of issues relating
to these models:
In section two we derive the
partition functions of the $D=6$, $4$, and $3$ theories (corresponding to
$K=4$, $8$ and $16$ respectively), using the factorization method of Gepner
and Qiu\markup{[\putref{ref7}]},
as well as
demonstrating a new approach to obtaining the superstring partition function.
In section 3 we consider other necessary elements of string
theory. In particular, we propose a generalization of the GSO
projection that applies to the fractional superstring and we address
the question of whether similar theories at different Ka\v c-Moody levels
can be constructed. Finally, we comment on the presumed ghost system
and derive a linear, rather than quadratic, relationship between
the conformal anomaly
contribution from the fractional supercurrent ghosts and the conformal
dimensions of those ghosts.
Additionally, in this section, a comparison with the superstring is made and
we attempt to elucidate its features in the current, more general
context.

\sectionnumstyle{blank}
\section{\bf Section 2: Factorization of the Fractional Superstring Partition
Functions.}
\sectionnumstyle{arabic}\sectionnum=2
\equationnum=0

We now construct the level-$K$ partition functions of these theories from the
well understood characters of SU(2) primary fields.
These closed string partition functions
$Z_K$ have the
general form $$ Z_K = a_A\, \vert A_K\vert^2 + a_B\, \vert B_K\vert^2
+ a_C\, \vert C_K\vert^2\,\, , \eqno\eqnlabel{pf1}$$
where $a_A$, $a_B$, and $a_C$ are integer coefficients.
The $A_K$ term in eq.~(\puteqn{pf1}) contains the massless graviton and
gravitino.  These $D<10$ fractional superstrings have a new feature not
present in the standard $D=10$
superstrings (which correspond to $K=2$). This is the existence of the
massive $B_K$-- and $C_K$--sectors.  These additional
sectors were originally derived by the authors of
refs.~\putref{ref63,ref62,ref65} by applying $S$ transformations
to the $A_K$--sector and then demanding modular invariance of the
theory.  In this section, we will discuss (1) new aspects of the relationship
between
the $B_K$-- and $C_K$--sectors and the $A_K$--sector
and (2) the
presence of spacetime supersymmetry (SUSY) in all sectors.
(Specifically, these type II models have $N=2$ spacetime SUSY,
with the holomorphic and
antiholomorphic sectors each effectively contributing an $N=1$ SUSY.
Hence, heterotic fractional superstrings possess only $N=1$ SUSY.)

We will
demonstrate that spacetime SUSY results from
the action of a twist current used in the
derivation of these partition functions.  Only by this twisting can
cancellation between bosonic and fermionic terms occur at each mass level
in the $A_K$-- and $B_K$--sectors.  The same twisting
results in a
``self-cancellation'' of terms in the $C_K$--sector (which exists only in
the four- and three-dimensional models).  This self-cancellation
may suggest an anyonic
interpretation of the $C_K$--sector states.
That uncompactified spacetime
anyons can presumably exist only in three or less dimensions would seem to
contradict our claim that the $K=8$
($D=4$) model may contain spacetime anyons. We
will argue shortly that one dimension of the $K=8$ fractional string is
probably compactified. Examination of the $B_K$--sector in the $D=4$ model
further suggests this. Anyonic interpretation of the $C_K$--sector fields was
first proposed in ref.~[\putref{ref63}].

Before we systematically derive the fractional superstring partition functions
(FSPF's) for each critical dimension, we
will review the character, $Z(\phi^j_m(z))$, for the Verma
module, $[\phi^j_m]$,\footnote{From here on, we do not distinguish between
the primary field $\phi^j_m$ and its complete Verma module $[\phi^j_m]$.
Thus, $\phi^j_m$ can represent either, depending on the context.}
containing a single (holomorphic) parafermionic primary field
$\phi^j_m(z)$ and its descendents.  The form of the character is
\subequationnumstyle{alphabetic}
$$ \eqalignno{
Z(\phi^j_m(z))
&= q^{-c/24}\, {\rm tr}\, q^{L_0} &\eqnlabel{2a-a}\cr
&=\eta(\tau)c^{2j}_{2m}(\tau)\,\, , &\eqnlabel{2a-b}}$$
\subequationnumstyle{blank}
where $q= {\rm e}^{2\pi \tau}$ (with $\tau$ the one-loop modular parameter)
and $\eta$ is the
Dedekind eta-function,
$$ \eta= q^{1/24}\prod^{\infty}_{n=1}(1-q^n)
\eqno\eqnlabel{defeta}$$
with $q= e^{2\pi i \tau}$. $c^{2j}_{2m}$ is a
string function\markup{[\putref{ref8a, ref8b}]} defined by
\subequationnumstyle{alphabetic}
$$\eqalignno{
c^{2j}_{2m}
&= q^{h^j_m + {1\over 4(K+2)}}{1\over \eta^3}
   \sum^{\infty}_{r,s=0} (-1)^{r+s}
   q^{r(r+1)/2 + s(s+1)/2 + rs(K+1)}\times\cr
&\quad \left\{ q^{r(j+m) + s(j-m)} - q^{K+1-2j+r(K+1-j-m)+s(K+1-j+m)}\right\}
&\eqnlabel{cfn-a}\cr
&= q^{h^j_m - {c(SU(2)_K)\over 8}}(1 + \cdots)
& \eqnlabel{cfn-b}\cr}$$
\subequationnumstyle{blank}
In this notation $h^j_m \equiv h(\phi^j_m)$ and $c(SU(2)_K)= {3K\over K+2}$.
Also, as per standard convention, the level
of the string function is suppressed.
These
string functions obey the same equivalences as their associated primary
fields $\phi^j_m$:
\subequationnumstyle{alphabetic}
$$c^{2j}_{2m}= c^{2j}_{2m+2K} = c^{K-2j}_{2m-K}\, .\eqno\eqnlabel{cid-a}$$
Additionally
$$c^{2j}_{2m}= c^{2j}_{-2m}\, .\eqno\eqnlabel{cid-b}$$
\subequationnumstyle{blank}

Since the $K=2$ theory
is the standard Type II superstring
theory,\footnote{The $K=2$ parafermion model is a
$c={1\over2}$ CFT that
corresponds to a critical Ising (free fermion) model.}
expressing its partition function
in terms of string functions rather than
theta-functions
can be accomplished simply using the following set of identities:
$$ {K=2: \cases{2\eta^2(c^1_1)^2 = \vartheta_2/\eta\, ;\cr
               \eta^2(c^0_0 + c^2_0)^2 = \vartheta_3/\eta\, ;\cr
               \eta^2(c^0_0 - c^2_0)^2 = \vartheta_4/\eta\, .}}
\eqno\eqnlabel{fpequiv}$$

For each spacetime dimension in these theories, a term in the
partition function  of the form (\puteqn{2a-b})
is tensored with the partition function $Z\left(X\right)$
for an uncompactified chiral boson $X(z)$. Since,
$$ Z\left(X(z)\right)\propto {1\over\eta(\tau)}\,\, ,
\eqno\eqnlabel{partstdim}$$
the $\eta(\tau)$ factors cancel out in
$Z(\phi^j_m(z))\times Z(X(z))$. Similar cancellation of
$\bar\eta(\bar\tau)$ occurs in the antiholomorphic sector.
In the
following partition functions, we generally suppress the trivial factor of
$({\rm Im}\, \tau)^{-8/K}$ contributed together by the $D-2$ holomorphic and
anti-holomorphic world sheet boson partition functions.

\sectionnumstyle{blank}
\subsectionnumstyle{blank}
\subsection{\bf 2.1: Derivation of the Partition Functions}
\sectionnumstyle{arabic}\sectionnum=2

By the string function equivalences, the partition functions for the
level-$K$ fractional superstrings in
refs.~[\putref{ref63,ref62,ref65,ref12}] in critical spacetime dimensions
\hbox {$D= 2 + {16\over K}= 10, 6,$} {$\,4,\, {\rm~and~}3$}
can be written (in light-cone gauge) as:

{\settabs 8 \columns
\+ \cr
\+ $D=10$ & $(K=2)$: &&$Z_{2} =  \vert A_2\vert^2$, ~where\cr}
$$\eqalignno{ A_2 &= {1\over 2}\left\{ (c^0_0 + c^2_0)^8 - (c^0_0 -
c^2_0)^8 \right\}_{\rm boson} - 8(c^1_1)^8_{\rm fermion}\cr
&= 8\left\{ (c^0_0)^7
c^2_0 + 7(c^0_0)^5(c^2_0)^3 +7(c^0_0)^3(c^2_0)^5 +
c^0_0(c^2_0)^7\right\}_{\rm boson}  -
8(c^1_1)^8_{\rm fermion} &\eqnlabel{part2}}$$ {\settabs 8 \columns
\subon
\+ $D=6$ & $(K=4)$: &&$Z_4= \vert A_4\vert^2 + 3\vert B_4\vert^2$, ~where\cr}
$$\eqalignno {A_4 &= {\rm\hskip .37 truecm}4\left\{(c^0_0 + c^4_0)^3
(c^2_0) - (c^2_0)^4\right\}\cr &\quad + 4\left\{(c^0_2 + c^4_2)^3 (c^2_2) -
(c^2_2)^4\right\}&\eqnlabel{part4-a}\cr B_4 &= {\rm\hskip
.37truecm}4\left\{(c^0_0 +
c^4_0)(c^0_2+c^4_2)^2(c^2_0) - (c^2_0)^2(c^2_2)^2\right\}\cr &\quad +
4\left\{(c^0_2 + c^4_2)(c^0_0+c^4_0)^2(c^2_2) -
(c^2_2)^2(c^2_0)^2\right\}&\eqnlabel{part4-b}\cr}$$
{\settabs 8 \columns
\suboff
\subon
\+ $D=4$ & $(K=8)$:
&&$Z_8= \vert A_8\vert^2 + \vert B_8\vert^2 + 2\vert C_8\vert^2$, ~where\cr}
$$\eqalignno { A_8 & = {\rm\hskip .37 truecm}2\left\{(c^0_0 + c^8_0)(c^2_0
+ c^6_0) - (c^4_0)^2\right\}\cr &\quad +2\left\{(c^0_4+ c^8_4)(c^2_4+c^6_4)
- (c^4_4)^2\right\}&\eqnlabel{part8-a}\cr B_8 &= {\rm\hskip .37
truecm}2\left\{(c^0_0+c^8_0)(c^2_4+c^6_4)-(c^4_0c^4_4)\right\}\cr &\quad +
2\left\{(c^0_4+c^8_4)(c^2_0+c^6_0)-(c^4_4c^4_0)\right\}&\eqnlabel{part8-b}\cr
C_8 &= {\rm
\hskip .37truecm}2\left\{(c^0_2 + c^8_2)(c^2_2 + c^6_2) -
(c^4_2)^2\right\}\cr &\quad +2\left\{(c^0_2 + c^8_2)(c^2_2 + c^6_2) -
(c^4_2)^2\right\}&\eqnlabel{part8-c}\cr}$$ {\settabs 8 \columns
\suboff
\subon
\+ $D=3$ & $(K=16)$:
&&$ Z_{16} = \vert A_{16}\vert^2 + \vert C_{16}\vert^2 $, ~where\cr}
$$\eqalignno{A_{16} &= {\rm\hskip .3 truecm}\left\{(c^2_0 + c^{14}_0) -
c^8_0\right\}\cr &\quad +\left\{(c^2_8 + c^{14}_8) - c^8_8\right\}
&\eqnlabel{part16-a}\cr
C_{16} &= {\rm\hskip .3 truecm}\left\{(c^2_4 + c^{14}_4) -
c^8_4\right\}\cr &\quad +\left\{(c^2_4 + c^{14}_4) -
c^8_4\right\}\,\, .&\eqnlabel{part16-b}\cr}$$
\suboff

The $D=10$ partition function, written in string function format,
was obtained by the authors of refs.~[\putref{ref63,ref65}] as a check of
their program, both by computer generation and by the
$K=2$ string functions/Jacobi $\theta$-functions equivalences.
In each model, the massless spin-2 particle and its supersymmetric partner
arise from the $A_K$--sector. The $B_K$-- and $C_K$--sectors were obtained by
acting on the $A_K$--sector with
the $SL(2,\Z)$ modular group generators, $S: \tau\rightarrow -1/\tau$,
and $T:\tau\rightarrow \tau +1$.
At each level of $K$, the contribution of each sector is separately zero.
This is
consistent with spacetime SUSY and suggests cancellation
between bosonic and fermionic terms at each mass level.
This leads to the following
identities\markup{[\putref{ref63}]}:
$$A_2=A_4=B_4=A_8=B_8=C_8=A_{16}=C_{16}=0\,\, .\eqno\eqnlabel{partident}$$

The factorization method of Gepner and Qiu\markup{[\putref{ref7}]} for
string function partition functions allows us to rederive the above
partition functions systematically.
Gepner and Qiu have shown that we can express any
general modular invariant parafermionic partition function,
\subequationnumstyle{alphabetic}
$$ Z= \vert \eta\vert^2 \sum N_{l,n,\bar l,\bar n} c^l_n \bar c^{\bar
l}_{\bar n}\, ,\eqno\eqnlabel{partfn2-a}$$
in the form
$$Z= \vert\eta\vert^2\sum
{1\over 2}L_{l,\bar l}M_{n,\bar n}c^l_n \bar c^{\bar l}_{\bar n}\,
,\eqno\eqnlabel{partfn2-b}$$
\subequationnumstyle{blank}
(with $c^{l=2j}_{n=2m}=0$ unless $l-n\in 2\Z$ since $\phi^j_m=0$ for
$j-m\not\in\Z$).
This factorization of
$N_{l,n,\bar l,\bar n}$
results from the transformation properties of
$c^l_n$ under the modular group generators $S$ and
$T$:
\subequationnumstyle{alphabetic}
$$\eqalignno{S: c^l_n &\rightarrow {1\over \sqrt{-i\tau K(K+2)}}
\sum_{l'=0}^{K}\sum_{n'= -K+1 \atop l'-n`\in 2Z}^K \exp\left\{{i\pi n
n'\over K}\right\} \sin\left\{{\pi (l+1)(l'+1)\over K+2}\right\} c^{l'}_{n'}
{\rm ~~~~~~~~~} &\eqnlabel{ctrans-a}\cr
T: c^l_n &\rightarrow \exp\left\{2\pi
i\left({l(l+2)\over 4(K+2)} - {n^2\over 4K} -
{K\over 8(K+2)}\right)\right\} c^l_n\,\, ,
&\eqnlabel{ctrans-b}\cr}$$
\subequationnumstyle{blank}
which are an effect of the
definition of string functions $c^l_n$ (at level-$K$) in terms of the
$SU(2)_K$ affine characters $\chi_l$ and the Jacobi theta-function
$\theta_{n,K}$:\footnote{The associated relationship between the level-$K$
$SU(2)$ primary fields $\Phi^j$ and the parafermionic $\phi^j_m$ is
$$\Phi^j= \sum_{m=-j}^j \phi^j_m\,  :\exp\left\{ {i{m\over \sqrt{K}}
\varphi}\right\}:$$
where $\varphi$ is the $U(1)$ boson field of the $SU(2)$ theory.}
$$\chi_l(\tau)
= \sum^K_{n= -K+1} c^l_n(\tau)\theta_{n,K}(\tau)\,
.\eqno\eqnlabel{partfn4}$$

Gepner and Qiu proved that as a result of the factorization,
$$N_{l,n,\bar l,\bar n} = {1\over 2}L_{l,\bar l}\, M_{m,\bar m}\,\, ,
\eqno\eqnlabel{Nfactorization}$$
we can construct all modular invariant partition functions
(MIPF's) for parafermions from a product
of modular invariant solutions for the $(l,\bar l)$ and
$(n,\bar n)$ indices separately. That is,
eq.~(\puteqn{partfn2-b}) is modular invariant if and only if the
$SU(2)$ affine partition function
\subequationnumstyle{alphabetic}
$$ W=
\sum_{l,\bar l= 0}^K L_{l,\bar l}\chi_l(\tau)\bar\chi_{\bar
l}(\bar\tau)\eqno\eqnlabel{pfq-a}$$
and the $U(1)$ partition function
$$ V= {1\over\vert\eta(\tau)\vert^2}
\sum_{n,\bar n= -K+1}^K M_{n,\bar n}\theta_{n,K}\bar\theta_{\bar
n,K}\eqno\eqnlabel{pfq-b}$$
\subequationnumstyle{blank}
 are simultaneously modular invariant;
{\it i.e.} $N_{l,n,\bar l,\bar n}= {1\over 2}L_{l,\bar l}M_{n,\bar n}$ belongs
to a
MIPF (\puteqn{partfn2-a}) if and only if $L_{l,\bar l}$ and $M_{n,\bar n}$
correspond to MIPF's of the forms (\puteqn{pfq-a}) and
(\puteqn{pfq-b}), respectively.

\subequationnumstyle{alphabetic}
The proof of Gepner and Qiu can also be applied to show that
for modular invariance of
a $d$-dimensional (where $d=D-2$) parafermion tensor product theory,
$$Z =
\vert \eta\vert^2 \sum {\bmit N}_{\vec l,\vec n,\vec {\bar l},\vec {\bar n}}
c^{l_1}_{n_1}c^{l_2}_{n_2}\cdots c^{l_{d}}_{n_{d}}
\bar c^{\bar l_1}_{\bar n_1}\bar c^{\bar l_2}_{\bar n_2}\cdots
\bar c^{\bar l_{d}}_{\bar n_{d}}\,\, ,
\eqno\eqnlabel{tppf-a}$$
is necessary that there be
affine$\,\times\, U(1)$ factorization,
$$Z = \vert\eta\vert^2\sum
{1\over 2}{\bmit L}_{\vec l,\vec {\bar l}}{\bmit M}_{\vec n,\vec {\bar n}}
c^{l_1}_{n_1}c^{l_2}_{n_2}\cdots c^{l_{d}}_{n_{d}}
\bar c^{\bar l_1}_{\bar n_1}\bar c^{\bar l_2}_{\bar n_2}\cdots
\bar c^{\bar l_{d}}_{\bar n_{d}}\,\, ,
\eqno\eqnlabel{tppf-b}$$
with $\bmit L$, and $\bmit M$ corresponding to $d$-dimensional
modular invariant generalizations of eqs. (\puteqn{pfq-a}) and
(\puteqn{pfq-b}).
\subequationnumstyle{blank}

Due to the nature of a $U(1)$ CFT, it is obvious that
a tensor $\bmit M$ corresponding to a MIPF for $d$-factors of $U(1)$
CFT's can be written as a tensor product of $d$-independent matrix $\bmit M$
solutions to (\puteqn{pfq-b}) ``twisted'' by simple
currents $\cal J$.\footnote{A simple
current, $\cal J$, is a primary field of a CFT
which, when fused with any other primary field, $\Phi_l$, (including itself)
in the CFT produces only a single primary field as a product state:
${\cal J}\times\Phi_l = \Phi_{l'}$.}

In the following pages we demonstrate that the factorization approach to
deriving
the FSPF's,
suggests much about the meaning of the different sectors in
fractional superstrings, the
related ``projection'' terms,
the origin of spacetime supersymmetry,
and the significance of a special $U(1)$ twist current.

\subsectionnumstyle{blank}
\subsection{{\bf 2.2}: \it Affine Factor and ``W'' Partition Function}
\subsectionnumstyle{blank}

In the $A_K$--sectors defined by eqs.~(\puteqn{part4-a}, \puteqn{part8-a},
\puteqn{part16-a}),
the terms inside the first (upper) set of brackets,
carry ``$n\equiv 2m=0$'' subscripts and can be shown to
correspond to spacetime bosons;
while the terms inside the second (lower) set carry ``$n= K/2$'' and
correspond to spacetime fermions.  Expressing the $A_K$--sectors in this form
makes a one--to--one correspondence between bosonic and fermionic states in the
$A_K$--sector manifest. If we remove the subscripts on the string functions
in the bosonic and fermionic subsectors
(which is parallel to replacing $c^l_n$ with
$\chi_l$) we find the subsectors become equivalent. In fact, under this
operation of removing the ``$n$'' subscripts and replacing each string function
by its
corresponding affine character
(which we will denote by $\buildrel {\rm
affine}\over\Longrightarrow$), all sectors become equivalent up to an
integer coefficient:
\subon
{\settabs 8 \columns\+ $D=6$ &$(K=4)$:\cr} $$A_4,B_4 \hbox to
1cm{\hfill}{\buildrel {\rm
affine}\over\Longrightarrow}\hbox to 1cm{\hfill}
A_4^{\rm aff}\equiv (\chi_0+\chi_K)^3\chi_{2}-(\chi_{K/2})^4
\eqno\eqnlabel{affine-a}$$
{\settabs 8 \columns\+ $D=4$ &$(K=8)$:\cr} $$A_8,B_8,C_8
\hbox to 1cm{\hfill}{\buildrel {\rm affine}\over\Longrightarrow}
\hbox to 1cm{\hfill}
A_8^{\rm aff}\equiv (\chi_0+\chi_K)(\chi_2+\chi_{K-2})
- (\chi_{K/2})^2 \eqno\eqnlabel{affine-b}$$ {\settabs 8 \columns\+ $D=3$
&$(K=16)$:\cr} $$A_{16},C_{16}
\hbox to 1cm{\hfill}{\buildrel {\rm affine}\over\Longrightarrow}\hbox to
1cm{\hfill}
A_{16}^{\rm aff}\equiv (\chi_2 + \chi_{K-2}) -\chi_{K/2}\,\,
.\eqno\eqnlabel{affine-c}$$
\suboff

\noindent
We see that the $B$-- and $C$--sectors are not arbitrary additions,
necessitated only by modular invariance, but rather are naturally
related to the physically motivated $A$--sectors: the
corresponding affine partition function is the same for each sector.
Further, the affine characters, $A^{\rm aff}_K$ in
eqs.~(\puteqn{affine-a}, \puteqn{affine-b}, \puteqn{affine-c}) all have the
general form
$$ A_K^{\rm aff} \equiv (\chi_0 +\chi_K)^{D-3}(\chi_2+\chi_{K-2}) -
(\chi_{K/2})^{D-2}\,\, . \eqno\eqnlabel{affall}\,\, ,$$
(with $(\chi_2+\chi_{K-2})$ replaced by $\chi_2$ for $K=16$,
since then $\chi_2=\chi_{K-2}$).
The corresponding affine partition function is
$$ W_K = \vert A^{\rm aff}_K \vert^2\, .\eqno\eqnlabel{affine2}$$

This class of partition functions (\puteqn{affine2}) is indeed modular
invariant and possesses special qualities.
(Note that the modular invariance of $W$ requires $A_K^{\rm aff}$ to
transform back into itself under $S$.)
This is easiest to show for $K=16$.
The $SU(2)_{16}$ MIPF's for $D=3$ are trivial to classify since
at this level the A--D--E classification forms a complete basis set of modular
invariants, even for MIPF's containing terms with negative coefficients. The
only free parameters in $K=16$ affine partition functions $Z\left(
SU(2)_{16}\right)$ are integers $a$, $b$, and $c$ where $$Z\left(
SU(2)_{K=16}\right) =
a\, Z({\rm A}_{17})+b\, Z({\rm D}_{10})+c\, Z({\rm E}_7)\,\,
.\eqno\eqnlabel{affine3}$$

Demanding that neither a left- nor a right-moving tachyonic state be in
the Hilbert space of states in the $K=16$
fractional superstring when the intercept $v$, defined by $$L_0\vert
{\rm physical}\rangle = v\vert {\rm physical}\rangle\, ,
\eqno\eqnlabel{intercept}$$
is positive, removes these degrees of freedom and requires $a= -(b+c)=0$,
independent of the possible $U(1)$ partition
functions.
These specific values for $a$, $b$, and $c$ give us (\puteqn{affine2})
for this level:
$$W_{16} = Z({\rm D}_{10})-Z({\rm E}_{7})
= \vert A^{\rm aff}_{16}\vert^2
\,\, .\eqno\eqnlabel{16deriv}$$

The corresponding partition functions for $K= 8$ and $4$
can also be expressed as the difference of two
known partition functions:
$W_8$ is the difference between a ${\rm D}_6\otimes{\rm D}_6$
MIPF and an exceptional MIPF derived using the conformal
embedding
$$SU(2)^{(1)}_{K=8}\otimes SU(2)^{(1)}_{K=8}
\otimes SO(8)^{(1)}_{K=4}\subset SO(32)^{(1)}_{K=1}
\eqno\eqnlabel{8derive}$$
and the triality of $SO(8)$.\markup{[\putref{new1,new2}]}
$W_4$ is similarly the difference between a
${\rm D}_4\otimes{\rm D}_4\otimes{\rm D}_4\otimes{\rm D}_4$
MIPF and a simple current invariant of
${\rm D}_4\otimes{\rm D}_4\otimes{\rm D}_4\otimes{\rm D}_4$.
Although it is not possible to create the needed simple currents
from $SU(2)_4$ fields, this may be realized using those of $SU(3)_1$.
This is possible because the ${\rm D}_4$ invariant of $SU(2)_4$
is equivalent to the diagonal invariant of $SU(3)_1$. When rewritten
in $SU(3)_1$ language, the $[SU(2)_4]^4$ tensor product field
$(\Phi^0 + \Phi^K)^{3}\Phi^{2}(\bar\Phi^{K/4})^{4}$
becomes the simple current,
$(\Phi^0_{SU(3)})^{3}\Phi^{\pm 3}_{SU(3)}(\bar\Phi^{\pm 3}_{SU(3)})^{4}$
of $[SU(3)_1]^4$, where $\Phi^0_{SU(3)}$ is the $SU(3)_1$ identity field, and
$\Phi^{+3}_{SU(3)}$ and $\Phi^{-3}_{SU(3)}$ denote the 3 and $\bar 3$
representation fields.

\subequationnumstyle{alphabetic}
The parafermion partition function corresponding to
eqs.~(\puteqn{affall}) and (\puteqn{affine2}) is
$$ Z_K({\rm affine ~factor})
= \vert (c^0_0 +c^K_0)^{D-3}(c^2_0+c^{K-2}_0) - (c^{K/2}_0)^{D-2}\vert^2 \,\, ,
\eqno\eqnlabel{affinepk-a}$$ representing the states created by the fields
$$\left(
(\phi^0_0 +\phi^{K/2}_0)^{D-3}(\phi^1_0+\phi^{(K-2)/2}_0)
- (\phi^{K/4}_0)^{D-2}
\right)
\left(
(\bar\phi^0_0 +\bar\phi^{K/2}_0)^{D-3}(\bar\phi^1_0+\bar\phi^{(K-2)/2}_0)
- (\bar\phi^{K/4}_0)^{D-2}
\right)
\eqno\eqnlabel{affinepk-b}$$
\subequationnumstyle{blank}
acting on the parafermion vacuum.
In ``factoring'' the parafermion partition functions
(\puteqn{part4-a},b), (\puteqn{part8-a}-c), and (\puteqn{part16-a},b),
into separate partition functions for the affine and $U(1)$ contributions
we do not intend to imply that the string functions
can actually be factored into $c^l_0\times c^0_n=c^l_n$, nor do we
imply that $Z_K({\rm affine ~factor})$ above or $Z_K(U(1){\rm ~factor})$
presented in the following subsection are modular invariant.
Rather, we mean to use partition functions
(\puteqn{affinepk-a}) and (2.30b, 2.32b, 2.34b)
only as an artificial construct for
developing a deeper understanding of the
function of the parafermion primary fields $\phi^j_0$
and $\phi^0_m$ in these models.
Though not true for the partition functions,
factorization is, indeed, valid for the primary fields, $\phi^j_m$:
$\phi^j_0\otimes\phi^0_m=\phi^j_m$ (for integer $j,\, m$).

\subsection{{\bf 2.3}: \it $U(1)$ Factor and the ``$V$'' Partition Function}

We now consider the $U(1)$ factor, $\bmit M$,
carrying the $(n,\bar n)$-indices in the FSPF's.
Since all
$A_K$--, $B_K$--, and $C_K$--sectors in the level-$K$ fractional
superstring partition
function (and even the boson and fermion subsectors separately in $A_K$)
contain the
same affine factor, it is clearly the choice of the $U(1)$
factor which determines the spacetime supersymmetry of the
fractional superstring theories. That is, spacetime spins of particles in the
Hilbert space of states depend upon the
${\bmit M}$'s
that are allowed in tensored versions of eq.~(\puteqn{pfq-b}).
In the case of matrix $\bmit M$ rather than a more
complicated tensor, invariance of (\puteqn{pfq-b})
under $S$ requires that the components $M_{n\bar n}$
be related by
\subon
$$ M_{n',\bar n'} = {1\over 2K}\sum_{n,\bar n}M_{n,\bar n}
{\rm e}^{i\pi nn'/K} {\rm e}^{i\pi \bar n \bar n'/K}\,\, ,
\eqno\eqnlabel{m-a}$$
and $T$ invariance
demands that $$ {n^2 - \bar n^2\over 4K}\in \Z\,\, ,
{\rm ~~if~} M_{n,\bar n}\neq 0\,\,. \eqno\eqnlabel{m-b}$$
\suboff
At every level-$K$ there is a unique modular
invariant function corresponding to
each factorization\markup{\putref{ref7}]},
$\alpha\times\beta=K$, where $\alpha,\,\, \beta\in\Z$.
Denoting the matrix elements of ${\bmit M}^{\alpha,\beta}$ by
$M^{\alpha,\beta}_{n,\bar n}$,
they are given by
$$M^{\alpha,\beta}_{n,\bar n} = {1\over 2}
\sum_{x\in {\Z}_{2\beta}\atop y\in {\Z}_{2\alpha}}
\delta_{n,\alpha x +\beta y}\delta_{\bar n,\alpha x -\beta y}\,\, .
\eqno\eqnlabel{m-c}$$
By (\puteqn{m-c}),
$M^{\alpha,\beta}_{n,\bar n}= M^{\beta,\alpha}_{n,-\bar n}$.
Hence, ${\bmit M}^{\alpha,\beta}$ and ${\bmit M}^{\beta,\alpha}$
result in equivalent FSPF's.
To avoid this redundancy, we demand that $\alpha\leq\beta$.

Thus,
for $K=4$ the two distinct choices for the matrix ${\bmit M}^{\alpha,\beta}$
are ${\bmit M}^{1,4}$ and ${\bmit M}^{2,2}$; for
$K=8$, we have ${\bmit M}^{1,8}$ and ${\bmit M}^{2,4}$; and
for $K=16$, the three alternatives are ${\bmit M}^{1,16}$,
${\bmit M}^{2,8}$, and ${\bmit M}^{4,4}$.
${\bmit M}^{1,K}$ represents the level-$K$ diagonal, $n=\bar n$,
partition function.
${\bmit M}^{\alpha, \beta={K\over\alpha}}$
corresponds to the
diagonal partition function twisted by a $\Z_{\alpha}$ symmetry.
(Twisting by $\Z_{\alpha}$ and $\Z_{K/\alpha}$ produce isomorphic
models.)
Our investigations revealed that all possible
simple
tensor product combinations of
these ${\bmit M}^{\alpha,\beta}$ matrices are insufficient for
producing fractional superstrings with spacetime SUSY (and, thus, no
tachyons).
We have found that twisting by a special simple $U(1)$ current
(shown below) is required to achieve this.
Of the potential choices of ${\bmit M}$ from a $U(1)$ MIPF
that could be combined with ${\bmit L}$ from an affine MIPF,
one can show (as indicated from string function identities)
that the following are the only ones
producing numerically zero FSPF's:

{\settabs 8 \columns
\+ $D=6$ & $(K=4)$:\cr
\+ \cr}

The ${\bmit M}= {\bmit M}^{2,2}\otimes{\bmit M}^{2,2}\otimes{\bmit
M}^{2,2}\otimes{\bmit M}^{2,2}$ model twisted by the
simple $U(1)$ current\footnote{Recall that
the parafermion primary fields $\phi^0_m$
have simple fusion rules,
$\phi^0_m\otimes\phi^0_{m'}=\phi^0_{m+m'\pmod{K}} $
(for $m,\, m'\in {\Z}$) and form a ${\Z}_K$ closed subalgebra.
This fusion rule, likewise, holds for
the $U(1)$ fields $:\exp \{i{m\over K}\varphi\} :.$  The isomorphism
makes it clear that any simple current, ${\cal J}_K$, in this subsection
that contains only integer $m$ can be
expressed equivalently either in terms of these parafermion fields
or in terms of $U(1)$ fields.
In view of the following discussion, we express all of the
simple twist currents, ${\cal J}_K$, in parafermion language.}

$${\cal J}_4\equiv \phi_{K/4}^0\phi_{K/4}^0\phi_{K/4}^0\phi_{K/4}^0
\bar\phi_0^0\bar\phi_0^0\bar\phi_0^0\bar\phi_0^0 \eqno\eqnlabel{j4}$$
has the following $U(1)$ partition function:
\subon
$$\eqalignno{V_4 &= \hbox to .25cm{\hfill}[(\t_{0,4}+\t_{4,4})^4(\bt_{0,4}
 + \bt_{4,4})^4 +
(\t_{2,4} +\t_{-2,4})^4(\bt_{2,4}+ \bt_{-2,4})^4\cr
&\hbox to 1em{\hfill}
+ (\t_{0,4} +\t_{4,4})^2(\t_{2,4}+\t_{-2,4})^2(\bt_{
0,4} +\bt_{4,4})^2(\bt_{2,4} + \bt_{-2,4})^2\cr
&\hbox to 1em{\hfill}
+ (\t_{2,4}+\t_{-2,4})^2(\t_{0,4}+\t_{4,4})^2(\bt_{2,4} +\bt_{-2,4})^2
(\bt_{0,4} + \bt_{4,4})^2]_{\rm untwisted}\cr
&&\eqnlabel{nn4-a}\cr
&\hbox to 1em{\hfill}
+ [(\t_{2,4}+\t_{-2,4})^4(\bt_{0,4}+\bt_{4,4})^4 + (\t_{4,4}+\t_{0,4})^4
(\bt_{2,4}+\bt_{-2,4})^4\cr
&\hbox to 1em{\hfill}
+ \hbox to
.15cm{\hfill}(\t_{2,4}+\t_{-2,4})^2(\t_{4,4}+\t_{0,4})^2(\bt_{0,4}+\bt_{4,4})^2
(\bt_{2,4}+\bt_{-2,4})^2\cr
&\hbox to 1em{\hfill}
+ \hbox to .15cm{\hfill}(\t_{4,4}+\t_{0,4})^2(\t_{2,4}+\t_{-2,4})^2
(\bt_{2,4} + \bt_{-2,4})^2(\bt_{0,4} +\bt_{4,4})^2]_{\rm twisted}\,\, .}$$

Writing this in parafermionic form, and then
using string function identities, followed by
regrouping according to $A_4$ and $B_4$ components,
results in
$$Z_4(U(1){\rm ~ factor})= \vert (c^0_0)^4 +
(c^0_2)^4\vert^2_{_{(A_4)}} + \vert (c^0_0)^2(c^0_2)^2 + (c^0_2)^2(c^0_0)^2
\vert^2_{_{(B_4)}}\,\, ,
\eqno\eqnlabel{nn4-b}$$
which represents the tensor product primary fields
$$\left\{
\left[ (\phi^0_0)^4 + (\phi^0_1)^4 \right]
\left[ (\bar\phi^0_0)^4 + (\bar\phi^0_1)^4 \right]
+
\left[ (\phi^0_0)^2(\phi^0_1)^2 + (\phi^0_1)^2(\phi^0_0)^2\right]
\left[ (\bar\phi^0_0)^2(\bar\phi^0_1)^2 +
(\bar\phi^0_1)^2(\bar\phi^0_0)^2\right]
\right\}
\eqno\eqnlabel{nn4-c}$$
\suboff
acting on the parafermion vacuum.

{\settabs 8\columns
\+ \cr
\+ $D=4$ & $(K=8)$:\cr
\+\cr}

The ${\bmit M}= {\bmit M}^{2,4}\otimes{\bmit M}^{2,4}$ model twisted by the
simple $U(1)$ current $${\cal J}_8\equiv
\phi_{K/4}^0\phi_{K/4}^0\bar\phi_0^0\bar\phi_0^0\eqno\eqnlabel{j8}$$
corresponds to
\subon
$$\eqalignno{V_8 &= \hbox to .35cm{\hfill}[(\t_{0,8}+\t_{8,8})(\bt_{0,8}
+ \bt_{8,8}) +
(\t_{4,8}+\t_{-4,8})(\bt_{4,8}+\bt_{-4,8})]^2_{\rm untwisted}\cr
&\hbox to 1em{\hfill}
+ [(\t_{2,8}+\t_{-6,8})(\bt_{2,8}+\bt_{-6,8})
+(\t_{-2,8}+\t_{6,8})(\bt_{-2,8}+\bt_{6,8})]^2_{\rm untwisted}\cr
&&\eqnlabel{nn8-a}\cr
&\hbox to 1em{\hfill}
+ [(\t_{4,8}+\t_{-4,8})(\bt_{0,8} + \bt_{8,8})
+ (\t_{0,8}+\t_{8,8})(\bt_{4,8} + \bt_{-4,8})]^2_{\rm twisted}\cr
&\hbox to 1em{\hfill} +
[(\t_{6,8}+\t_{-2,8})(\bt_{2,8}+\bt_{-6,8})
+(\t_{2,8}+\t_{-6,8})(\bt_{-2,8}  +\bt_{6,8})]^2_{\rm twisted}\,\, .}$$
Hence, we have
$$ Z_8(U(1){\rm ~factor})=
\vert (c^0_0)^2 +
(c^0_4)^2\vert^2_{_{(A_8)}} + \vert (c^0_0)(c^0_4)
+(c^0_4)(c^0_0)\vert^2_{_{(B_8)}} + 4\vert
(c^0_2)^2\vert^2_{_{(C_8)}}\,\, , \eqno\eqnlabel{nn8-b}$$
with the related tensor product primary fields
$$
\left\{
\left[ (\phi^0_0)^2 +(\phi^0_2)^2 \right]
\left[ (\bar\phi^0_0)^2 +(\bar\phi^0_2)^2 \right]
+
\left(\phi^0_0 \phi^0_2 +\phi^0_2\phi^0_0 \right)
\left(\bar\phi^0_0 \bar\phi^0_2 +\bar\phi^0_2\bar\phi^0_0 \right)
+
4(\phi^0_1)^2(\bar\phi^0_1)
\right\}
\,\, . \eqno\eqnlabel{nn8-c}$$
\suboff

{\settabs 8\columns
\+ \cr
\+ $D=3$ & $(K=16)$:\cr
\+\cr}

The ${\bmit M}= {\bmit M}^{4,4}$ model twisted by the simple $U(1)$ current
$${\cal J}_{16}\equiv \phi_{K/4}^0\bar\phi_0^0\eqno\eqnlabel{j16}$$
produces,
\subon
$$\eqalignno{ V\left(K=16\right) &= \hbox to .38cm{\hfill}\vert (\t_{0,16} +
\t_{16,16}) +
(\t_{8,16} +\t_{-8,16})\vert^2_{\rm untwisted}\cr
&\cr
&\hbox to 1em{\hfill}
+ \vert (\t_{4,16} + \t_{-4,16}) + (\t_{12,16} +\t_{-12,16})
\vert^2_{\rm untwisted}\,\, .
 &\eqnlabel{nn16-a}}$$
Thus, the corresponding parafermion partition function is
$$Z_{16}(U(1){\rm ~factor})=
\vert c^0_0 + c^0_8\vert^2_{_{(A_{16})}} +
4\vert c^0_4\vert^2_{_{(C_{16})}}\,\, ,
\eqno\eqnlabel{nn16-b}$$
and its primary fields are
$$\left\{
\left(\phi^0_0 + \phi^0_4\right)
\left(\bar\phi^0_0 + \bar\phi^0_4\right)
+ 4
\left(\phi^0_2\right)
\left(\bar\phi^0_2\right)
\right\}\,\, .
\eqno\eqnlabel{nn16-c}$$
\suboff
(In this case the twisting is trivial since ${\cal J}_{16}$ is already
present in the initial untwisted model.)

We wish to point out that
the partition function for the standard $D=10$ superstring can also be
factored into affine and $U(1)$ parts:
{\settabs 8\columns
\+ \cr
\+ $D=10$ & $K=2$:\cr}
\subon
$$ A_2 {\buildrel {\rm affine}\over\Longrightarrow}
\sum_{i {\rm ~odd~}= 1}^7 {8\choose i}(\chi_0)^{i}(\chi_K)^{8-i} -
(\chi_{K/2})^8 \eqno\eqnlabel{affine10-a}$$
The accompanying $U(1)$ factor is
$$\eqalignno{
V_2&=
\phantom{35\times}\vert (\vartheta_{0,2})^8 +
(\vartheta_{1,2})^8 + (\vartheta_{-1,2})^8 + (\vartheta_{2,2})^8\vert^2\cr
&\phantom{= } + 35\vert(\vartheta_{0,2} + \vartheta_{2,2})^4
(\vartheta_{1,2} + \vartheta_{-1,2})^4\vert^2\cr
&\phantom{= } + 35[(\vartheta_{1,2}+ \vartheta_{-1,2})^4
(\vartheta_{0,2}+\vartheta_{2,2})^4]
[(\bar\vartheta_{0,2}+\bar\vartheta_{2,2})^4
 (\bar\vartheta_{1,2}+\bar\vartheta_{-1,2})^4]\,\, .
&\eqnlabel{affine10-b}}$$
\suboff
This originates from the
$${\bmit M}= {\bmit M}^{2,1}\otimes{\bmit M}^{2,1}\otimes{\bmit M}^{2,1}
\otimes{\bmit M}^{2,1}\otimes{\bmit M}^{2,1}\otimes{\bmit M}^{2,1}
\otimes{\bmit M}^{2,1}\otimes{\bmit M}^{2,1}$$
model being twisted by the simple $U(1)$ current
$${\cal J}^{\rm theta}_2\equiv ( :\exp\{i\varphi/2 \}:)^8
\eqno\eqnlabel{j2}$$
The difference between the factorization for $K=2$
and those for $K>2$ is that
here we cannot define an actual parafermion twist current $(\phi_{K/4}^0)^8$
since $\phi^0_{K/4}=0$ for $K=2$.
Relatedly, the effective
$Z_2(U(1) {\rm~factor})$ contributing to eq.~(\puteqn{part2})
reduces to just the first mod-squared term in eq.~(\puteqn{affine10-b}) since
$c^l_n\equiv 0$ for $l-n\neq 0 \pmod{2}$.

All of the above simple twist currents for $K>2$ are of the general form
$${\cal J}_K =
(\phi^0_{K/4})^{D-2}(\bar\phi^0_0)^{D-2}\,\, .
\eqno\eqnlabel{gensc}$$
(Note that $\bar {\cal J}_K\equiv
(\phi^0_0)^{D-2}(\bar\phi^0_{K/4})^{D-2}$ is
automatically generated as a twisted field also.)
We believe this specific class of twist currents
is the key to spacetime
supersymmetry in the parafermion models.
Without the twisting effects of ${\cal J}_K$,
numerically zero modular invariant FSPF's in three, four, and six
dimensions  cannot be formed and thus spacetime SUSY would be impossible.
This twisting also reveals much about the necessity of
non-$A_K$--sectors.  Terms from the twisted and untwisted
sectors of these models become equally mixed in the $\vert A_K\vert^2$,
$\vert B_K\vert^2$, and $\vert C_K\vert^2$ contribution to the level-$K$
partition function.
Further, this twisting keeps the string functions with $n\not\equiv 0,\, K/2
\pmod{K}$ from mixing with those possessing $n\equiv 0,K/2 \pmod{K}$.  This is
especially significant since we believe the former string functions
in the $C_K$--sector likely
correspond to spacetime fields of fractional spin-statistics ({\it i.e.,}
anyons)
and the latter in both $A_K$ and $B_K$ to spacetime bosons and
fermions. If mixing were allowed, normal spacetime SUSY would be broken and
replaced by a fractional supersymmetry, most-likely ruining Lorentz
invariance for $D>3$.

Since in the antiholomorphic sector ${\cal J}_K$ acts as the identity, we
will focus on its effect in the holomorphic sector.
In the $A_K$--sector the operator
$(\phi_{K/4}^0)^{D-2}$ transforms the bosonic (fermionic)
nonprojection\footnote{We use the same language as the authors of
refs.~[\putref{ref12}].  Nonprojection refers to the
bosonic and fermionic fields in the $A^{\rm boson}_K$ and $A^{\rm fermion}_K$
subsectors, respectively, corresponding to string functions with positive
coefficients, whereas projection fields refer to those
corresponding to string functions with negative signs.  With this definition
comes an overall minus sign coefficient on $A_K^{\rm fermion}$, as shown in
eq.~(2.39a).  For example,
in (2.39b), the bosonic non-projection fields are
\hbox {$(\phi^0_0 + \phi^2_0)^3(\phi^1_0)$}
 and the bosonic projection is
$(\phi^1_0)^4$. Similarly, in (2.39c) the fermionic non-projection
field is $(\phi^1_1)^4$ and the projections are
$(\phi^0_1 +\phi^2_1)^3(\phi^1_1)$.}
fields into the
fermionic (bosonic) projection fields and vice-versa. For example, consider
the effect of this twist current on the fields represented in
\subon
$$ A_4\equiv A^{\rm boson}_4 - A^{\rm fermion}_4\,\,
,\eqno\eqnlabel{abosferm-a}$$
where
$$\eqalignno{A^{\rm boson}_4 &= 4\left\{ (c^0_0 + c^4_0)^3(c^2_0) -
                                          (c^2_0)^4\right\}
                         &\eqnlabel{abosferm-b}\cr
             A^{\rm fermion}_4 &= 4\left\{
                             (c^2_2)^4 - (c^0_2 + c^4_2)^3(c^2_2)\right\}\,\, .
                         &\eqnlabel{abosferm-c}\cr}$$
\suboff
Twisting by $(\phi^0_{K/4})^{D-2}$ transforms the related fields as
\subon
$$\eqalignno{ (\phi^0_0 + \phi^2_0)^3(\phi^1_0) &\hbox to 1cm{\hfill}
{\buildrel {(\phi^0_{K/4})^{D-2}}\over\Longleftrightarrow}\hbox to 1cm{\hfill}
(\phi^2_1 + \phi^0_1)^3 (\phi^1_1)
&\eqnlabel{phitwist-a}\cr (\phi^1_0)^4 &\hbox to 1cm{\hfill}
{\buildrel {(\phi^0_{K/4})^{D-2}}\over\Longleftrightarrow}\hbox to 1cm{\hfill}
(\phi^1_1)^4\,\, .
&\eqnlabel{phitwist-b}\cr}$$
\suboff

Although the full meaning of the projection fields is not yet understood,
the authors of refs.~[\putref{ref62}] and [\putref{ref12}] argue that the
corresponding string functions
should be interpreted as ``internal'' projections, {\it i.e.},
cancellations of degrees of freedom in the fractional superstring models.
See also [\putref{ad93}], [\putref{ref68}], and [\putref{cd}]).
Relatedly, the authors show that when the $A_K$--sector is
written as $A^{\rm boson}_K - A^{\rm fermion}_K$,
as done above, the $q$-expansions of
both $A^{\rm boson}_K$ and $A^{\rm fermion}_K$ are all positive.
Including the
fermionic projection terms results in the identity
\subon
$$\eta^{D-2} A_K^{\rm fermion} = (D-2)\left( {(\theta_2)^4\over
16\eta^4}\right)^{{D-2\over 8}}\,\, .
\eqno\eqnlabel{af-a}$$
Eq.~(\puteqn{af-a}) is the standard theta-function expression for $D-2$
world sheet Ramond Majorana-Weyl fermions.
Further, $$\eta^{D-2} A_K^{\rm boson} = (D-2)\left(
{(\theta_3)^4 - (\theta_4)^4\over 16\eta^4}\right)^{{D-2\over 8}}\,\, .
\eqno\eqnlabel{af-b}$$
\suboff

Now consider the $B_K$--sectors. For $K=4$ and $8$ the operator
$(\phi^0_{K/4})^{D-2}$ transforms the primary fields corresponding to the
partition functions terms in the first set of brackets on the RHS of
eqs.~(\puteqn{part4-b},\puteqn{part8-b})
into the fields represented by the partition functions terms in the
second set. For example, in the $K=4$ ($D=6$) case
\subon
$$\eqalignno{(\phi^0_0 + \phi^2_0)(\phi^1_0)(\phi^0_1 + \phi^2_1)^2
&\hbox to 1cm{\hfill}
{\buildrel {(\phi^0_{K/4})^{D-2}}\over\Longleftrightarrow}\hbox to 1cm{\hfill}
(\phi^2_1 +
\phi^0_1)(\phi^1_1)(\phi^2_0+\phi^0_0)^2 & \eqnlabel{phib-a}\cr
(\phi^1_0)^2(\phi^1_1)^2 &\hbox to 1cm{\hfill}
{\buildrel {(\phi^0_{K/4})^{D-2}}\over\Longleftrightarrow}\hbox to 1cm{\hfill}
(\phi^1_1)^2(\phi^1_0)^2\,\, .&
\eqnlabel{phib-b}\cr}$$
\suboff

Making an analogy with what occurs in the $A_K$--sector, we suggest that
$(\phi^0_{K/4})^{D-2}$ transforms bosonic (fermionic)
nonprojection fields into fermionic (bosonic) projection fields and
vice-versa in the $B_K$--sector also.
Thus, use of the twist current ${\cal J}_K$ allows for bosonic and fermionic
interpretation of these
fields\footnote{Similar conclusions have been reached by K. Dienes and P.
Argyres for different reasons. They have, in fact, found theta-function
expressions for the $B_K^{\rm boson}$-- and $B_K^{\rm fermion}$--
subsectors.\markup{[\putref{ref68},\putref{ad93}]}}:
\subequationnumstyle{alphabetic}
$$B_4\equiv B^{\rm boson}_4 - B^{\rm fermion}_4\,\, ,
\eqno\eqnlabel{bbf-a}$$ where
$$\eqalignno{B^{\rm boson}_4
&= 4\left\{(c^0_0 + c^4_0)(c^2_0)(c^0_2+c^4_2)^2 -
(c^2_0)^2(c^2_2)^2\right\}&\eqnlabel{bbf-b}\cr
B^{\rm fermion}_4 &=
4\left\{(c^2_2)^2(c^2_0)^2 - (c^0_2+c^4_2)(c^2_2)(c^0_0
+c^4_0)^2\right\}\,\, .&\eqnlabel{bbf-c}\cr}$$
\subequationnumstyle{blank}
What appears as the projection term, $(c^2_0)^2(c^2_2)^2$, for the
proposed bosonic part acts as the nonprojection term for the fermionic half
when the subscripts are reversed.  One interpretation is
this implies a
compactification of two transverse dimensions.\footnote{This was also
suggested in ref.~[\putref{ref62}] working from a different approach.}
Let us choose the compactified directions to correspond to the last 2
strings functions in a term.
Thus, the
spin-statistics of the physical states of the $D=6$
model as observed in four-dimensional uncompactified spacetime is
determined
by the (matching) $n$ subscripts of the first two string
functions
(corresponding to the two uncompactified transverse dimensions) in each term
of four string functions, $c^{l_1}_n c^{l_2}_n c^{l_3}_{n'} c^{l_4}_{n'}$ .
(Assuming instead that the first two string functions corresponded to
the compactified dimensions, means interchanging the definitions
of $B^{\rm boson}_4$ and $B^{\rm fermion}_4$ above.)
The $B_8$ terms can
be interpreted similarly when one dimension is compactified.

However, the
$C_K$--sectors are harder to interpret. Under
$(\phi^0_{K/4})^{D-2}$ twisting, string functions with $K/4$ subscripts
are invariant, transforming back into themselves.  Thus, following
the pattern of $A_K$ and $B_K$ we would end up writing, for example,
$C_{16}$ as
\subequationnumstyle{alphabetic}
$$C_{16}= C_{16}^a - C_{16}^b \eqno\eqnlabel{cspin-a}$$ where,
$$\eqalignno{C^a_{16}
&= (c^2_4 + c^{14}_4) - c^8_4 & \eqnlabel{cspin-b}\cr C^b_{16} &= c^8_4 -
(c^2_4 + c^{14}_4)\,\, . & \eqnlabel{cspin-c}\cr}$$
\subequationnumstyle{blank}

The transformations of the corresponding primary fields are not quite as
trivial, though.
$(\phi^1_2 + \phi^7_2)$ is transformed into its conjugate field
$(\phi^7_{-2} + \phi^1_{-2})$ and likewise $\phi^4_2$ into
$\phi^4_{-2}$,
suggesting that $C^a_{16}$ and $C^b_{16}$ are the partition functions
for conjugate fields. Remember, however, that $C_{16}=0$. Even though we may
interpret this sector as containing two conjugate spacetime fields, this
(trivially) means that the partition function for each is identically zero.
We refer to
this effect in the $C_K$--sector as ``self-cancellation.'' One
interpretation is that there are no states in the $C_K$ sector of the Hilbert
space that survive all of the internal projections. If this is correct,
a question may arise as to the consistency of the $K=8$ and
$16$ theories.  Alternatively, perhaps anyon statistics allow two
(interacting?)  fields of either identical fractional spacetime spins
$s_1=s_2={2m\over K}$, or spacetime spins related by $s_1={2m\over K}= 1- s_2$,
where in both cases $0<m<{K\over 2} \pmod{1}$,
to somehow cancel each other's contribution to the partition function.

Using the $\phi^j_m\equiv\phi^j_{m+K}\equiv\phi^{{K\over
2}-j}_{m-{K\over 2}}$ equivalences at level-$K\in 4\Z$, a PCFT has $K/2$
distinct classes of integer $m$ values. If one associates these classes with
distinct spacetime spins (statistics) and assumes $m$ and $-m$ are also
in the same classes since
$(\phi^0_m)^{\dagger}= \phi^0_{-m}$, then the
number of spacetime spin classes reduces to ${K\over 4} +1$. Since $m=0$
$(m= {K\over 4})$ is
associated with spacetime bosons (fermions),
we suggest that general $m$ correspond to particles of
spacetime spin ${2m\over K}$,
${2m\over K} + \Z^+$, or $\Z^+ - {2m\over K}$.
If this is so, most likely
 spin$(m)\in \left\{{2m\over K},\, \Z^+ + {2m\over K}\right\}$
for $0< m< K/4 \pmod{K/2}$
and spin$(m)\in\Z^+ - {2m\over K}$ for $-K/4< m<0 \pmod{K/2}$.
This is one of the few
spin assignment rules that maintains the equivalences of the fields $\phi^j_m$
under
$(j,\, m)\rightarrow({k\over 2}-j,\, m-{K\over 2})\rightarrow(j,\, m+K)$
transformations.  According
to this rule, the fields in the $C_K$--sectors have quarter spins
(statistics),
which agrees with prior claims\markup{[\putref{ref63,ref62,ref65}]}.

Also, we do not believe
products of primary fields in different $m$ classes in the $B_K$--sectors
correspond to definite spacetime spin states unless some dimensions are
compactified. Otherwise by our interpretation of $m$ values above, Lorentz
invariance in uncompactified spacetime would be lost.
In particular, Lorentz invariance requires that either all or none of the
transverse modes in uncompactified spacetime be fermionic spinors.
Further, $B$--sector particles cannot
correspond to fractional spacetime spin particles for a
consistent theory. Thus, the $D=6\, (4)$ model must have two (one) of its
dimensions compactified. (This implies that the $D=6,\, 4$ partition
functions are incomplete: momentum (and winding) factors
for the two compactified dimensions would have to be added, while maintaining
modular invariance.)

Note that the $B_8$--sector of the $D=4$ model
appears necessary for more reasons than just modular invariance of
the theory. By the above spacetime
spin assignments, this model suggests massive
spin-quarter states (anyons) in the $C_K$--sectors,
which presumably cannot exist in $D>3$ uncompactified dimensions.
However, the $B_K$--sector, by forcing compactification to three dimensions
where anyons are allowed, would save the model, making it self-consistent.
Of course, anyons in the  $K=16$ theory with $D=3$ are physically
acceptable. (Indeed, no $B_K$--sector
is needed and none exists, which would otherwise reduce the theory to zero
transverse dimensions.) Thus,
$K=8$ and $K=16$ models are probably both allowed solutions for three
uncompactified spacetime dimensional models.
If this interpretation is correct, then it is
the $B_K$--sector for $K=8$ which makes that theory self-consistent.

An alternative, less restrictive, assignment of spacetime spin
is possible. Another view is that the $m$ quantum number is not
fundamental for determining spacetime spin. Instead, the
transformation of states under $(\phi^{0}_{K/4})^{D-2}$ can be considered to
be what divides the set of states into spacetime bosonic and fermionic
classes. With this interpretation, compactification in the $B_K$--sector
is no more necessary than in the $A_K$--sector. Unfortunately, it is not
{\it a priori} obvious, in this approach, which group of states is bosonic,
and which fermionic. In the $A_K$--sector, the assignment can  also be
made phenomenologically.
In the $B_K$--sector, we have no such guide.
Of course, using the $m$ quantum number to determine spacetime spin does
not truly tell us which states have bosonic or fermionic statistics,
since the result depends on the arbitrary choice of which of the
two (one) transverse dimensions to compactify.

A final note of caution involves multiloop modular invariance.
One-loop modular invariance amounts
to invariance under $S$ and $T$ transformations. However modular
invariance at higher orders requires an additional
invariance under $U$ transformations: Dehn twists mixing loops of
neighboring tori of $g>1$ Riemann
surfaces.\markup{[\putref{ref9a,ref9b,ref9c,ref9d}]}
We believe neither our new method of generating the one-loop
partitions, nor the original method of Argyres {\it et al.} firmly prove the
multiloop modular invariance that is required for a truly consistent
theory.

\sectionnumstyle{blank}
\section{\bf Section 3: Beyond the Partition Function: Additional Comments}
\sectionnumstyle{arabic}\sectionnum=3\equationnum=0

The previous discussion of the FSPF's
in section two does not fully demonstrate the consistency of the
fractional superstrings, nor does it sufficiently compare them to the $K=2$
superstring. In this section, we now comment further on these aspects of
potential string theories.
We consider the analog of the GSO
projection, the uniqueness of the ``twist'' field $\phi^{K/4}_{K/4}$ for
producing spacetime fermions,
and the ghost system for the fractional supercurrent.

\subsectionnumstyle{blank}
\subsection{{\bf 3.1}: Generalized Commutation Relations and the GSO
Projection}
\subsectionnumstyle{blank}

One of the major complications of
generalizing from the $K=2$ fermion case to $K>2$ is that the parafermions
(and bosonic field representations) do not have simple commutation
relations\markup{[\putref{ref4}]}.
What are the commutation relations for non-(half) integral spin particles?
Naively,
the first possible generalization of standard (anti-)commutation
relations for two
fields $A$ and $B$ with fractional spins seems to be:
$$AB- {\rm e}^{[i4\pi ~{\rm spin} (A) ~ {\rm spin} (B)]}BA
=0\eqno\eqnlabel{wrong}$$
(which reduces to the expected result for bosons and fermions). This is too
simple a generalization, however\markup{[\putref{ref11}]}.
Fractional spin particles must be representations
of the braid group. Zamolodchikov and Fateev\markup{[\putref{ref4}]} have
shown that world sheet parafermions (of fractional spin) have complicated
commutation relations that
involve an infinite number of modes of a given field. For example:
\subon
$$\eqalignno{\sum_{l=0}^{\infty} C^{(l)}_{(-1/3)}&\left[
A_{n+(1-q)/3-l} A^{\dagger}_{m-(1-q)/3 +l} +
A^{\dagger}_{m- (2-q)/3-l}A_{n-(2-q)/3 +l}
\right]=\cr
&-{1\over 2}\left( n-{q\over 3}\right) \left( n+1 -{q\over 3}\right)
\delta_{n+m,0} + {8\over 3c}L_{n+m}&\eqnlabel{commut-a}}$$
and
$$\sum_{l=0}^{\infty} C^{(l)}_{(-2/3)}
\left[
A_{n-q/3-l}A_{m+(2-q)/3+l}-
A_{m-q/3-l}A_{n+(2-q)/3+l}\right]
= {\lambda\over 2}\left( n-m \right)A^{\dagger}_{(2-2q)/3 +n+m}\,\, .
\eqno\eqnlabel{commut-b}$$
\suboff
where $A$ is a parafermion field, and $L_n$ are the generators of the
Virasoro algebra.
$\lambda$ is a real coefficient,
$n$ is integer and $q=0,1,2 \pmod{3}$ is a $\Z_3$ charge of
Zamolodchikov and Fateev that can be
assigned to each primary field in the $K=4$ model.
The coefficients $C^{(l)}_{(\alpha)}$ are determined by
the power expansion
$$(1-x)^{\alpha}= \sum_{l=0}^{\infty}
C^{(l)}_{(\alpha)}x^l\,\, .\eqno\eqnlabel{cla}$$
As usual, $c= {2(K-1)\over K+2}$ is the central charge of the level-$K$
 PCFT.

These commutation relations are derived from the OPE of the related
fields.\markup{[\putref{ref4}]}
(Hence more terms in a given OPE should result in more complicated
commutation relations.)
Similar relations between the modes of two different primary
fields should also be derivable from their OPE's.
The significance of these commutation relations is that they
severely reduce the number of
distinct physical states in parafermionic models. There are
several equivalent ways of creating a given physical state from the
vacuum using different mode excitations from different parafermion
primary fields
in the same CFT. Thus, the actual Hilbert space of states for this $K=4$
model will be much reduced compared to the space prior to moding
out by these equivalences.\footnote{These equivalences
have subsequently been explicitly shown and the distinct low mass$^2$
fields determined in Argyres {\it et al}.\markup{[\putref{ref12}]}}
Although the fields in the parafermion CFT do not (anti-)commute,
but instead have complicated commutation relations, some insight can be
gained by comparing the $D=6$, $K=4$ FSC model to the standard $D=10$
superstring. We can, in fact, draw parallels between $\epsilon$ and the
standard fermionic superpartner, $\psi$, of an uncompactified boson X.
In the free fermion approach, developed both by Kawai, Lewellen and Tye
and by
Antoniadis, Bachas and Kounnas, generalized GSO projections based on boundary
conditions of the world sheet fermions are
formed.\mpr{ref9a, ref9b, ref9c, ref9d} Fermions with
half-integer modes (NS-type) are responsible for $\Z_1$ (trivial)
projections; fermions
with integer modes (R-type) induce $\Z_2$ projections. In the non-Ramond
sectors these $\Z_2$ projections remove complete states, while in the
Ramond sector itself,
remove half of the spin modes, giving chirality. Fermions with general
complex boundary conditions,
$$\psi (\sigma=2\pi)=-e^{i\pi {a\over b}}\psi (\sigma=0)\,\, ,
\eqno\eqnlabel{fbc}$$
form in the non-Ramond sector $\Z_{2b}$ projections if $a$ is odd and $\Z_b$
projections if $a$ is even (with $a$ and $b$ coprime and chosen in the range
$-1\leq a/b< 1$).
For free-fermionic models, the GSO operator, coming
from a sector where the set of world sheet
fermions $\{\psi^i\}$ have  boundary conditions
\subon
$$\psi^i(2\pi)= -{\rm e}^{i\pi x^i}\psi^i(0)\, ,\eqno\eqnlabel{fbc-a}$$
and acting on a physical state $\vert {\rm phys}\big>_{\vec y}$
in a sector where the same
fermions have boundary conditions
$$\psi^i(2\pi)= -{\rm e}^{i\pi y^i}\psi(0)\, ,\eqno\eqnlabel{fbc-b}$$
\suboff
takes the form,
$$\left\{ {\rm e}^{i\pi {\vec x}\cdot {\vec F}_{\vec y}}=\delta_{\vec y}
C({\vec y}\vert {\vec x})\right\}\vert {\rm phys}\big>
\eqno\eqnlabel{gso}$$
for states surviving the projection.
Those states not satisfying the demands of the GSO operator for at least
one sector $\vec x$ will not appear in the partition function of the
corresponding model. In eq.~(\puteqn{gso}),
${\vec F}_{\vec y}$ is the (vector) fermion number
operator for states in sector $\vec y$. $\delta_{\vec y}$
is $-1$ if either the left-moving or right-moving $\psi^{\rm spacetime}$
are periodic and 1 otherwise.
$C(\vec y\vert\vec x)$ is a phase with value chosen from an allowed set
of order $g_{\vec y,\vec x} = GCD(N_{\vec y},N_{\vec x})$, where
$N_{\vec y}$ is the lowest positive integer such that
$N_{\vec y}\times \vec y = \vec 0 \pmod{2}\,\, .$

Now consider the $\epsilon$ fields in the $K=4$ parafermion theory.
The normal untwisted, ({\it i.e.}, Neveu-Schwarz) modes of $\epsilon$ are
$\epsilon^+_{-{1\over3}-n}$ and $\epsilon^-_{{1\over3}-n}$
where $n\in\Z$. That is, untwisted $\epsilon= \epsilon^+ + \epsilon^-$
has the following normal-mode expansions.
{\settabs 3\columns
\+ \cr
\+ \cr
\+ N-S Sector:\cr}
\subon
$$ \eqalignno{
\epsilon^+ (\sigma_1, \sigma_2) &=
\sum_{n= 1}^{\infty}
[
\epsilon_{n-1/3}
\,
\exp\left\{ -i(n - 1/3) (\sigma_1 +\sigma_2) \right\} \cr
&\phantom{\sum_{n= 1}^{\infty}}\hbox to .2truecm{\hfill}
 + \bar\epsilon_{2/3 - n }
\,
\exp\left\{-i(2/3 - n )(\sigma_1+\sigma_2 )\right\}
]
&\eqnlabel{emodes-a}\cr
\epsilon^- (\sigma_1, \sigma_2) &=
\sum_{n= 1}^{\infty}[
 \epsilon_{1/3 - n }\, \exp\left\{-i(1/3 - n  )(\sigma_1+\sigma_2)\right\}\cr
&\phantom{\sum_{n= 1}^{\infty}}\hbox to .2truecm{\hfill}
+ \bar\epsilon_{n - 2/3 }\,
\exp\left\{-i( n - 2/3 )(\sigma_1 +\sigma_2)\right\}]
&\eqnlabel{emodes-b}}$$
\suboff
(where $\epsilon^{\dag}_r=\epsilon_{-r}$ and
$\bar\epsilon^{\dag}_r=\bar\epsilon_{-r}$).
The associated boundary conditions in this sector are
\subon
$$ \eqalignno{
\epsilon^+(\sigma_1+2\pi)&=
   {\rm e}^{+i 2\pi/3}\, \epsilon^+(\sigma_1)
&\eqnlabel{bcns-a}\cr
\epsilon^-(\sigma_1+2\pi)&=
   {\rm e}^{-i 2\pi/3}\, \epsilon^-(\sigma_1)\,\, .
&\eqnlabel{bcns-b}}$$
\suboff

Like the standard fermion, the $\epsilon$ operators at level-four
can be in twisted sectors, where the normal-mode
expansions have the following form.
\hfill\vfill\eject

{\settabs 3\columns
\+ General Twisted Sector:\cr}
\subon
$$\eqalignno{
\epsilon^+ (\sigma_1, \sigma_2) &= \sum_{n= 1}^{\infty}[
\epsilon_{n - {1/3} - {a/2b} }\,
    \exp\left\{ -i(n - {1/3} - {a/2b} )(\sigma_1 +\sigma_2)\right\}\cr
&\phantom{\sum_{n= 1}^{\infty}}\hbox to .2truecm{\hfill}
+ \bar\epsilon_{{2/3} - n - {a/2b}  }\,
\exp\left\{-i({2/3} - n - {a/2b} )(\sigma_1 +\sigma_2)\right\} ]
&\eqnlabel{twistmodes-a}\cr
\epsilon^- (\sigma_1, \sigma_2) &=
\sum_{n= 1}^{\infty}[
\epsilon_{{1/3} - n +  {a/2b} }\,
\exp\left\{-i({1/3} - n + {a/2b} )(\sigma_1 +\sigma_2)\right\}\cr
&\phantom{\sum_{n= 1}^{\infty}}\hbox to .2truecm{\hfill}
+ \bar\epsilon_{n - {2\over 3} + {a\over 2b}}\,
\exp\left\{ -i( n - {2/3} + {a/2b})(\sigma_1 +\sigma_2)\right\} ]
&\eqnlabel{twistmodes-b}}$$
\suboff
The associated boundary conditions are
\subon
$$\eqalignno{
\epsilon^+(\sigma_1+2\pi)&=
{\rm e}^{+i 2\pi (1/3)}\,  {\rm e}^{i\pi {a/b}}\, \epsilon^+(\sigma_1)
&\eqnlabel{twistbcns-a}\cr
\epsilon^-(\sigma_1+2\pi)&=
{\rm e}^{-i 2\pi(1/3)}\,  {\rm e}^{-i\pi {a/b}}\, \epsilon^-(\sigma_1)\,\, .
&\eqnlabel{twistbcns-b}}$$
\suboff

{}From the analogy of free-fermion models, we suggest that in $K=4$ parafermion
models the presence of a sector containing twisted $\epsilon$ fields
with boundary conditions
(\puteqn{twistbcns-a}) or (\puteqn{twistbcns-b}) will result in
$\Z_{b}$ or $\Z_{2b}$ GSO projections,
depending on whether $a$ is even or odd respectively. (We assume $a$ and
$b$ are relative primes and $-2/3\leq a/b< 4/3$.)

Zero modes correspond to ${a/b}= -2/3$.
Thus, we conjecture that the presence of these (twisted) zero modes
$\epsilon_n$, $n\in\Z$ in a model, results in a
generalized $\Z_3$ GSO projection.
Admittedly, this is suggested with the hindsight of having the partition
function for this theory.  Nevertheless, we mention this in attempting to
give more physical meaning to the partition function.
Likewise for $K=8$ and $16$,
one might expect $\Z_5$ and
$\Z_9$ projections, respectively. Such projections for
$K=8$ and $16$ could be significantly altered though, by the effects of the
non-Abelian braiding of the non-local interactions.

One other aspect to notice is that within the range $-2/3\leq a/b<4/3$
there are actually two distinct N-S sectors, corresponding not just to
$a/b=0$, but also $a/b= 2/3$.  This corresponds to  $\Z_2$ symmetry
$\epsilon^+\leftrightarrow\epsilon^-$.  Though this symmetry may be obvious,
it could explain the origin of the additional $\Z_2$ projection we will
shortly discuss.

For the $K=4$ FSC model, one expects a GSO projection to depend on a
generalization of fermion number. However, the naive generalization to
parafermion number, $F(\phi^1_0)$, is insufficient. We find that we must also
consider the multiplicities of the other two ``physically
distinguished'' fields, the twist field, $\phi^1_1$ and  the field
$\phi^0_1$, which raises the $m$ quantum number.

In order to derive the MIPF we discovered that, indeed, a  $\Z_3$ projection
must be applied to both the left-moving modes
(LM) and right-moving modes (RM) independently.
Survival of a physical state, $\vert {\rm phys}\big>$, in the Hilbert space
under this $\Z_3$ projection requires
\subon
$$\left\{{\rm e}^{\left\{i\pi \vec {2\over 3}\cdot\left[
{\vec F}_{{\rm LM}\, ({\rm RM})}(\phi^1_0)
+ {\vec F}_{{\rm LM}\, ({\rm RM})}(\phi^1_1)\right]\right\}}
= {\rm e}^{i\pi {2\over3}}\right\}
\vert {\rm phys}\big>\,\, ,\eqno\eqnlabel{gsoz3-a}$$
or equivalently
$$\left\{Q_{3,\, {\rm LM}\,({\rm RM})}\equiv\sum_i F_{i,\, {\rm LM} \,
({\rm RM})}(\phi^1_0)
+ \sum_i F_{i,\, {\rm LM} \, ({\rm RM})}(\phi^1_1) = 1 \pmod{3}
\right\} \vert {\rm phys}\big>\,\, , \eqno\eqnlabel{gsoz3-b}$$
where $F_i(\phi^j_m)_{{\rm LM}\, ({\rm RM})}$ is the number operator for the
field
$\phi^j_m$ along
the $i$ direction for left-moving (right-moving) modes.
One may note that this projection
does not prevent mixing holomorphic $A_4$--sector and antiholomorphic
$B_4$--sector
terms. However, it need not, for this is separately
prevented by the standard requirement that
$M^2_{\rm LM}=M^2_{\rm RM}$, {\it i.e.}, $L_0 = \bar L_0$,
which here results in the RM factors in the partition function being the
complex conjugates of the LM, giving only mod-squared terms in the
partition function.

Prior to projection by this extended GSO operator,
we consider all physical states associated with
the LM partition function terms in the expansion of
$(c^0_0 + c^4_0 + c^2_0)^4$ or $(c^2_2 + c^4_2)^4$
to be in the $A_4$--sector. (The RM physical states in the $A$ sector
have parallel association with the complex conjugates of these partition
function terms.)
Similarly, we initially place in the $B_4$--sector
all the LM physical states associated with the partition function terms in the
expansion of $(c^2_2 + c^4_2)^2(c^0_0 + c^4_0 +c^2_0)^2$ or
$(c^0_0 + c^4_0 + c^2_0)^2(c^2_2 + c^4_2)^2$.
There is however, a
third class of states; let us call this the ``$D_4$'' class. This latter
class would be present in the original Hilbert space if not for an additional
$\Z_2$ GSO projection.  Left moving states in
the $D_4$ class, would have partition
functions that are terms in the expansion of
$(c^0_0 + c^4_0 + c^2_0)^3(c^2_2 + c^4_2)$ or
$(c^2_2 + c^4_2)^3(c^0_0+c^4_0+c^2_0)$. The thirty-two
$D_4$ terms in the expansions
(pairing $c^0_0$ and $c^4_0$, rather than
expanding $(c^0_0 +c^4_0)^n$ as in Table 3.1)
are likewise dividable into classes based on their
associated $\Z_3$ charges, $Q_3$.
Twelve have charge $0 \pmod{3}$, twelve have charge $1 \pmod{3}$
and eight have charge $2 \pmod{3}$. Without the $\Z_2$ projection it
is impossible to
to keep just the correct terms in the $A_4$-- and $B_4$--sectors,
and also project away all of the $D_4$--sector terms.
Simple variations of the projection (\puteqn{gsoz3-a}) cannot do the job.
All $D_4$ terms can be eliminated, without further projections on the $A_4$
and $B_4$ terms, by an obvious $\Z_2$ projection,
$$\left\{\sum_i F_{i,\, {\rm LM} \, ({\rm RM})}(\phi^1_1)
+ \sum_i F_{i,\, {\rm LM} \, ({\rm RM})}(\phi^0_{\pm 1}) = 0 {\rm ~mod~} 2
\right\} \vert {\rm phys}\big>\,\, . \eqno\eqnlabel{gsoz3-c}$$
\suboff
(Note that for $K=2$,
$\phi^1_1$ is equivalent to the vacuum and $\phi^1_0$ is
indistinguishable from the usual fermion, $\phi^1_0$. Thus for $K=2$
there is no additional $\Z_2$ GSO projection.)

Consideration
of these $D$ class states reveals some physical meaning to
our particular $\Z_3$ charge and the additional $\Z_2$ projection.
First, in all sectors the charge $Q_3$ commutes with
$(\phi^0_{K/4})^{D-2}$, which transforms between non-projection and
projection states of opposite spacetime statistics in the
$A_4$-- and $B_4$--sectors.
Second, the values of this charge are also associated with specific
mass$^2\pmod{1}$ levels. Third, only for the
$A_4$-- and $B_4$--sector states does mass$^2\pmod{1}$ commute with the same
twist operator $(\phi^0_{K/4})^{D-2}$.
Recall, in section two we suggested that
twisting by this latter field was the key
to spacetime SUSY. Without any of our projections the
mass$^2$ levels $\pmod{1}$
of states present would be mass$^2= 0,~ {1\over 12},~ {2\over 12},~\dots
{11\over 12}$. When acting on  $D_4$--sector fields,
$(\phi^0_{K/4})^{D-2}$ transforms
\hbox {mass$^2={i\over 12} \pmod{1}$} states into
\hbox {mass$^2={ i +6\over 12} \pmod{1}$} states.
Thus, states in the $D_4$--sector paired by
supersymmetry would be required to appear in different sectors ({\it i.e.,}
different mod-squared terms) of the
partition function, in order to preserve $T$ invariance. As a result, the
paired contributions to the partition function cannot cancel, proving that
$D_4$
terms cannot be part of any supersymmetric theory.
Although mass$^2 \pmod 1$ commutes with $(\phi^0_{K/4})^{D-2}$ in the
$A_4(Q_3=0)$,
$A_4(Q_3=-1)$, $B_4(Q_3=0)$, $B_4(Q_3=-1)$  subsectors,
within these subsectors
(1) there is either a single bosonic state or fermionic state of lowest
mass without superpartner of equal mass, and/or
(2) the lowest mass states are tachyonic.
(See Table 3.1.)
Thus, our specific GSO
projections in terms of
our $\Z_3$ charge projection and our $\Z_2$ projection equate to spacetime
SUSY.

Our assignments of states as spacetime
bosons or fermions in the $B_4$--sector,
uses an additional projection that we believe distinguishes between the
two.  Following the pattern in
eqs.~(2.8b) with bosonic/fermionic assignment of related states defined in
eqs.~(2.43a-c), we suggest that for these states the two primary fields,
$\phi^{j_3}_{m_3}$ and $\phi^{j_4}_{m_4=m_3}$
(implicitly) assigned compactified spacetime
indices must be the same, {\it i.e.}, $j_3= j_4$,
or else must form  a term in the expansion of
$(\phi^0_0 + \phi^2_0)^2$. This second case is related to $\phi^0_0$ and
$\phi^2_0$ producing the same spacetime fermion field, $\phi^2_1$, when
separately twisted by $\phi^0_{K/4}$.
(Note however that $\phi^2_1\times\phi^0_{K/4}=\phi^0_0$ only.)
Following this rule, neither the states
corresponding to
$(c^2_0)(c^0_0)(c^2_2)(c^4_2)$ and $(c^2_2)(c^4_2)(c^2_0)(c^0_0)$,
(which transform between each other under twisting by
$\phi^0_{K/4}\phi^0_{K/4}\phi^0_{K/4}\phi^0_{K/4}$
nor those associated with
$(c^2_0)(c^4_0)(c^2_2)(c^4_2)$ and $(c^2_2)(c^4_2)(c^2_0)(c^4_0)$,
survive the projections as either spacetime bosons or fermions.
However, for completeness we include the partition functions
in the $B_4$--sector columns of Table 3.1. We define the associated states
as either
spacetime bosons or fermions simply by the value of $m_3=m_4$.
This is academic, though, because the states do not
survive the $\Z_3$
projections.
\hfill\vfill\eject

\centertext{Table 3.1 Masses of $K=4$ Highest Weight States\\
(Represented by Their Associated
Characters)}
{\settabs 8 \columns
\+ \cr
\+ & \hso $A_4$-Sector &&&Survives&&\hso $B_4$-Sector\cr
\+ \underline{\hbox to 6.6cm{\hfill}}&&&\hso $Q_3$&GSO &\underline{\hbox to
7.7cm{\hfill}}
\cr
\+ \cr
\+ {Boson} &\hso {Mass$^2$}
&\hso {Fermion} &
&
&{Boson} &\hsf {Mass$^2$}
&\hsf {Fermion}\cr
\+ $(c^4_0)^2(c^4_0)^2$ &\hso\h $3{2\over3}$ &&\hso\phantom{$-$}0
& No & $(c^4_0)^2(c^4_2)^2$
&\hsf\h $3{1\over6}$ &\hsf $(c^4_2)^2(c^4_0)^2$\cr
\+ $(c^2_0)\hof (c^4_0)(c^4_0)^2$ &\hso\h 3 &&\hso\phantom{$-$}1 & Yes
& $(c^2_0)\hof (c^4_0)\hof (c^4_2)^2$
&\hsf\h $2{1\over2}$ &\hsf $(c^2_2)\hof (c^4_2)(c^4_0)^2$\cr
\+ $(c^0_0)\hof (c^4_0)\hof (c^4_0)^2$ & $\hso\h 2{2\over 3}$
&\hso $(c^4_2)^2(c^4_2)^2$
&\hso\phantom{$-$}0 & No
& $(c^0_0)\hof (c^4_0)\hof (c^4_2)^2$ &\hsf\h $2{1\over6}$ &\hsf
$(c^4_2)^2(c^0_0)\hof (c^4_0)$ \cr
\+ $(c^4_0)^2(c^2_0)^2$ &\hso\h $2{1\over 3}$ && \hso $-1$ & No
& $(c^4_0)^2(c^2_2)^2$ &\hsf\h $1{5\over 6}$ &\hsf $(c^4_2)^2(c^2_0)^2$\cr
\+ $(c^2_0)^2(c^4_0)^2$ &&&&& $(c^2_0)^2(c^4_2)^2$ &&\hsf
$(c^2_2)^2(c^4_0)^2$\cr
\+ &&&&& $(c^2_0)\hof (c^4_0)\hof (c^2_2)\hof (c^4_2)$ & &\hsf $(c^2_2)\hof
(c^4_2)\hof (c^2_0)\hof (c^4_0)$\cr
\+ $(c^2_0)\hof (c^0_0)\hof (c^4_0)^2$ &\hso\h 2
&\hso $(c^2_2)(c^4_2)(c^4_2)^2$ &\hso\phantom{$-$}1 & Yes
& $(c^2_0)\hof(c^0_0)\hof  (c^4_2)^2$ &\hsf\h
$1{1\over 2}$ &\hsf $(c^2_2)\hof (c^4_2)\hof (c^0_0)\hof (c^4_0)$\cr
\+ $(c^2_0)\hof (c^4_0)\hof (c^2_0)^2$ &\hso\h $1{2\over 3}$
&&\hso\phantom{$-$}0 & No &
$(c^2_0)\hof (c^4_0)\hof (c^2_2)^2$
&\hsf\h $1{1\over6}$ & $\hsf (c^2_2)\hof (c^4_2)\hof (c^2_0)^2$\cr
\+ $(c^0_0)^2(c^4_0)^2$&&&&& $(c^0_0)^2(c^4_2)^2$\cr
\+ $(c^4_0)^2(c^0_0)^2$&&&&&&&\hsf $(c^4_2)^2(c^0_0)^2$\cr
\+ $(c^0_0)\hof (c^4_0)\hof (c^2_0)^2$ &\hso\h $1{1\over 3}$
&\hso $(c^2_2)^2(c^4_2)^2$ &\hso $-1$ &
No & $(c^0_0)\hof (c^4_0)\hof (c^2_2)^2$ &\hsf\h\hskip .22cm ${5\over6}$\cr
\+ && \hso $(c^4_2)^2(c^2_2)^2$
&&& $(c^0_0)\hof (c^2_0)\hof (c^2_2)\hof (c^4_2)$
&&\hsf $(c^2_2)\hof (c^4_2)\hof (c^0_0)\hof (c^2_0)$\cr
\+ $(c^2_0)^2(c^2_0)^2$ &\hso\h 1 & &\hso\phantom{$-$}1 & Yes
& $(c^2_0)^2(c^2_2)^2$ &
$\hsf\h\hskip .22cm {1\over 2}$ &\hsf $(c^2_2)^2(c^2_0)^2$\cr
\+ $(c^2_0)\hof(c^4_0)\hof (c^0_0)^2$ &&&&&&&\hsf $(c^2_2)\hof (c^4_2)\hof
 (c^0_0)^2$\cr
\+ $(c^2_0)\hof (c^0_0)\hof (c^2_0)^2$ &\hso\h\hskip .22cm ${2\over 3}$
&\hso $(c^2_2)(c^4_2)(c^2_2)^2$
&\hso\phantom{$-$}0 & No & $(c^2_0)\hof (c^0_0)\hof (c^2_2)^2$
&\hsf\h\hskip .22cm ${1\over6}$\cr
\+ $(c^0_0)\hof (c^4_0)\hof (c^0_0)^2$&\cr
\+ $(c^0_0)^2(c^2_0)^2$
&\hso\h\hskip .22cm ${1\over 3}$
&&\hso $-1$ &No & $(c^0_0)^2(c^2_2)^2$
&\hskip .57cm\h $-{1\over6}$ \cr
\+ $(c^2_0)^2(c^0_0)^2$&&&&&&&\hsf $(c^2_2)^2(c^0_0)^2$\cr
\+ $(c^2_0)\hof (c^0_0)\hof (c^0_0)^2$ &\hso\h\hskip .22cm 0
&\hso $(c^2_2)^2(c^2_2)^2$
&\hso\phantom{$-$}1 & Yes \cr
\+ $(c^0_0)^2 (c^0_0)^2$ &\hskip .9cm\h $-{1\over 3}$
&&\hso\phantom{$-$}0 & No\cr
\+ \cr}
\hfill\vfill\eject

\centertext{Table 3.2 Mass Sectors as Function of $\Z_3$ Charge}
{\settabs 7\columns
\+\cr
\+ Lowest $M^2$  & $M^2$ mod 1 & Sector & $\Z_3$ Charge & Sector
& $M^2$ mod 1 & Lowest $M^2$\cr
\+ \overline{\hbox to 6cm{\hfill}}&&&&\overline{\hbox to 6.7cm{\hfill}}\cr
\+ \htf 0 & 0 & $A_4$  & $Q_3={\hbox to .3cm{\hfill}}1$ & $B_4$ & ${6\over 12}$
& ${6\over 12}$\cr
\+\cr
\+ $-{1\over 12}$ &${11\over 12}$ & $D_4$ & $Q_3= {\hbox to .3cm{\hfill}}0$
& $D_4$  &${5\over 12}$&${5\over
12}$\cr
\+\cr
\+ $-{2\over 12}$ &${10\over 12}$ & $B_4$ & $Q_3= -1$ & $A_4$
&${4\over 12}$ &${4\over
12}$\cr
\+\cr
\+ $-{3\over 12}$ &${9\over 12}$ & $D_4$ & $Q_3= {\hbox to .3cm{\hfill}}1 $
& $D_4$ & ${3\over 12}$
&${3\over 12}$\cr
\+\cr
\+ $-{4\over 12}$ & ${8\over 12}$ & $A_4$ & $Q_3={\hbox to .3cm{\hfill}}0$
& $B_4$ & ${2\over 12}$ &${2\over 12}$\cr
\+\cr
\+\htf ${7\over 12}$ & ${7\over 12}$& $D_4$ & $Q_3= -1$& $D_4$
& ${1\over 12}$ & ${1\over 12}$\cr
\+ \cr}

\n (In Table 3.2, columns one and seven give the lowest
mass$^2$  of a state with center
column $\Z_3$ charge in the appropriate sector. For the $D_4$--sector states,
under $(\phi^0_{K/4})^{D-2}$ twistings, mass$^2$ values in column two
transform into mass$^2$ values in column six of the same row and vice-versa.)

In the $K=4$ case unlike $K=2$, we find that
the $\Z_3$ projection in the Ramond sector wipes out complete spinor fields,
not
just some of the modes within a given spin field. This type of projection
does not occur in the Ramond sector for $K=2$ since there are no fermionic
states with fractional mass$^2$ values in the $D=10$ model.
Note also that
our $\Z_3$ GSO projections relate to the $\Z_3$ symmetry
pointed out in
[\putref{ref4}] and briefly commented on after
eqs.~(\puteqn{commut-a},\puteqn{commut-b}).

For $K=8$, a more generalized $\Z_5$ projection
holds true for all
sectors.  For the $K=16$ theory, there are too few terms and products of
string functions to determine if a $\Z_9$ projection
is operative.  In the
$K=4$ case, the value of our LM (RM) $Q_3$ charges for states surviving the
projection is set by demanding that the massless spin-2 state
$\epsilon^{\mu}_{-{1\over3}}\bar\epsilon^{\bar\nu}_{-{1\over3}}\left\vert 0
\right\rangle$ survives.  In the $A$, $B$, (and $C$ for $K=8,16$) sectors,
these
projections result in states with squared masses of $0+$ integer,
${1\over 2} +$ integer, and ${3\over 4} +$ integer, respectively.

\sectionnumstyle{blank}
\section{\bf 3.2: The Unique Role of the Twist Field, $\phi_{K/4}^{K/4}$.}
\sectionnumstyle{arabic}
\sectionnum=3
\subsectionnumstyle{arabic}\subsectionnum=2
\equationnum=0

As Table 3.1 indicates for the particular case of $K=4$,
the massless $A$-sector spacetime fermion in the fractional superstring theory
is created in light-cone gauge by a $(D-2)$--dimensional tensor product of
$(\phi^{K/4}_{K/4})$ fields
(with associated string function character $(c^{K/2}_{K/2})^d$)
acting on the vacuum.
In this section
we examine whether other consistent models are possible if one
generalizes the twist field (as $\phi^{K/4}_{K/4}$ is referred to in
parafermion models),
to another that could fulfill its role of creating massless spacetime fermions
or if $\phi^{K/4}_{K/4}$ uniquely qualifies for this task.
When it is demanded that
$\phi^{K/4}_{K/4}$ and the $\epsilon\equiv\phi^1_0$ field of reference
[\putref{ref63,ref62,ref65}] be used, we can
derive the critical dimensions of possible models by observing that
$K=2,\, 4,\, 8,$ and $16$  are  the only levels for which
$$h(\phi^1_0)/h(\phi^{K/4}_{K/4})\in\Z\,\, . \eqno\eqnlabel{stdim-a}$$
If we assume (as in [\putref{ref63}]) that the
operator $(\phi_{K/4}^{K/4})^{\mu} $ acting on the (tachyonic) vacuum produces
a massless spacetime
spinor vacuum along the direction $\mu$, and $(\phi^1_0)^{\mu}$ produces a
massless spin-1 state, then for spacetime supersymmetry
(specifically $N=2$ SUSY for fractional type II theories and $N=1$ SUSY for
fractional heterotic)
$h(\phi^1_0)/h(\phi^{K/4}_{K/4})$  must equal the number
of  transverse
spin modes, {\it i.e.,}
$$\eqalignno{ h(\phi^1_0) &= (D-2) h(\phi^{K/4}_{K/4})\cr
{2\over K+2}&= (D-2) {K/8\over K+2}\,\, .&\eqnlabel{stdim-b}\cr}$$
Hence,
$$ D= 2 + {16\over K}\in\Z\,\, .\eqno\eqnlabel{dvsk}$$
Thus, from this one assumption, the possible integer spacetime
dimensions are determined along with the possible levels $K$.
Perhaps not coincidentally, the allowed dimensions are precisely the ones
in which classical supersymmetry is possible. This is clearly a
complementary method to the approach
for determining $D$ followed in [\putref{ref63,ref62,ref65}].

Demanding eq.~(\puteqn{stdim-a}) guarantees
\hbox{spin-1} and \hbox{spin-1/2}
 superpartners at
$${\rm mass}^2= {\rm mass}^2({\rm vacuum}) + h(\phi^1_0)
= {\rm mass}^2({\rm vacuum}) +
(D-2)\times h(\phi^{K/4}_{K/4})\, .\eqno\eqnlabel{mass1}$$
{\it A priori} simply demanding the ratio
be integer is not
sufficient to guarantee spacetime supersymmetry. However, in the previous
subsections it proves to be; the masslessness of the
\hbox{(spin-1, spin-1/2) pair occurred automatically.}
\vskip 1cm
\centertext{Figure 3.1 Supersymmetry of Lowest Mass States of the
Fractional Open String}
\vskip .5cm
{\centertext{\underline{\hbox to 4in{$m^2({\rm spin}-1)\hfill =\hfill
m^2({\rm spin}-1/2)$}}}}\\
{\centertext{\hbox to 1in{\hfill}}}\\
{\centertext{\hbox to 4in{$h(\phi^1_0)$ $\big\Uparrow$\hfill
$(D-2)\times h(\phi^{K/4}_{K/4})$ $\big\Uparrow$}}}\\
{\centertext{\hbox to 1 in{\hfill}}}\\
{\centertext{\underline{\hbox to 4 in{\hfill $m^2({\rm vacuum})$\hfill}}}}

\vskip 1cm
In fractional superstrings, the primary field
$\phi^{K/4}_{K/4}\equiv\phi^{K/4}_{-K/4}$ for $K=4,\, 8,\, {\rm and~} 16$,
with related partition function
\hbox{$Z^{K/4}_{K/4}=\eta c^{K/2}_{K/2}$}, is
 viewed as the generalization of
$\phi^{1/2}_ {1/2}$ at $K=2$.
Are there any other parafermion operators at
additional levels-$K$ that could be used to transform the bosonic vacuum
into a massless fermionic vacuum
and bring about local spacetime supersymmetric models? The
answer is that by demanding masslessness of the (spin-1, spin-1/2)
pair,\footnote{Masslessness of
at least the left- and right-moving the spin-1 spacetime fields (whose
tensor product forms the massless spin-2 graviton in a closed string)
is of course required
for a consistent string theory.  Consistent two-dimensional field
theories with
$$\eqalign{
{\rm ~lowest~mass}&{\rm ~of~left-~(right-)moving~spacetime~spin-1~fields}
=\cr
& {\rm ~lowest~mass~of~left-~(right-)moving~spacetime~spin-1/2~fields}
\equiv M_{\rm min} > 0} $$
may exist (as we discuss below) but, the physical interpretation of such
models is not clear, (other than to say they would not be theories with
gravity.)}
there is clearly no other choice for $K<500$.

The proof is short. We do not assume first that the
massless spin-1 fields are a result of the $\phi^1_0$ fields.  Rather, the
necessity of choosing $\phi^1_0$
appears to be the result of the uniqueness of
$\phi^{K/4}_{K/4}$.
{\settabs 2\columns
\+ \cr}

{\it Proof}:
Assume we have a consistent (modular invariant) closed fractional
superstring theory
at level-$K$ with supersymmetry in $D$ dimensional spacetime,
($N=2$ for fractional type II theories and $N=1$ SUSY for
fractional heterotic).
Let the massless left (right) spin-1 field be
$(\phi^{j_1}_{m_1})^{\mu} \vert {\rm vacuum}>$. This requires that
$\phi^{j_1}_{m_1}$  have conformal dimension
$$h(\phi^{j_1}_{m_1})= c_{\rm eff}/24= (D-2) {K\over 8(K+2)}\,\,
.\eqno\eqnlabel{spin1cd}$$
Thus, the twist field $\phi^{j_2}_{m_2}$
 that produces the spinor vacuum along one of the $D-2$
transverse dimensions must have conformal dimension
$$h(\phi^{j_2}_{m_2})= {K\over 8(K+2)}\,\, . \eqno\eqnlabel{spin1half}$$
For $K<500$ the only primary fields with dimension $K\over 8(K+2)$
are the series
of $\phi^{K/4}_{K/4}$ for $K\in 2\Z$, and the accidental solutions
$\phi^2_0$ for $K=48$, $\phi^3_0$ for $K=96$,
and $\phi^{9/2}_{7/2}$ for $K=98$. With $m=0$, it is clear that the $K=48$
and $96$ fields could not be used to generate spacetime fermions. The
$K=98$ case could not be used because there is no
candidate field at that level whose
conformal dimension is a multiple of (\puteqn{spin1half})
(and thus no replacement for $\epsilon\equiv\phi^1_0$).
(A proof of the uniqueness of $\phi^{K/4}_{K/4}$ for all
$K$ is being prepared by G.C.)

Confirmation of $\phi^{K/4}_{K/4}$ as the spin-1/2 operator, though, does not
immediately lead one to conclude that $\phi^1_0$ is the only
possible
choice for producing massless boson fields. Table 3.3 shows alternative
fields at new levels-$K\not= 2,\, 4,\, 8,$ or $16$ whose conformal
dimension is one, two, or four times the conformal dimension of
$\phi^{K/4}_{K/4}$.
(Note that successful alternatives to
$\phi^1_0$ would lead to a  relationship between level and
spacetime dimension differing from eq.~(\puteqn{dvsk}).) However, nearly
all alternatives are of the form $\phi^{j>1}_0$ and we would expect that
modular invariant models using
$\phi^{j>1}_0$ to create massless bosons, would necessarily include
tachyonic  $(\phi^1_0)^{\mu}\vert {\rm vacuum}>$ states.
That is, (although we have not proven this yet), we do not believe valid GSO
projections
exist which can project away these tachyons while simultaneously
keeping the massless graviton and gravitino and giving modular invariance.
Further, the remaining fields on the list have $m\not= 0\pmod{K}$.
Each of these
would not have the correct fusion rules with itself, nor with
$\phi^{K/4}_{K/4}$ to be a spacetime boson.
\hfill\vfill\eject

\centertext{Table 3.3 Fields $\phi^{j_1}_{m_1}\neq\phi^1_0$
with Conformal Dimensions in Integer Ratio with
$h(\phi^{K/4}_{K/4})$}
{\settabs 5 \columns
\+\cr
\+ & {$K$} & {$\phi^j_m$} &
{$h(\phi^j_m)/h(\phi^{K/4}_{K/4})$}\cr
\+ & $\overline{\hbox to 9.3cm{\hfill}}$\cr
\+ & 12 & $\phi^2_0$ & \hskip 1.3cm 4\cr
\+ & 24 & $\phi^2_0$ & \hskip 1.3cm 2\cr
\+ &   & $\phi^3_0$ & \hskip 1.3cm 4\cr
\+ & 36 & $\phi^7_6$ & \hskip 1.3cm 4\cr
\+ & 40 & $\phi^4_0$ & \hskip 1.3cm 4\cr
\+ & 48 & $\phi^2_0$ & \hskip 1.3cm 1\cr
\+ &    & $\phi^3_0$ & \hskip 1.3cm 2\cr
\+ & 60 & $\phi^5_0$ & \hskip 1.3cm 4\cr
\+ & 80 & $\phi^4_0$ & \hskip 1.3cm 2\cr
\+ & 84 & $\phi^6_0$ & \hskip 1.3cm 4\cr
\+ & 96 & $\phi^3_0$ & \hskip 1.3cm 1\cr
\+ & 112 & $\phi^7_0$ & \hskip 1.3cm 4\cr
\+ & 120 & $\phi^5_0$ & \hskip 1.3cm 2\cr
\+ & $\vdots$ & $\vdots$ & \hskip 1.3cm $\vdots$\cr
\+\cr}

Lastly, we want to consider the possibility that there is meaning
to (non-stringy) two-dimensional field theories that
contain neither supergravity nor even gravity. Instead let a
model of this type contain only a global supersymmetry. The
lowest mass spin-1
and spin-1/2
left- or right-moving fields,
$(\phi^1_0)^{\mu}\vert {\rm vacuum}>$ and
$(\phi^{j_3}_{m_3})^{D-2}\vert {\rm vacuum}>$, respectively,
would be related by
$${\rm mass}^2({\rm vacuum}) + h(\phi^1_0) = {\rm mass}^2({\rm vacuum}) +
(D-2)\times h(\phi^{j_3}_{m_3})\, .\eqno\eqnlabel{massive}$$
In parafermion CFT's there
is only a very small number (12) of potential candidates
for $\phi^{j_3}_{m_3}$.
(Like $\phi^{K/4}_{K/4}$ these twelve are all of the form
$\phi^{j_3}_{\pm j_3}$.)  We are
able to reduce the number of candidates down to this finite number very
quickly by proving no possible candidate could have $j_3>10$, independent of
the level-$K$. We demonstrate this as follows:

Any potential level-$K$ candidate $\phi^{j_3}_{m_3}$ must satisfy the
condition of
$${K\over K+2}\left[ j_3(j_3+1) -2\right] \leq (m_3)^2 \leq (j_3)^2
\leq K^2/4 \,\, .\eqno\eqnlabel{constraint}$$
By parafermion equivalences (1.1),
$\vert m\vert\leq j\leq K/2$ can be required for any level-$K$
fields.
The other  half of the inequality,
${K\over K+2}\left[ j_3(j_3+1) -2 \right]\leq (m_3)^2$
results from the weak requirement that the
conformal dimension of the candidate spin-1/2 field, $\phi^{j_3}_{m_3}$,
creating the fermion ground state
along one spacetime direction cannot be greater than the conformal
dimension of $\phi^1_0$, {\it i.e.},
$h(\phi^{j_3}_{m_3})\leq h(\phi^1_0)$.

 From eq.~(\puteqn{constraint}),
we can determine both the minimum and maximum values
of $K$, for a given $j$, (independent of $m$).
These limits are \hbox{$K_{\rm min}= 2j_3$} and
\hbox{$K_{\rm max}= {\rm ~int}( {2(j_3)^2\over j_3-2})$.}
Thus the number of different levels-$K$ that can correspond to the field
$\phi^{j_3}_{m_3}$ is int$({{5j_3-2}\over {j_3-2}})$.
This number quickly decreases to six as $j_3$ increases to $10$ and
equals $5$ for $j_3 > 10$.
For a given $j_3$, we will express the levels-$K$ under consideration as
\hbox{$K_i= 2j_3 +i$}.
Also, we find that from \hbox{$K_{\rm min}=2j_3$},
the weak constraint on $m_3$
implies that we need only consider $\phi^{j_3}_{m_3=\pm j_3}$ fields.

Thus, our search reduces to finding fields $\phi^{j_3}_{j_3}$ whose
conformal dimensions satisfy
$${{h(\phi^1_0)\over h(\phi^{j_3}_{\pm j_3})} =
  {{2\over K_i +2}\over {j_3(j_3+1)\over K_i+2}
- {(j_3)^2 \over K_i}}\in\Z} \,\, .
  \eqno\eqnlabel{ratioj}$$
It is easily shown that there are no solutions to
eq.~(\puteqn{ratioj}) for $i=0 {\rm ~~to~~} 4$ and $j_3>10$.
As a result, we have reduced our search for possible alternative sources of
fermionic
ground states to only $\phi^{j_3}_{\pm j_3}$ with $0<j_3\leq 10$.  Within this
range of $j_3$, a computer
search reveals the following complete set of $\phi^{j_3}_{\pm j_3}$
fields that
obey eq.~(\puteqn{ratioj}), as shown in Table 3.4.
\hfill\vfill\eject
\centertext{Table 3.4 Potential Alternatives, $\phi^{j_3}_{m_3}$,
to $\phi^{K/4}_{K/4}$ for Spin Fields}
{\settabs 7 \columns
\+ \cr
\+ {$j_3$}     & {$\pm m_3$}     & {$K$}
& {$i$} & {$h(\phi^1_0)$} & {$h(\phi^{j_3}_{m_3})$}
& {$D= {h(\phi^1_0)\over h(\phi^{j_3}_{m_3})} +2$}\cr
\+ $\overline{\hbox to 16.8cm{\hfill}}$\cr
\+ 1/2 & 1/2 & 2  & 1  & 1/2 & 1/16 & \hskip 1.1cm 10 **\cr
\+     &     & 3  & 2  & 2/5 & 1/15 & \hskip 1.1cm 8\cr
\+     &     & 5  & 4  & 2/7 & 2/35 & \hskip 1.1cm 7 \cr
\+ \cr
\+ 1   & 1   & 3  & 1  & 2/5 & 1/15 & \hskip 1.1cm 8 \cr
\+     &     & 4  & 2  & 1/3 & 1/12 & \hskip 1.1cm 6 **\cr
\+     &     & 6  & 4  & 1/4 & 1/12 & \hskip 1.1cm 5 \cr
\+ \cr
\+ 3/2 & 3/2 & 9  & 6  & 2/11 & 1/11 & \hskip 1.1cm 4 \cr
\+ \cr
\+ 2   & 2   & 5  & 1  & 2/7 & 2/35 & \hskip 1.1cm 7\cr
\+     &     & 6  & 2  & 1/4 & 1/12 & \hskip 1.1cm 5 \cr
\+     &     & 8  & 4  & 1/5 & 1/10 & \hskip 1.1cm 4 **\cr
\+ \cr
\+ 5/2 & 5/2 & 25  & 20 & 2/27 & 2/27 & \hskip 1.1cm 3 \cr
\+ \cr
\+ 3   & 3   & 9  & 3  & 2/11 & 1/11 & \hskip 1.1cm 4 \cr
\+     &     & 18 & 12 & 1/10 & 1/10 & \hskip 1.1cm 3 \cr
\+ \cr
\+ 4   & 4   & 16 & 8  & 1/9  & 1/9  & \hskip 1.1cm 3 **\cr
\+ \cr
\+ 6   & 6   & 18 & 6  & 1/10 & 1/10 & \hskip 1.1cm 3 \cr
\+ \cr
\+ 10  & 10  & 25 & 5  & 2/27 & 2/27 & \hskip 1.1cm 3 \cr
\+ \cr}
\hve

The sets of solutions for
$j_3= {1\over 2},\, 1,\, {\rm and~} 2$
are related. The existence of a set of solutions,
\hbox{$\{i=1,\, 2,\, {\rm and~} 4\}$}, for any
one of these $j_3$ implies identical sets \hbox{$\{ i\}$} for the remaining
two
$j_3$ as well.
The known $\phi^{K/4}_{K/4}$ solutions (marked with a **)
correspond to the \hbox{$i=1,\, 2,\, {\rm and~} 4$} elements in the
\hbox{$j_3= {1\over 2},\, 1,\, {\rm and~}2$} sets respectively. Whether this
pattern
suggests anything about the additional related $\phi^{j_3}_{\pm j_3}$ in
these
sets, other than explaining their appearance in the above table, remains
to be seen.

The set of distinct, physically relevant fields can be further reduced.
There is
a redundancy in the above list. Among this list,
for all but the standard $\phi^{K/4}_{K/4}$ solutions,
there are two fields at each level, with distinct values of $j_3$.
These pairs are related by the field equivalences (\puteqn{phidents}):
\subon
$$\eqalignno{
\hbox to 1cm{$\phi^{1/2}_{\pm 1/2}$\hfill}
&\equiv\hbox to 1cm{$\phi^1_{\mp 1}$\hfill} {\rm~~at~level-} K=3
&\eqnlabel{id-a}\cr
\hbox to 1cm{$\phi^{1/2}_{\pm 1/2}$\hfill}
&\equiv\hbox to 1cm{$\phi^2_{\mp 2}$\hfill} {\rm~~at~level-} K=5
&\eqnlabel{id-b}\cr
\hbox to 1cm{$\phi^1_{\pm 1}$\hfill}
&\equiv\hbox to 1cm{$\phi^2_{\mp 2}$\hfill} {\rm~~at~level-} K=6
&\eqnlabel{id-c}\cr
\hbox to 1cm{$\phi^{3/2}_{\pm 3/2}$\hfill}
&\equiv\hbox to 1cm{$\phi^3_{\mp 3}$\hfill} {\rm~~at~level-} K=9
&\eqnlabel{id-d}\cr
\hbox to 1cm{$\phi^3_{\pm 3}$\hfill}
&\equiv\hbox to 1cm{$\phi^6_{\mp 6}$\hfill} {\rm~~at~level-} K=18
&\eqnlabel{id-e}\cr
\hbox to 1cm{$\phi^{5/2}_{\pm 5/2}$\hfill}
&\equiv\hbox to 1cm{$\phi^{10}_{\mp 10}$\hfill} {\rm~~at~level-} K= 25\,\, .
&\eqnlabel{id-f}\cr}$$
\suboff
Because $\phi^j_m$ and $\phi^j_{-m}$ have identical  partition functions
and $\phi^j_{-m}\equiv(\phi^j_m)^{\dagger}$
we can reduce the number of possible alternate fields in half, down to
six. (Note that we have not been distinguishing between $\pm$ on $m$
anyway.)

If we want models with {\it minimal} super Yang-Mills Lagrangians
we can reduce the number of the fields to investigate further.
Such theories
exist classically only in $D_{SUSY}=10,\, 6,\, 4,\, 3,\, ({\rm and~} 2)$
spacetime. Thus we can consider only
those $\phi^{j_3}_{\pm j_3}$ in the above list that have integer
conformal dimension
ratios of $D_{SUSY}-2 = h(\phi^1_0)/h(\phi^{j_3}_{\pm j_3})=
8,\, 4,\, 2,\, {\rm and~}
1$.
This would reduce the fields to consider to just the three
new possibilities for $D= 4$, and $3$ since there
are no new additional for $D=10$ or $6$.

\sectionnumstyle{blank}
\section{\bf 3.3: Ghost Conformal Anomalies}
\sectionnumstyle{arabic}
\sectionnum=3
\subsectionnumstyle{arabic}\subsectionnum=3
\equationnum=0

Whatever the conformal dimension $h$ of a general local (holomorphic)
symmetry
current $U(z)$ is, the
conformal dimension of its associated ghost $\gamma (z)$ must be $(1-h)$.
This is most
easily seen from the BRST approach. Recall that the BRST current $j(z)$ must
have
conformal dimension 1, if the nilpotent BRST charge, $Q$, defined by
$${Q= {1\over 2\pi i}\oint dz\, j(z)}\,\, ,\eqno\eqnlabel{brst}$$
is to  be conformally invariant.
The symmetry current $U$ appears in the BRST current via the term
$U\gamma$, thus requiring $h(\gamma)= 1- h(U)$.

The standard ghost action for the local Virasoro current
(energy--momentum tensor) $T_{\rm em}$,
with conformal dimension $h(T_{\rm em})=2$, is
$$S_{\rm Virasoro~ghosts}=
{ 1\over \pi}\int d^2z (b\bar\partial c + \bar b\partial\bar c)
\,\, , \eqno\eqnlabel{ga1}$$
where $c$ $(\bar c)$ is the (anti)holomorphic
ghost and $b$  $(\bar b)$ the related (anti)holomorphic antighost.
 From this action we get the conformal ghost terms in the total
energy--momentum tensor
(henceforth we only consider the holomorphic sector)
$$T_{\rm Virasoro~ghosts}(z) = c\partial b + 2(\partial c)b\,\, .
\eqno\eqnlabel{vemgh}$$
Thus, the  conformal dimension, $h(b)$,  of the antighost $b$ is
$$h(b)= 1-h(c)=h(T_{V.G.}=2)\,\, .\eqno\eqnlabel{hbem}$$
Together the ghost system $(c$, $b)$ has the well-known conformal anomaly
$c({\rm ghosts})=-26$, as apparent from the O.P.E. of $T_{\rm V.G.}$
with itself:
$$T_{V.G.}(z)T_{V.G.}(w)= {1\over 2}{-26\over (z-w)^4} +
{2\over (z-w)^2} T_{V.G.}(z) + {1\over z-w}\partial T_{V.G.}(z) + \cdots
\eqno\eqnlabel{vemgope}$$

The standard holomorphic supersymmetry current, $J$,
for $K=2$ (with conformal dimension $h(J)=3/2$)
has the associated ghost/antighost
combination $\gamma(z)$ and $\beta(z)$ with conformal dimensions
$h(\gamma)= 1-h(J)= -{1\over 2}$ and $h(\beta)= 1-h(\gamma)=
h(J)={3\over2}$,
respectively.
The total conformal anomaly of the ghost system is $c(\beta,\gamma) =+11$.
This is similarly found from the supersymmetry ghost action
$$S_{\rm SUSY~ghosts}= {1\over \pi}\int d^2z \beta\bar\partial\gamma\,\, ,
\eqno\eqnlabel{susyghaction}$$
and associated ghost energy momentum tensor
$$T_{\rm SUSY~ghosts}= -{1\over 2}\gamma\partial\beta
-{3\over 2}(\partial\gamma) \beta\,\, . \eqno\eqnlabel{emsusygh}$$

These examples demonstrate the generic properties of free ghost actions,
energy-momen-
tum tensors and conformal anomalies for commuting or
anticommuting ghost/anti-ghost
systems $(\gamma,\beta)$ resulting from a
local fermionic or bosonic, respectively, symmetry current $U$ with
conformal dimension $h(U)$:
\subon
$$\eqalignno{S &= {1\over \pi}\int d^2 z \,\beta\bar\partial\gamma
&\eqnlabel{gen-a}\cr
T_{\beta,\gamma} &= -h(\beta)\beta\partial\gamma
+ h(\gamma)(\partial\beta)\gamma
&\eqnlabel{gen-b}\cr
c(\beta,\gamma) &= 2\delta\left( -6h(\gamma)h(\beta) +1\right)
&\eqnlabel{gen-c}\cr
                &= 2\delta\left( -6h(\gamma)(1-h(\gamma)) +1\right)
\,\, , &\eqnlabel{gen-d}\cr}$$
\suboff $\delta$ equals $-1\, (+1)$ a for
bosonic (fermionic) current. Thus, $c$ appears as a quadratic function
of $h(\gamma)$.

This pattern does not (as one might expect)
continue for the presumed ghosts, $\gamma$, associated with the
fractional supersymmetry current, $J^K_{\rm FSC}$, with conformal dimension
$h(J^K_{\rm FSC})= 1+{2\over K+2}$.
This can be determined even without knowledge of
the exact ghost system. For a conformally invariant theory,
the total conformal anomaly from the fractional spin ghosts must be
$$\eqalignno{c({\rm fract.~ghosts})
&= -\left( c({\rm matter}) + c({\rm Virasoro})\right)\cr
&= -(D\times {3K\over K+2}) + 26\cr
&= -(2+ {16\over K})({3K\over K+2}) +26\,\, .
&\eqnlabel{cghostis}\cr}$$
Now we express (\puteqn{cghostis}) as a function of
$h(\gamma)=  1-h(J^K_{\rm FSC})$.
The result is,
$$c({\rm fract.~ghosts})= 20 +18h(\gamma)\,\, .\eqno\eqnlabel{ghlinear}$$
We have expressed the ghost anomaly in terms of the dimension of the
$\gamma$ ghost
because this is the only one we know with certainty (from the
BRST argument).  Due to the nonlocality (Abelian or otherwise) of the
fractional currents, it is difficult to
determine
what the complete ghost system is and what the related conformal dimensions
are. The non-quadratic form of the conformal anomaly strongly suggests that
this is not a free ghost system.

Recently, there have been hints that the assumed
$c({\rm matter})= D\times 3K/(K+2)$ contribution
may be incorrect when the effects of the projection terms are correctly taken
into account.\markup{[\putref{ref63}]}  It has been pointed out that the
effective central charge
for the light-cone degrees of freedom may actually be
$$c_{\rm eff}=(D-2)\times{3\over 2}\,\, .\eqno\eqnlabel{ceffnew}$$
(For $K>2$ this is obviously less than $(D-2){3K\over K+2}$.)  When
we add this to
the anomaly contribution from the longitudinal modes, we get
$$c({\rm matter})_{\rm new}
={24\over K}+{6K\over K+2}\,\, .\eqno\eqnlabel{cmatnew}$$
Using this new value of $c({\rm matter})$ to determine $c({\rm fract.~ghost})$
results in
$$c({\rm fract.~ghost})_{\rm new}= -h(\gamma)\times{2(K-1)(5K+12)\over K}
\eqno\eqnlabel{cmatnewh}$$
in place of eq.~(\puteqn{ghlinear}).
If this is, indeed, the correct $c({\rm fract.~ghost})$
{\it and if we are justified in expecting that}
$c({\rm fract.~ghost})$ {\it should depend upon} $K$ {\it in a continuous way},
then the coefficient of
$h(\gamma)$ in (\puteqn{cmatnewh}) should reveal something about the
complex ghost action.

\sectionnumstyle{blank}
\section{\bf Section IV: Conclusions}
\subsectionnumstyle{blank}

A viable and consistent generalization of the superstring would be an
important
development. Our work has shown that the fractional
superstring has many intriguing features that merit further study. The
partition functions for these theories are found to have simple
origins when derived systematically
through the factorization approach of Gepner and Qiu.
Furthermore, using this affine/theta-function factorization of the
parafermion partition  functions, we have related the $A_K$--sector
containing the graviton and
gravitino with the massive sectors, $B_K$ and $C_K$. A bosonic/fermionic
interpretation
of the $B_K$--subsectors was given.  Apparent ``self-cancellation'' of the
$C_K$--sector was shown, the meaning of which is under further
investigation by G.~C. A possible GSO projection was found, adding
hope that the partition functions have a natural physical
interpretation. Nevertheless, fundamental questions
remain concerning the ghost system and current algebra, which prevent a
definite conclusion as to whether or not
these are consistent theories. However,
even if  the theories
are ultimately shown to be inconsistent, we believe that this program will at
least provide interesting identities and new insight into
the one case we know is consistent, $K=2$. In other words, viewed in this
more general context, we may better understand what is special about
the usual superstring.

Acknowledgements:
G. C. and P.R. wish to thank S.-H. Tye, P. Argyres, J. Grochocinski, and
Costas
Efthimiou for their hospitality
and many useful discussions during a recent visit to Cornell, and likewise
Keith Dienes for helpful discussions.

\sectionnumstyle{blank}
{\bf\section{References:}}

\begin{putreferences}

\reference{ref9d}{Alvarez-Gaum\' e, L. et al.:
Comm. Math. Phys. {\bf 106} 1 (1986)}

\reference{ref9b}{Antoniadis, I. et al.:
Nucl. Phys. {\bf B289} 87 (1987)}

\reference{ref15a} {Antoniadis, I., and Bachas, C.:
Nucl. Phys. {\bf B278}  343 (1986)}

\reference{ref9c}{Antoniadis, I. and Bachas, C.:
Nucl. Phys. {\bf B298} 87 (1987)}

\reference{ref3}{Ardalan, F., and Mansouri, F.:
 Phys. Rev. {\bf D9} 3341 (1974);
Phys. Rev. Lett. {\bf 56} 2456 (1986);
Phys. Lett. {\bf B176} 99 (1986)}

\reference{ref61}{Argyres, P. et al.:
Phys. Lett. {\bf B235} (1991)}

\reference{ref62}{Argyres, P., and Tye, S.H.:
Phys. Rev. Lett. {\bf 67} 3339 (1991)}

\reference{argyres91f}{Argyres, P., et al.:
Nucl. Phys. {\bf B367} 217 (1991)}

\reference{ref64}{Argyres, P. et al.
Nucl. Phys. {\bf B391} 409 (1993)}

\reference{ref65}{Argyres, P. et al.,
Comm. Math. Phys. {\bf 154} 471 (1993)}

\reference{ref12} {Argyres, P. et al.,
Phys. Rev. {\bf D46} 4533 (1992).
This paper was brought to our attention while our work was
in progress and parallels several of our lines
of thought.}

\reference{ref14d}{Bhattacharyya, A., et. al.,
Mod. Phys. Lett. {\bf A4} 1121 (1989):
Phys. Lett. {\bf B224} 384 (1989)}

\reference{ref19}{Capelli, A. et al.:
Comm. Math. Phys. {\bf 113} 1 (1987).}

\reference{ref16c}{Delius, G.: ITP-SB-12 (1989)}

\reference{ref10}{Cleaver, G.,  and Lewellen, D.:
Phys. Rev. {\bf D300} 354 (1993).}

\reference{thesis}{Cleaver, G.:
Ka\v c-Moody Algebras and String Theory
(Ph.D. Thesis) 1993.}

\reference{cd}{Cleaver, G., and Dienes, K.:
``Internal Projection Operators for
Fractional Superstrings'', OHSTPY-HEP-T-93-022,
McGill/93-43. To Appear.}

\reference{ad93}{Argyres, P., and Dienes, K.:
Phys. Rev. Lett. {\bf 71} 819 (1993).}

\reference{ref68}{Dienes, K.:
Nucl. Phys. {\bf B413} 103 (1994).}

\reference{ref63}{Dienes, K., and Tye, S.H.:
Nucl. Phys. {\bf B376} 297 (1992)}

\reference{ref5}{Frampton, P., and Ubriaco, M.:
Phys. Rev. {\bf D38} 1341 (1988)}

\reference{new1}{Fuchs, J. et al.: Int. Jour. of Mod. Phys.
{\bf A7} 2245 (1992)}

\reference{ref17c} {Gato-Rivera, B., and Schellekens, A.N.:
Nucl. Phys. {\bf B353} 519 (1991)}

\reference{ref7}{Gepner, D.,  and Qiu, Z.:
Nucl. Phys. {\bf B285} 423 (1987)}

\reference{ref13} {Green,  H.S.:
Phys. Rev. {\bf 90} 270 (1953)}

\reference{ref1a}{Green, M., Schwarz, J., and Witten, E.:
Superstring Theory {\bf Vols. I \& II}. Cambridge:
University Press 1987}

\reference{ref15b}{Hama, M. et al.: RUP1 (1992)}

\reference{ref1d}{Hull, C.:
A Review of W Strings. CTP TAMU-30/92 (1992).}

\reference{ref8b}{Ka\v c, V.,  and Peterson, D.:
Bull. AMS {\bf 3} 1057 (1980);
Adv. Math. {\bf 53} 125 (1984)}

\reference{ref8a}{Ka\v c, V.:
Adv. Math. {\bf 35} 264 (1980)}

\reference{ref1c}{Kaku, M.: Strings, Conformal Fields and Topology.
New York: Springer-Verlag 1991}

\reference{ref9a} {Kawai, H. et al.:
Nucl. Phys.
{\bf B288} 1 (1987) 1}

\reference{ref16a} {Li, K.,  and Warner, N.:
Phys. Lett. {\bf B211} 101 (1988)}

\reference{ref14a} {Mansouri, F.,  and Wu, X.:
Mod. Phys. Lett. {\bf A2} 215 (1987);
Phys. Lett. {\bf B203} 417 (1988);
J. Math. Phys. {\bf 30} (1989) 892}

\reference{ref1b}{Schellekens, B., ed.: Superstring Construction.
Amsterdam: North Holland 1989}

\reference{ref17b} {Schellekens, A.N.:
Phys. Lett. {\bf 244} 255 (1990)}

\reference{ref17a} {Schellekens, A.N., and Yankielowicz, S.:
Nucl. Phys. {\bf B327} 3 (1989)}

\reference{new2} {Verstegen, D.: Phys. Lett. {\bf B326}
567 (1990)}

\reference{ref11} {Wilczek, F., ed.:
Fractional Statistics and Anyon Superconductivity.
Singapore: World Scientific 1990}

\reference{ref4}{Zamolodchikov, A., and Fateev, V.:
Sov. Phys. JETP {\bf 62} 215
(1985); Teor. Mat. Phys. {\bf 71} 163 (1987)}

\end{putreferences}

\bye